\documentclass[aps,showpacs,twocolumn,superscriptaddress,preprintnumbers,bibnotes]{revtex4}
\usepackage{ifpdf}
\usepackage{graphicx}
\usepackage{color}
\usepackage{dcolumn}
\usepackage{bm}
\usepackage{mathrsfs}
\usepackage{amsfonts}
\usepackage{amssymb}
\usepackage{amsmath}
\usepackage{dsfont}
\usepackage{longtable}
\usepackage{fixmath}
\usepackage{upgreek}
\usepackage[bookmarks,colorlinks]{hyperref}
\hypersetup{colorlinks=true,%
            citecolor=blue,%
            filecolor=black,%
            linkcolor=blue,%
            urlcolor=blue}
\def\t#1{\tilde{#1}}
\def\wh#1{\widehat{#1}}
\def\h#1{\hat{#1}}
\def\b#1{\bar{#1}}

\def\ul#1{\,\underline{\!#1\!}\,}

\def\uul#1{\,\underline{\phantom{{}_{\!}}\!#1\!}\,}

\def\Sc#1{\textsc{#1}}

\DeclareMathOperator{\re}{Re}
\DeclareMathOperator{\im}{Im}
\DeclareMathOperator{\e}{e}
\DeclareMathOperator{\rd}{d\!}
\DeclareMathOperator{\sgn}{sgn}
\begin{document}
\ifx\href\undefined\else\hypersetup{linktocpage=true}\fi
\bibliographystyle{apsrev}

\preprint{Preprint number: ITP-UU-2008/23}

\title{Comment on ``Breakdown of the Luttinger sum rule within the Mott-Hubbard insulator'', by J. Kokalj and P. Prelov\v{s}ek, \href{http://link.aps.org/doi/10.1103/PhysRevB.78.153103}{\emph{Phys.~Rev.} B~{\bf 78}, 153103 (2008)}}

\author{Behnam Farid}
\affiliation{Institute for Theoretical Physics, Department of Physics and Astronomy, University of Utrecht, \\
Leuvenlaan 4, 3584 CE Utrecht, The Netherlands }
\email{B.Farid@phys.uu.nl}
\author{Alexei M. Tsvelik}
\affiliation{Department of Physics, Brookhaven National Laboratory, Upton, New York 11973-5000, USA}
\email{tsvelik@bnl.gov}

\date{\today}

\begin{abstract}
On the basis of an analysis of the numerical results corresponding to the half-filled one-dimensional $t$-$t'$-$V$ model on some finite lattices, Kokalj and Prelov\v{s}ek (KP) have in a recent paper [\emph{Phys. Rev.} B~{\bf 78}, 153103 (2008)] concluded that the Luttinger theorem does not apply for the Mott-Hubbard (MH) insulating phase of this model (corresponding to the region $V\gg t$) in the thermodynamic limit; KP even suggested, incorrectly, that failure of the Luttinger theorem were apparent for a half-filled finite system consisting of $N=26$ lattice sites. By employing a simple model for the self-energy $\Sigma$ of a MH state, we show that the finite-size-scaling approach of the type utilised by KP is \emph{not} reliable for the system sizes considered by KP: although the Luttinger theorem is \emph{exactly} satisfied for our model, by employing the latter finite-size-scaling approach \emph{and} the same lattice sizes as employed by KP, we obtain comparable quantitative amounts of ``violation'' of the Luttinger theorem as reported by KP for various values of $V/t$ in the range $[3.6, 10]$. On the basis of the equivalence of the model under consideration (at half-filling and for small values of $t'/t$ --- ideally, for $t'=0$) and the $XXZ$ spin-chain Hamiltonian for SU(2) spins, we further show that for $V$ greater than a critical value $V_{\rm c}(t,t')$, where $V_{\rm c}(t,0) = 2t$, the system under consideration has a charge-density-wave (CDW) ground state \emph{in the thermodynamic limit}, corresponding to a doubling of the unit cell in comparison with that specific to the underlying lattice. Although this ground state is also insulating, its spectral gap is due to the broken translational symmetry; it is \emph{not} a correlation-induced MH gap. The Luttinger theorem (as generalised by Luttinger for inhomogeneous ground states) is therefore \emph{a priori} valid for this broken-symmetry ground state. This fact rigorously establishes that the conclusion by KP is indeed erroneous. Finally, we present a heuristic argument due to Volovik that sheds light on the mechanism underlying the robustness of the Luttinger theorem. In an extensive appendix, we present the details of the calculation of the single-particle Green function of the broken-symmetry ground state of the model under consideration (for $V \downarrow V_{\rm c}$) by means of bosonization and in terms of the form factors of a class of soliton-generating non-local fields pertaining to the quantum sine-Gordon Hamiltonian, as determined by Lukyanov and Zamolodchikov (LZ); the latter fields can be shown to represent the slowly-varying chiral fields of our bosonized problem.
\end{abstract}


\maketitle

{\footnotesize{\tableofcontents}}
\section{Introduction}
\label{s1}

Considering for simplicity a system of spin-less electrons, for the uniform $N_{\rm e}$-particle ground state (GS) of this system one defines the Luttinger number
\begin{equation}
N_{\Sc l} \doteq \sum_{\bm k} \Theta(G({\bm k};\mu)), \label{e1}
\end{equation}
where $\Theta(x)$ is the unit-step function, $G({\bm k};\varepsilon)$ the zero-temperature limit of the thermal single-particle Green function in the grand-canonical ensemble in which the mean value of the number of particles is equal to $N_{\rm e}$, and $\mu$ the zero-temperature limit of the associated chemical potential (see later). Thus $G({\bm k};\varepsilon)$ corresponds to the $N_{\rm e}$-particle \emph{GS} of the Hamiltonian $\wh{H}$ of the system. The summation with respect to ${\bm k}$ in Eq.~(\ref{e1}) is over the entire available wave-vector space; for systems defined on a Bravais lattice, the latter region consists of the corresponding first Brillouin zone ($\mathrm{1BZ}$). The Luttinger theorem under consideration states that \cite{LW60,JML60,IED03,AMT03,ET05,KRT06,BF07a,BF07b}:
\begin{equation}
N_{\Sc l} = N_{\rm e}. \label{e2}
\end{equation}
There are some aspects to be taken into account (in particular in dealing with finite systems) concerning the definition of $\Theta(x)$, specifically at $x=0$, which have been elaborated on in Ref.~\cite{BF07a} and which we shall not discuss here.

Inspecting the proof of the Luttinger theorem \cite{BF07a}, one concludes that the validity of this theorem is dependent on the equality of the value of the chemical potential $\mu$ on the right-hand side (RHS) Eq.~(\ref{e1}) with that of the thermodynamic variable $\b{\mu}$ specific to the grand-canonical ensemble to which $G({\bm k};\varepsilon)$ corresponds. With $E_{M;0}$ denoting the energy of the $M$-particle GS of $\wh{H}$, one readily verifies that the implicit dependence of $G({\bm k};\varepsilon)$ on the thermodynamic variable $\b{\mu}$ is \emph{solely} through the requirements
\begin{equation}
E_{N_{\rm e};0} - E_{N_{\rm e}-1;0} < \b{\mu} < E_{N_{\rm e}+1;0} - E_{N_{\rm e};0}. \label{e3}
\end{equation}
For \emph{metallic} $N_{\rm e}$-particle GSs, where $N_{\rm e}$ is necessarily macroscopically large, the deviation of $E_{N_{\rm e}+1;0}-E_{N_{\rm e};0}$ from $E_{N_{\rm e};0}-E_{N_{\rm e}-1;0}$ is microscopically small, of the order of $1/N_{\rm e}$ (see Appendix B in Ref.~\cite{BF07a}). For these states, the value of $\b{\mu}$ is therefore up to an infinitesimal correction uniquely determined and is equal to $E_{N_{\rm e}+1;0}-E_{N_{\rm e};0}$. The latter quantity coincides, up to an infinitesimal correction, with the zero-temperature \emph{limit} of the chemical potential $\mu_{\beta}$, where $\beta$ is the inverse temperature, satisfying the equation of state corresponding to $N_{\rm e}$ as the mean value of the number of particles in the grand-canonical ensemble under consideration. Consequently, neglecting infinitesimal corrections, for metallic GSs the value of the $\mu$ on the RHS of Eq.~(\ref{e1}) is bound to be equal to the zero-temperature limit of $\mu_{\beta}$.

For \emph{insulating} $N_{\rm e}$-particle GSs, $E_{N_{\rm e}+1;0}-E_{N_{\rm e};0}$ deviates from $E_{N_{\rm e};0}-E_{N_{\rm e}-1;0}$ by a finite amount, equal to the magnitude of the fundamental spectral gap. Since for these states, and for $\b{\mu}$ satisfying the inequalities in Eq.~(\ref{e3}), the value of $G({\bm k};\varepsilon)$, $\forall {\bm k},\varepsilon$, is independent of the precise value of $\b{\mu}$, one naturally concludes that the Luttinger theorem in Eq.~(\ref{e2}) should apply for \emph{all} values of $\mu$ satisfying the same inequalities as $\b{\mu}$. That this is not necessarily the case, was first discovered by Rosch \cite{AR06} in his study of a simple model of a MH insulator.

In Ref.~\cite{BF07a} (see also Ref.~\cite{BF07b}) it is unequivocally demonstrated that the possibility of the breakdown of the Luttinger theorem for MH insulators has its root in the possibility of the existence of a false \cite{Note1} zero-temperature limit of a sum over Matsubara frequencies, and that this false limit is eliminated through identifying the $\mu$ on the RHS of Eq.~(\ref{e1}) with the zero-temperature limit of $\mu_{\beta}$. Thus, insofar as the $\mu$ on the RHS of Eq.~(\ref{e1}) is concerned, insulating $N_{\rm e}$-particle GSs are to be dealt with in exactly the same way as their metallic counterparts (see the previous paragraph). In this connection, it is important to realise that from the standpoint of the Lehmann representation of the thermal single-particle Green function $\mathscr{G}$ (of which $G$ is the zero-temperature limit), metallic and insulating systems are fundamentally similar at non-vanishing temperatures \cite{BF07a}.

We should emphasise that were it not for the possibility of the existence of the above-mentioned false limit, the Luttinger theorem would apply for \emph{all} values of $\mu$ satisfying the same inequalities as $\b{\mu}$. Neither Stanescu, Phillips and Choy \cite{SPC07} nor KP \cite{KP08} appear to have appreciated this fundamental aspect.

\section{The $t$-$t'$-$V$ model in one space dimension}
\label{s2}

The system considered by KP \cite{KP08} is defined on a regular one-dimensional lattice and is described by the Hamiltonian
\begin{equation}
\wh{H} = \sum_{j=1}^{N} \big[ -(t\, \h{c}_{j}^{\dag} \h{c}_{j+1} + t'\, \h{c}_{j}^{\dag} \h{c}_{j+2}) + \mathrm{H.c.} + V\, \h{n}_j \h{n}_{j+1}\big], \label{e4}
\end{equation}
where
\begin{equation}
\h{n}_{j} \doteq \h{c}_j^{\dag} \h{c}_j \label{e5}
\end{equation}
is the occupation-number operator of site $j$. In Eq.~(\ref{e4}), the site with index $j=N+j'$ is identified with that with index $j=j'$, where $j'=1, 2$. Unless we indicate otherwise, in what follows we shall use the lattice constant $a$ as the unit of length so that the set of ${\bm k}$ (hereafter $k$) points relevant to the sum in Eq.~(\ref{e1}) is a subset of the interval $(-\pi,\pi]$ (the points $-\pi$ and $\pi$ are identified along the $k$ axis). Using the periodic boundary condition, for the lattice under consideration, and with $N$ even, one has:
\begin{equation}
\mathrm{1BZ} \doteq \Big\{ \frac{2\pi l}{N}\,\|\, l = -\frac{N}{2}+1, -\frac{N}{2} +2 , \dots, \frac{N}{2} \Big\}. \label{e6}
\end{equation}
For the half-filled case, considered by KP \cite{KP08},
\begin{equation}
N_{\rm e} = \frac{1}{2} N. \label{e7}
\end{equation}

\subsection{Discussion of the numerical results by KP concerning finite systems}
\label{s2a}

The finite systems for which KP \cite{KP08} have carried out explicit numerical calculations, correspond to
\begin{equation}
N = 14, 18, 22, 26, 30, \label{e8}
\end{equation}
accommodating respectively $N_{\rm e} = 7, 9, 11, 13, 15$ spin-less electrons. For these systems, KP \cite{KP08} employed exact diagonalization of $\wh{H}$, making use of the Lanczos algorithm. We shall later show that in the thermodynamic limit, and for $V > V_{\rm c}(t,t')$, the half-filled GS of the $\wh{H}$ in Eq.~(\ref{e4}) is not uniform, but a non-uniform CDW state \cite{HSSB82} (see Sec.~III herein). This broken-symmetry GS \emph{cannot} be reached by extrapolating the finite-size numerical results, which by necessity all correspond to uniform GSs.

In Fig.~2 of Ref.~\cite{KP08}, $t (1 + V/t)\, \re[G(k;\mu)]$ is presented as function of the discrete values of $k$, namely $k = 2\pi l/N$, $l=0,1,\dots,N/2$ (in Ref.~\cite{KP08}, $\mu$ is chosen as the origin of the energy axis). We note that $G(k;\mu) \in \mathds{R}$, $\forall k$ \cite{BF07a}, so that the explicit use of $\re[G(k;\mu)]$ may imply that in the calculations reported in Ref.~\cite{KP08}, $\im[G(k;\mu)] \not\equiv 0$. This is relevant, since the numerical artifact $\im[G(k;\mu)] \not\equiv 0$ can in principle give rise to a false violation of the Luttinger theorem \cite{BF07a} (see in particular Sec.~6.2 herein).

From the data in Fig.~2 of Ref.~\cite{KP08} one observes that for $N= 26$ the Luttinger theorem, Eq.~(\ref{e2}), indeed applies for \emph{all} $V/t \in [0,8]$: at $k = 2\pi l/N$, for $l=-6,-5,\dots,0,1,\dots,6$ (that is, for $13$ out of $26$ distinct $k$ points of which the underlying $\mathrm{1BZ}$ consists), one has $G(k;\mu) > 0$ (cf. Eq.~(\ref{e1})). Notwithstanding this fundamental fact, KP remarked that \cite{KP08}:
\begin{quote}
``we note that $k_{\Sc l}$ is indeed near $\pi/2$, however, even without finite-size scaling a small deviation $k_{\Sc l} \not= \pi/2$ may be observed for $V > 4t$.''
\end{quote}
Here $k_{\Sc l}$ is the Luttinger wave-vector, defined by KP \cite{KP08}, and others, as the solution of
\begin{equation}
G(k_{\Sc l};\mu) = 0. \label{e9}
\end{equation}
In Ref.~\cite{BF07a} (see in particular Sec.~2.4 herein) it has been emphasised that this and similar equations for higher space dimensions (defining the so-called `Luttinger surface') must be used with caution, for these equations may not have solutions. In the particular case at hand, where the underlying $\mathrm{1BZ}$ consists of a finite number of points, and where specifically $\pi/2 \not\in \mathrm{1BZ}$ for $N=26$ (cf. Eq.~(\ref{e6})), it should be abundantly clear that ``deviation $k_{\Sc l} \not= \pi/2$'' is devoid of any meaning, whatever. It seems, rather surprisingly, that the criterion adopted by KP \cite{KP08} for the validity of the Luttinger theorem amounts to the requirement of Eq.~(\ref{e9}) having the exact solution $k_{\Sc l} = \pi/2$ (and, by symmetry, $k_{\Sc l} = -\pi/2$), \emph{even} in the cases where $\pi/2$ does not belong to the underlying $\mathrm{1BZ}$! We repeat that, for the reasons spelled out in Ref.~\cite{BF07a}, even in the limit $N =\infty$, for which the $\mathrm{1BZ}$ consists of the continuum set $(-\pi,\pi]$, absence of a solution to Eq.~(\ref{e9}) \emph{cannot} be held as signifying breakdown of the Luttinger theorem.

One general technical remark is in order. In Ref.~\cite{KP08} KP indicated that
\begin{quote}
``it has been realised that the LSR [Luttinger sum rule] should as well apply to finite systems [10,11]'',
\end{quote}
where ``[10]'' refers to Ref.~\cite{KP07b} and ``[11]'' to Ref.~\cite{OBP07}. In this connection, we emphasise that \emph{this result is already implicit in the original paper by Luttinger and Ward \cite{LW60}}; nothing in the proof by Luttinger and Ward is dependent on whether the sums with respect to ${\bm k}$ in the underlying expressions run over a finite or a macroscopically-large number of ${\bm k}$ points, to be approximated by a continuum set; the only significant aspect associated with these sums is that they include the \emph{complete} set of the ${\bm k}$ points that are relevant to the uniform GS under consideration --- otherwise, by using an incomplete set of ${\bm k}$ points, the invariance property of the trace operator under the permutation of the product of the operators on which it operates, which is vital to the proof of the Luttinger-Ward identity, is violated \cite{BF07a}. What in our opinion has contributed to the obscuring of this very basic fact with regard to the applicability of the Luttinger theorem to the uniform GSs of \emph{finite} systems, is the explicit reliance by Luttinger and Ward \cite{LW60} on a continuum model; in contrast to lattice models (more explicitly, those without physical boundaries), the uniformity of the GS of the continuum model adopted by Luttinger and Ward is intimately connected with this model being infinitely large.

Summarising, on the basis of the data as presented in Fig.~2 of Ref.~\cite{KP08} we conclude that: \emph{for at least $N=26$ and $V/t \in [0,8]$, the explicit calculations performed by KP \cite{KP08} undisputedly confirm the validity of the Luttinger theorem, Eq.~(\ref{e2}), specifically in the MH insulating phase}. If this observation is applicable to all values of $N$ presented in Eq.~(\ref{e8}) above, it follows that the main conclusion arrived at by KP in Ref.~\cite{KP08}, that the Luttinger theorem broke down in the MH insulating phase, must entirely rest on the outcome of the finite-size-scaling scheme adopted by the authors. Following this observation, below we critically examine the last-mentioned scheme.

\subsection{An analysis of the finite-size-scaling scheme employed by KP}
\label{s2b}

In order to examine the reliability of the finite-size-scaling method adopted by KP \cite{KP08}, we employ the following model for the self-energy $\Sigma(k;\varepsilon)$ at $\varepsilon=\mu$:
\begin{equation}
\Sigma(k;\mu) = A + \frac{B}{k^2-\pi^2/4},\;\;\; k \in (-\pi,\pi], \label{e10}
\end{equation}
where $A$ and $B$ are constants to be determined. Evidently, by definition $A$ and $B$ are functions of $\mu$, however for our following considerations knowledge of the functional forms of these parameters will not be necessary. We emphasise that the expression in Eq.~(\ref{e10}) is hypothetical and is \emph{not} based on a microscopic calculation.

With $\varepsilon_k$ denoting the non-interacting single-particle energy dispersion, for which we choose the tight-binding expression
\begin{equation}
\varepsilon_{k} = -2 t \cos(k), \label{e11}
\end{equation}
for $G(k;\mu)$ one has (in the units where $\hbar=1$):
\begin{equation}
G(k;\mu) = \frac{1}{\mu - E_k},\label{e12}
\end{equation}
where
\begin{equation}
E_k \doteq \varepsilon_{k} + \Sigma(k;\mu). \label{e13}
\end{equation}
We remark that use of a more general expression for the energy dispersion $\varepsilon_k$, one involving $t'$ in addition to $t$, would only increase the number of parameters in our considerations without affecting the main conclusion of this section.

\begin{figure}[t!]
\centering
\includegraphics[angle=0, width=0.43\textwidth]{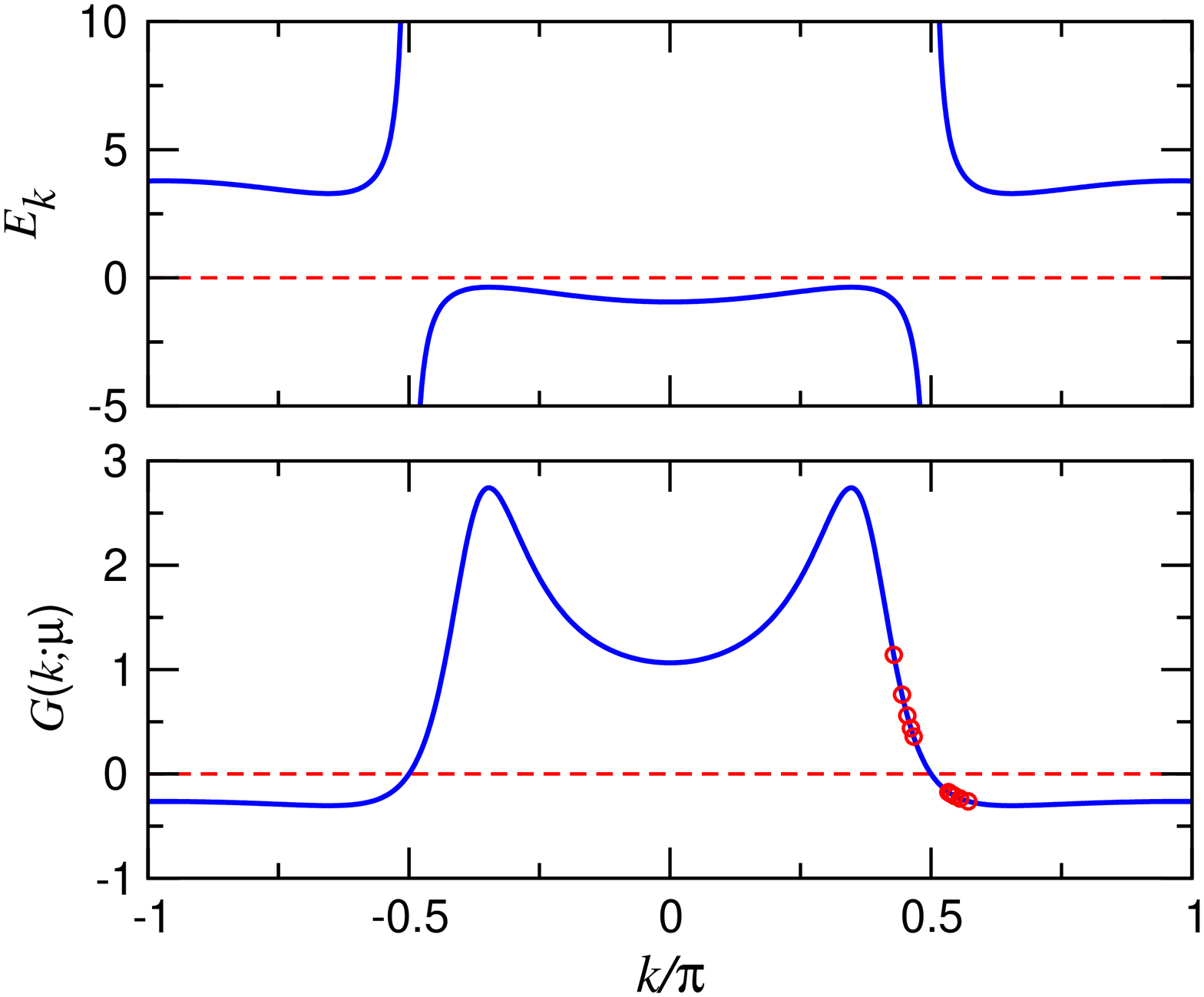}
\caption{(Colour online) The correlated single-particle energy dispersion $E_k$, Eq.~(\protect\ref{e13}), and the associated $G(k;\mu)$, Eq.~(\protect\ref{e12}), both in the units where $a=1$ and $\hbar=1$. The parameters used are, in the latter units, $t=1$, $A=1.6$, $B=1.33$ and $\mu=0$. From the behaviour of $E_k$ one observes that for a finite range of values of $\mu$ indeed the underlying GS is insulating; this is the case for $\mu=0$ (broken horizontal line). In the lower panel, the small circles in the vicinity of $k/\pi = 1/2$ mark the values of $G(k;\mu)$ at $k = \pi/2 \pm \pi/N$ for the values of $N$ listed in Eq.~(\protect\ref{e8}). Although for the model considered here, $k_{\Sc l}/\pi$ is exactly equal to $1/2$, the `finite-size scaling' scheme based on the last-mentioned values of $N$ yields $k_{\Sc l}/\pi = 0.500742$. See Table~\protect\ref{t1}. Note that indeed $G(k;\mu) >0$ for $k \in (-\pi/2,\pi/2)$ (cf. Eq.~(\protect\ref{e1})).}\label{f1}
\end{figure}

By employing a graphical presentation of $E_k$, one can determine a continuum of values for $A$, $B$, $t$ and $\mu$ for which the GS corresponding to the Green function in Eq.~(\ref{e12}) is a half-filled MH insulating GS. In Fig.~\ref{f1} we present the $E_k$ and the corresponding $G(k;\mu)$ for the specific values $A=1$, $B=1.33$, $t=1$ and $\mu=0$. Note that our specific choice $\mu=0$ is consistent with the explicit particle-hole \emph{asymmetry} in the calculations by KP \cite{KP08}, where $t'\not=0$ (see the diagram in the upper panel of Fig.~\ref{f1}).

Before proceeding with our analysis, two remarks are in order. Firstly, for the reasons indicated in Ref.~\cite{KP08}, for the system considered by KP, $G(k;\mu)$ must be small, $\forall k \in \mathrm{1BZ}$, for large values of $V/t$, scaling like $t/V^2$. This is clearly not the case for the function $G(k;\mu)$ displayed in Fig.~\ref{f1}. One can directly convince oneself that this aspect is of no consequence to our conclusions to be arrived at in this section. To this end, one should consider the fact that $k=\pm\pi/2$ remain the exact zeros of $G(k;\mu)$ on replacing $E_k$ and $\mu$ by $\alpha E_k$ and $\alpha \mu$, where $\alpha$ is a non-vanishing but otherwise arbitrary positive constant. This implies that for in particular $\mu=0$, through a simple re-scaling of the parameters $A$, $B$ and $t$, one will be able to render $G(k;\mu)$ as small as desired for all $k \in \mathrm{1BZ}$. One can readily verify that such uniform scaling of the above-mentioned parameters \emph{cannot} lead to any change in the conclusions of our analysis to be presented below (with the aid of the details that we present in appendix \ref{saa}, the reader may verify this statement through explicit calculations). One should in addition note that since the smallest $N$ that we consider in our following investigations is equal to $14$, Eq.~(\ref{e8}), we never probe $G(k;\mu)$ farther from $k=\pi/2$ than $\pi/14 \approx 0.22$ on either side of this point. Consequently, for our specific considerations in this paper it is entirely irrelevant what non-vanishing magnitude $G(k;\mu)$ has outside the interval $[\pi/2-\pi/14,\pi/2+\pi/14]$ (and by symmetry, also outside the interval $[-\pi/2-\pi/14,-\pi/2+\pi/14]$).

Secondly, concentrating on the half-filled GSs (see Eq.~(\ref{e7})), we assume that the expression for $\Sigma(k;\mu)$ in Eq.~(\ref{e10}) applies for \emph{all} $N$, so that by increasing the value of $N$ one merely samples the \emph{same} function at a more dense set of $k$ points of which the $\mathrm{1BZ}$ in Eq.~(\ref{e6}) consists. Denoting the $\mathrm{1BZ}$ corresponding to a system consisting of $M$ lattice sites by $\mathrm{1BZ}(M)$, with $k \in \mathrm{1BZ}(M)$ \emph{and} $k \in \mathrm{1BZ}(M')$, $M\not=M'$, in reality the value of $\Sigma(k;\mu)$ (and thus of $G(k;\mu)$) pertaining to the $M$-site system is different from that pertaining to the $M'$-site system. Since in the $t$-$t'$-$V$ model direct hoppings take place between nearest and next-nearest neighbours, in the absence of long-range order one may assume that for $M, M' > M_{\rm c}$, where $M_{\rm c}$ is a \emph{finite} integer, the last-mentioned difference is insignificantly small. The quantitative similarity between the results that we deduce below and those reported by KP in Ref.~\cite{KP08}, leads us to believe that the values of $N$ in Eq.~(\ref{e8}) satisfy the inequality $N > M_{\rm c}$.

Following KP \cite{KP08}, we define
\begin{equation}
\b{G}(N) \doteq \frac{1}{2} \big(G(\frac{\pi}{2}+ \frac{\pi}{N};\mu) + G(\frac{\pi}{2}-\frac{\pi}{N};\mu)\big), \label{e14}
\end{equation}
and
\begin{equation}
\Delta G(N) \doteq G(\frac{\pi}{2}+ \frac{\pi}{N};\mu) - G(\frac{\pi}{2}-\frac{\pi}{N};\mu). \label{e15}
\end{equation}
For the smallest and the largest values of $N$ given in Eq.~(\ref{e8}), one has respectively $\pi/N \approx 0.224$ and $\pi/N \approx 0.105$ (see the small circles in the lower panel of Fig.~\ref{f1}). Using the scaling relationships \cite{KP08}
\begin{equation}
\b{G}(N) = a_1 + \frac{b_1}{N} + \frac{c_1}{N^2}, \label{e16}
\end{equation}
\begin{equation}
\Delta G(N) = \frac{b_2}{N} + \frac{c_2}{N^2}, \label{e17}
\end{equation}
KP \cite{KP08} obtained:
\begin{equation}
\left. \frac{\partial G(k;\mu)}{\partial k}\right|_{k=\pi/2} = \lim_{N\to\infty} \frac{\Delta G(N)}{2\pi/N} = \frac{b_2}{2\pi}. \label{e18}
\end{equation}
By employing the explicitly-calculated values of $\b{G}(N)$ and $\Delta G(N)$, corresponding to the values of $N$ in Eq.~(\ref{e8}), KP \cite{KP08} deduced the values of the parameters $\{a_1, b_1, c_1\}$ and $\{ b_2, c_2\}$ with the aid of a least-squares fitting method.

Taylor expanding $G(k_{\Sc l};\mu)$ (for $N=\infty$) to \emph{linear order} at $k=\pi/2$, imposing the condition in Eq.~(\ref{e9}), KP \cite{KP08} obtained the \emph{approximate} expression:
\begin{equation}
k_{\Sc l} = \frac{\pi}{2} - \frac{G(\pi/2;\mu)}{\partial G(k;\mu)/\partial k\vert_{k=\pi/2}} = \frac{\pi}{2} -2\pi\,\frac{a_1}{b_2}. \label{e19}
\end{equation}
Clearly, the accuracy of this expression diminishes for increasing values of $\vert k_{\Sc l}-\pi/2\vert$. On the basis of the observation that $a_1/b_2 \not= 0$ for $V/t \in [3.6, 10]$, KP \cite{KP08} arrived at the conclusion that for the latter range of values of $V/t$, the Luttinger theorem broke down. See Fig,~5 in Ref.~\cite{KP08}.

\begin{table}[t!]
\caption{The parameters $\{a_1,a_2,a_3\}$ and $\{a_2,b_2\}$ (in the units where $a=1$ and $\hbar=1$), Eqs.~(\protect\ref{e16}) and (\protect\ref{e17}), for various values of the parameter $A$, Eq.~(\protect\ref{e10}), obtained by means of a least-squares fitting approach described in appendix \protect\ref{saa}. In all cases, $\mu=0$, $t=1$ and $B=1.33$, in the above-mentioned units. For fitting we have used exact data corresponding to the values of $N$ listed in Eq.~(\protect\ref{e8}). In the last column we present $k_{\Sc l}/\pi$, with $k_{\Sc l}$ as calculated according to the last expression in Eq.~(\protect\ref{e19}). The values for $k_{\Sc l}/\pi$ should be compared with the data for the same quantity displayed in Fig.~5 of Ref.~\protect\cite{KP08}.} \label{t1}
\begin{ruledtabular}
\begin{tabular}{c|cccccc}
$A$ & $a_1$ & $b_1$ & $c_1$ & $b_2$ & $c_2$ & $k_{\Sc l}/\pi$ \\
\hline\\
$1.6$  & $0.0046$ & $-0.4255$ & $91.080$ & $-12.509$ & $-99.290$ & $0.5007$ \\
$1.7$  & $0.0455$ & $-2.6752$ & $129.57$ & $-11.087$ & $-145.89$ & $0.5082$ \\
$1.8$  & $0.1145$ & $-6.3715$ & $187.00$ & $-9.0193$ & $-209.58$ & $0.5254$ \\
$1.87$ & $0.1900$ & $-10.340$ & $244.79$ & $-6.9890$ & $-269.15$ & $0.5544$ \\
$1.9$  & $0.2325$ & $-12.544$ & $275.86$ & $-5.9145$ & $-299.84$ & $0.5786$ \\
\end{tabular}
\end{ruledtabular}
\end{table}

By employing the expressions in Eqs.~(\ref{e16}) and (\ref{e17}) and the Green function in Eq.~(\ref{e12}) as determined in terms of the model self-energy in Eq.~(\ref{e10}) and the $\varepsilon_k$ in Eq.~(\ref{e11}), on the basis of the same approach as KP \cite{KP08} (specifically, by making use of the values of $N$ given in Eq.~(\ref{e8})), we have calculated the parameters $\{a_1, b_1, c_1\}$ and $\{ b_2, c_2\}$ (see appendix \ref{saa} for some relevant technical details). We present these parameters corresponding to $\mu=0$, $t=1$, $B=1.33$ and some values of $A$ in Table~\ref{t1}. In this table we also present the values for $k_{\Sc l}/\pi$ as calculated on the basis of the \emph{approximate} expression in Eq.~(\ref{e19}). One observes that for increasing values of $A$, the deviation of $k_{\Sc l}/\pi$ from the expected value $1/2$ increases. It is seen that in our model calculations, the increase in the value of $A$ brings about a similar change in $k_{\Sc l}$ as does the increase in the value of $V/t$ in the calculations by KP \cite{KP08}. Since for the model under consideration the zero of $G(k;\mu)$ at $k=\pi/2$ does \emph{not} depend on the value of $A$, it follows that the deviation of the calculated values for $k_{\Sc l}/\pi$ from $1/2$ is entirely bogus and a direct consequence of not employing sufficiently large values of $N$. \emph{This undisputedly demonstrates that the conclusion arrived at by KP \cite{KP08} is devoid of any significance as regards the validity or otherwise of the Luttinger theorem.} In contrast, if KP \cite{KP08} had not disputed the validity of the Luttinger theorem and imposed the condition $G(\pi/2;\mu) = 0$ as the result to be expected for $N\to\infty$, they would have gained some valuable information regarding the behaviours of $G(k;\mu)$ and $\Sigma(k;\mu)$ for $k$ in the close vicinity of $k=\pi/2$. For the reasons presented in Sec.~2.4 of Ref.~\cite{BF07a}, we should emphasise that satisfaction of the condition $G(\pi/2;\mu) = 0$ is \emph{not} a prerequisite for the validity of the Luttinger theorem at half-filling. Nonetheless, as we shall see later (see, e.g., Eq.~(\ref{ec242})), $G(\pi/2;\mu) = 0$ is exactly satisfied at half-filling and under the conditions for which the bosonization scheme is exact for the model under consideration.

\section{A heuristic argument in support of the Luttinger theorem}
\label{s3}

The reasoning that we present in this section is due to Volovik \cite{GEV07} and is related here with his kind permission.

According to Volovik \cite{GEV07}, a metal and an insulator, as characterised by respectively a pole and a zero in their corresponding single-particle Green functions, belong to the same topological universality class, or have the same topological invariant in the momentum space (Chap.~2 in Ref.~\cite{GEV06}). That is to say, the transformation of the last-mentioned pole into the last-mentioned zero at the metal-insulator transition does not change the topology of the Green function \cite{GEV06}. This implies that the adiabatic transition between the two states, as well as the adiabatic transformation between non-interacting and interacting fermions (Sec.~8.1.6 in Ref.~\cite{GEV03}), cannot give rise to `spectral flow': the state without excitation, the vacuum state, transforms into a state which is equally free of excitations. Consequently, as in the case of non-interacting fermions, for interacting fermions the number of particles in the GS remains to be determined by the `volume' of the Fermi sea in the case of metallic states and by that of the Luttinger sea \cite{BF07a} in the case of insulating states.

For completeness, \emph{topological stability} is defined (Sec.~8.1.6 in Ref.~\cite{GEV03}) as signifying the property by which any continuous change in a system leaves the system in the same topological universality class (see the previous paragraph). A continuous change in the system may be brought about by an adiabatic change in the coupling constant of the interaction potential.

\section{The CDW ground state of the $t$-$t'$-$V$ model in one space dimension, and the Luttinger theorem}
\label{s4}

In the thermodynamic limit, the half-filled GS of the one-dimensional $t$-$t'$-$V$ model  is a broken-symmetry state for $V > V_{\rm c}(t,t')$ (to be precise, there are two such GSs, differing only by a translation over an inter-site distance). For $V_{\rm c}(t,t')$ one specifically has $V_{\rm c}(t,0) = 2t$ (appendix \ref{sac}, in particular Eq.~(\ref{ec98})) \cite{HSSB82}. With $\langle\wh{O}\rangle$ denoting the GS expectation value of $\wh{O}$ per site, for the half-filled broken-symmetry GS one has (see Eq.~(\ref{ec63}) and the subsequent remarks)
\begin{equation}
\langle \h{n}_j\rangle = \frac{1}{2} + (-1)^j\, \delta n, \label{e20}
\end{equation}
where $\delta n$ is the CDW order parameter. In the uniform GS, $\delta n=0$. Thus, the primitive unit cell of the CDW GS consists of two sites, leading to the $\mathrm{1BZ}$ corresponding to this state, which we denote by $\mathrm{1BZ}'$, Eq.~(\ref{eb7}), having half the size of that corresponding to the half-filled \emph{uniform} GS, Eq.~(\ref{eb11}). The CDW GS is an insulator whose gaps in the single-particle excitation spectrum is located at the boundaries of the $\mathrm{1BZ}'$, exactly as in the case of ordinary semiconductors. In contrast to the latter systems, whose discrete translation symmetry is due to a periodic ionic potential (an \emph{external} potential), the discrete translation symmetry of the CDW state under consideration (a $Z_2$ symmetry) is due to a spontaneous symmetry breaking (see the second paragraph of Sec.~\ref{sac.4} on page~\pageref{chiralsb}). This breaking of the translation symmetry of the underlying Hamiltonian by the GS being possible only in the thermodynamic limit, for any finite realisation of the one-dimensional $t$-$t'$-$V$ model the corresponding GS is a uniform one. The half-filled GSs that KP \cite{KP08} have determined numerically are therefore all uniform, characterised by $\delta n = 0$.

The Luttinger theorem specific to inhomogeneous GSs is explicitly discussed by Luttinger in Sec.~V of Ref.~\cite{JML60}, under the heading ``The Band Case''. The role played by $G({\bm k};\mu)$ in the case of uniform GSs (see Eq.~(\ref{e1})), is in ``the band case'' played by the eigenvalues of the Hermitian single-particle Green matrix $\mathbb{G}({\bm k};\mu)$, which is the exact representation of the single-particle Green \emph{operator} $\wh{G}(\varepsilon)$ at $\varepsilon = \mu$ in the momentum space. For the CDW state under discussion, $\mathbb{G}(k;\mu)$ is a $2\times 2$ matrix (appendix \ref{sab}). For this state, the statement of the Luttinger theorem is that in Eq.~(\ref{e2}), in which the Luttinger number $N_{\Sc l}$ is defined according to (cf. Eq.~(\ref{e1}) above; see Eq.~(95) in Ref.~\cite{JML60} and note that $\upgamma_{\varsigma}(k;\mu) \equiv \big(\mu - Q_{k\varsigma}(\mu)\big)^{-1}$)
\begin{equation}
N_{\Sc l} \doteq \sum_{\varsigma \in \{ -, +\}} \sum_{k \in \mathrm{1BZ}'} \Theta\big(\upgamma_{\varsigma}(k;\mu)\big), \label{e21}
\end{equation}
where $\{ \upgamma_{\varsigma}(k;\mu)\}$ are the (real) eigenvalues of the Green matrix $\mathbb{G}(k;\mu)$. Owing to the insulating nature of the CDW state, in the light of Eq.~(\ref{e7}) one immediately observes that at half-filling the Luttinger theorem is trivially satisfied for this state. To appreciate this fact, one should realise that for all $k \in \mathrm{1BZ}'$, $\upgamma_{\pm}(k;\mu) \gtrless 0$ \emph{and} that, since each unit cell for the CDW state under consideration consists of \emph{two} lattice sites, for this state the number of $k$ points of which the $\mathrm{1BZ}'$, Eq.~(\ref{eb7}), consists is half that of the points of which the $\mathrm{1BZ}$ of the uniform GS consists.

\section{The inhomogeneous ground states of the $t$-$t'$-$V$ model in one space dimension}
\label{s5}

In this section we explicitly consider the broken-symmetry GS of the $t$-$t'$-$V$ model in two limits, one corresponding to $V \downarrow V_{\rm c}(t,t')$ (for $t'$ sufficiently small in comparison with $t$), and the other to $V \gg t \approx t'$. In appendix \ref{sac} we present the details underlying our calculations corresponding to the former limit.

\subsection{The case of $V \gtrapprox V_{\rm c}(t,t')$}
\label{s5a}

For the spectral gap $\mathcal{E}_{\Sc g}$ corresponding to $V$ larger than and sufficiently close to $V_{\rm c}(t,t')$, one has (Eqs.~(\ref{ec190}), (\ref{ec104}), (\ref{ec226}), (\ref{ec212}) and (\ref{ec214})):
\begin{equation}
\mathcal{E}_{\Sc g} \doteq 2 \mathcal{M} \sim 4\sqrt{2}\, t\, \exp\!\Big[\!-\frac{\pi^2}{2\sqrt{2 (V/V_{\rm c} -1)}}\Big], \label{e22}
\end{equation}
where $\mathcal{M}$ is the mass gap, Eq.~(\ref{e35}). Owing to this exponentially diminishing $\mathcal{E}_{\Sc g}$ as $V \downarrow V_{\rm c}$, for $V$ sufficiently close to $V_{\rm c}$ one can employ a continuum description for the calculation of the correlation functions of the system under considerations at distances that are large in comparison with the lattice constant $a$ of the underlying lattice. With $x \doteq j a$, in the framework of this description one expresses the lattice annihilation operator $\h{c}_j$ as (Eq.~(\ref{ec5})):
\begin{equation}
\h{c}_j \simeq \sqrt{a}\, \big\{ \e^{+i k_{\Sc f} (j a)}\, \h{R}(x) + \e^{-i k_{\Sc f} (j a)}\, \h{L}(x) \big\}, \label{e23}
\end{equation}
where the slowly-varying chiral field operators $\h{R}(x)$ and $\h{L}(x)$ can be represented in terms of the bosonic field operators $\h{\Pi}(x)$ and $\h{\Phi}(x)$, satisfying (in the thermodynamic limit --- see Eqs.~(\ref{ec18}) and (\ref{ec19}) --- and in the units where $\hbar=1$):
\begin{equation}
\big[\h{\Phi}(x), \h{\Pi}(x')\big]_{-} = i\, \delta(x-x'), \label{e24}
\end{equation}
as follows (Eqs.~(\ref{ec6}) and (\ref{ec27})):
\begin{widetext}
\begin{equation}
\left.\begin{array}{c} \h{R}(x)\\ \h{L}(x)\end{array} \right\}
\simeq \pm\frac{\h{U}_{\pm}}{\sqrt{2\pi a}}\, \e^{\mp i\pi x/L} \exp\!\big[\! -i\sqrt{\pi} \int_{-\infty}^x \rd x'\; \h{\Pi}(x') \pm i\sqrt{\pi}\, \h{\Phi}(x)\big], \label{e25}
\end{equation}
\end{widetext}
where $\h{U}_{\pm}$ are Klein factors \cite{FDMH81} (see Eq.~(3.14) herein) which ensure that $\h{c}_j$, as expressed according to Eq.~(\ref{e23}), appropriately removes one particle from any eigenstate of the number operator, in the relevant Fock space, on which it operates. As we shall indicate in appendix \ref{sac}, the signs `$\pm$' with which the RHS of Eq.~(\ref{e25}) is pre-multiplied (which equally could have been chosen to be `$\mp$') is of significance for the preservation of causality in calculating the single-particle Green function. At half-filling, where $k_{\Sc f} = \pi/(2a)$, Eq.~(\ref{ec3}), the phase factors $\e^{\pm k_{\Sc f} (j a)}$ in Eq.~(\ref{e23}) reduce to $(\pm i)^{j}$. We point out that the symbol `$\simeq$' in Eq.~(\ref{e25}) signifies two facts: first, that the expression in Eq.~(\ref{e25}) applies for $a\to 0$, or $\vert x/a\vert \gg 1$, and, second, that it is a leading-order term of an infinite Fourier-type series the formal form of which is presented in Eq.~(14.30) of Ref.~\cite{SS99}. Further, the `$\simeq$' in Eq.~(\ref{e23}) becomes a strict equality provided that the Fourier spectrum of $\h{c}_j$, $\forall j$, Eq.~(\ref{ec4}), is limited to two intervals of width $2 \Lambda$ centred at $\pm k_{\Sc f}$, where $\Lambda$ is equal to a fraction of $\pi/a$.

The dynamics of the slowly-varying chiral field operators $\h{R}(x)$ and $\h{L}(x)$ is governed by the quantum sine-Gordon Hamiltonian \cite{SC75} in Eq.~(\ref{ec76}) below, in which the parameters $g$, $u$ and $K$ are fully determined by $t$, $t'$ and $V$ (Eqs.~(\ref{ec77}), (\ref{ec123}) and (\ref{ec116})); provided that $t'$ is sufficiently small, at half-filling the low-lying excitations are affected only perturbatively by $t'$ (see Eqs.~(\ref{ec35}), (\ref{ec41}) and the remarks following these equations, in particular those in the paragraph following the latter equation, on page~\pageref{zerot1}).

Let $K_{\ell}'$ and $K_{\ell'}'$, Eq.~(\ref{eb6}), denote two reciprocal-lattice vectors corresponding to the Bravais lattice of which the underlying unit cells consist of two lattice points, and $k \in \mathrm{1BZ}'$, Eq.~(\ref{eb7}). For the $(\ell,\ell')$ element of $\mathbb{G}(k;z)$, where $\ell, \ell' \in \{0,1\}$, $k \in \mathrm{1BZ}'$ and $z \in \mathds{C}$, one has the expression in Eq.~(\ref{eb9}), in which
\begin{equation}
G_{j,j'}(z) \equiv \lim_{i\omega_n \to z} G_{j,j'}(i\omega_n). \label{e26}
\end{equation}
Here (Eq.~(25.14) in Ref.~\cite{FW03})
\begin{equation}
G_{j,j'}(i\omega_n) = \lim_{\beta \to\infty} \int_0^{\beta} \rd \tau\; \e^{i\omega_n \tau}\, \mathscr{G}_{j,j'}(\tau,0), \label{e27}
\end{equation}
in which (in the units where $\hbar=1$) \[\omega_n \doteq \frac{(2n+1) \pi}{\beta},\;\; n=0,\pm 1,\dots,\] is a fermion Matsubara frequency, with $\beta$ the inverse temperature (in the units where $k_{\Sc b} =1$), and (Eqs.~(23.3) and (23.6) in Ref.~\cite{FW03})
\begin{equation}
\mathscr{G}_{j,j'}(\tau,\tau') \doteq -\mathrm{Tr}\big\{\h{\rho}_{\Sc g}\, T_{\tau} [\h{c}_j(-i\tau) \h{c}_{j'}^{\dag}(-i\tau')]\big\}, \label{e28}
\end{equation}
where $\h{\rho}_{\Sc g}$ is the grand-canonical statistical operator, $T_{\tau}$ the $\tau$-ordering operator, and
\begin{equation}
\tau \doteq i t \label{e29}
\end{equation}
the imaginary time (similarly for $\tau'$); $\h{c}_j(-i\tau)$ is the imaginary-time counterpart of $\h{c}_j$ in the Heisenberg picture (similarly for $\h{c}_{j'}^{\dag}(-i\tau)$) (Eq.~(24.1) in Ref.~\cite{FW03}). In the zero-temperature limit, the trace on the RHS of Eq.~(\ref{e28}) reduces to the GS expectation value of the $\tau$-ordered product of $\h{c}_j(-i\tau)$ with $\h{c}_{j'}^{\dag}(-i\tau')$. We note that in Eq.~(\ref{e27}) we have identified $\tau'$ with zero on account of the fact that in equilibrium $\mathscr{G}_{j,j'}(\tau,\tau')$ is a function of $\tau-\tau'$.

In the CDW state, $G_{j,j'}(i\omega_n)$ is a non-trivial function of $j$ and $j'$, in contrast to the uniform state where it is a function of $j-j'$. With reference to Eq.~(\ref{e22}), for $V \downarrow V_{\rm c}$ the deviation of $G_{j,j'}(i\omega_n)$ from a function of $j-j'$ becomes exponentially small so that, for $V$ sufficiently close to $V_{\rm c}$, to an exponentially small error one can approximate $G_{j,j'}(i\omega_n)$ by a function of $j-j'$. On doing so, and by renumbering the site indices according to the scheme in Eq.~(\ref{eb1}), in what follows we shall restrict our considerations to the single-particle Green function $G(k;z)$, defined according to (appendix \ref{sab})
\begin{equation}
G(k;z) \doteq \sum_{j=-N/2+1}^{N/2} \e^{-i k (j a)}\, G_{j,0}(z), \;\;\; \mbox{\rm where}\;\;\; k \in \mathrm{1BZ}. \label{e30}
\end{equation}
Following the considerations in appendix \ref{sab}, the function $G(k;z)$ as defined here can be viewed as consisting of the union of $G_{0,0}(k;z)$ and $G_{1,1}(k;z)$, in which $k \in \mathrm{1BZ}'$.

Let $\h{A}(x,-i\tau)$ and $\h{B}(x,-i\tau)$ be the imaginary-time Heisenberg-picture counterparts of the Schr\"odingier-picture operators $\h{A}(x)$ and $\h{B}(x)$, where $\h{A}(x)$ and $\h{B}(x)$ may be one of $\h{L}(x)$ and $\h{R}(x)$. Defining, in analogy with Eq.~(\ref{e27}) (in view of the $G_{j,0}(z)$ in Eq.~(\ref{e30}), below we suppress $x'$ which would be the analogue of $j'$),
\begin{equation}
G_{\h{\Sc a},\h{\Sc b}}(x,i \omega_n) \doteq \lim_{\beta\to\infty} \int_0^{\beta} \rd \tau\; \e^{i \omega_n \tau}\, \mathscr{G}_{\h{\Sc a},\h{\Sc b}}(x,\tau), \label{e31}
\end{equation}
where (cf. Eq.~(\ref{e28}))
\begin{equation}
\mathscr{G}_{\h{\Sc a},\h{\Sc b}}(x,\tau) \doteq -\mathrm{Tr}\big\{\h{\rho}_{\Sc g}\, T_{\tau} [ \h{A}(x,-i\tau) \h{B}^{\dag}(0,0)]\big\}, \label{e32}
\end{equation}
by employing the expression in Eq.~(\ref{e23}), from Eq.~(\ref{e30}) we obtain that (as in Eq.~(\ref{e23}), below $x \doteq j a$):
\begin{widetext}
\begin{eqnarray}
G(k;z) &=& a\! \sum_{j=-N/2+1}^{N/2}\! \e^{-i k (j a)} \Big\{ \e^{i k_{\Sc f} (j a)}\, G_{\h{\Sc r},\h{\Sc r}}(x,z) + \e^{i k_{\Sc f} (j a)}\, G_{\h{\Sc r},\h{\Sc l}}(x,z) + \e^{-i k_{\Sc f} (j a)}\, G_{\h{\Sc l},\h{\Sc r}}(x,z) + \e^{-i k_{\Sc f} (j a)}\, G_{\h{\Sc l},\h{\Sc l}}(x,z) \Big\} \nonumber\\
&\simeq& \int_{-L/2}^{L/2} \rd x\; \e^{-i k x} \Big\{ \e^{i k_{\Sc f} x}\, G_{\h{\Sc r},\h{\Sc r}}(x,z) + \e^{i k_{\Sc f} x}\, G_{\h{\Sc r},\h{\Sc l}}(x,z) + \e^{-i k_{\Sc f} x}\, G_{\h{\Sc l},\h{\Sc r}}(x,z) + \e^{-i k_{\Sc f} x}\, G_{\h{\Sc l},\h{\Sc l}}(x,z)\Big\} \nonumber\\
&\equiv& \t{G}_{\h{\Sc r},\h{\Sc r}}(k - k_{\Sc f};z) + \t{G}_{\h{\Sc r},\h{\Sc l}}(k - k_{\Sc f};z) + \t{G}_{\h{\Sc l},\h{\Sc r}}(k +k_{\Sc f};z) + \t{G}_{\h{\Sc l},\h{\Sc l}}(k + k_{\Sc f};z) \nonumber\\
&\equiv& \frac{1}{u} \Big\{ \t{G}_{\h{\ul{\Sc r}},\h{\ul{\Sc r}}}(k - k_{\Sc f}; \frac{1}{u}\, z) + \t{G}_{\h{\ul{\Sc r}},\h{\ul{\Sc l}}}(k - k_{\Sc f};\frac{1}{u}\, z) + \t{G}_{\h{\ul{\Sc l}},\h{\ul{\Sc r}}}(k +k_{\Sc f};\frac{1}{u}\, z) + \t{G}_{\h{\ul{\Sc l}},\h{\ul{\Sc l}}}(k + k_{\Sc f};\frac{1}{u}\, z)\Big\}, \label{e33}
\end{eqnarray}
\end{widetext}
where $\h{\ul{L}}$ and $\h{\ul{R}}$ are the time-scaled counterparts of $\h{L}$ and $\h{R}$ respectively, Eq.~(\ref{ec234}), and $u$ is the renormalized Fermi velocity $v_{\Sc f}$, Eq.~(\ref{ec123}); in the limit $V = V_{\rm c}$, one has $u = \pi v_{\Sc f}/2 \approx 1.57\, v_{\Sc f}$, Eqs.~(\ref{ec40}) and (\ref{ec126}). In arriving at the last equality in Eq.~(\ref{e33}), we have made use of the equality in Eq.~(\ref{ec235}). The asymptotic equality `$\simeq$' in Eq.~(\ref{e33}) applies in the limit of $a\to 0$, where, with $L \equiv N a$, according to the Euler-Maclaurin summation formula one has (item 23.1.30 in Ref.~\cite{AS72})
\begin{eqnarray}
\sum_{j=-N/2+1}^{N/2} f(j a) &=& \frac{1}{a} \int_{-L/2+a}^{L/2} \rd x\; f(x) + \dots \nonumber\\
&\simeq& \frac{1}{a} \int_{-L/2}^{L/2} \rd x\; f(x).
\label{e34}
\end{eqnarray}
In appendix \ref{sac} we deal with $\t{G}_{\h{\ul{\Sc a}},\h{\ul{\Sc b}}}(k;z)$, where $\h{\ul{A}}$ and $\h{\ul{B}}$ stand for one of $\h{\ul{L}}$ and $\h{\ul{R}}$.

\begin{figure}[t!]
\includegraphics[angle=0, width=0.43\textwidth]{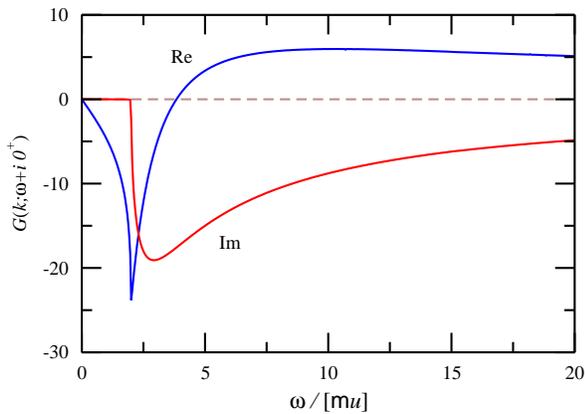}
\caption{(Colour online) The real and imaginary parts of $G(k;\omega+i 0^+)$ for $k=k_{\Sc f} \equiv \pi/2$ as function of $\omega$, calculated on the basis of the expression in Eq.~(\protect\ref{e33}), in which $\{\t{G}_{\h{\ul{\Sc a}},\h{\ul{\Sc b}}}(k;i\ul\omega_n)\}$ are evaluated according to Eqs.~(\protect\ref{ec239}) and (\protect\ref{ec272}); following Eq.~(\protect\ref{ec241}), $G(k_{\Sc f};\omega-i 0^+)$ is for $\omega <0$ exactly equal to $-G(k_{\Sc f};-\omega+i 0^+)$. We have employed $\upxi = 50$, which, following Eq.~(\protect\ref{ec253}), corresponds to $K = 25/51 < 1/2$, for which the GS is insulating (see the remarks centred at Eq.~(\protect\ref{ec97}), the second paragraph of Sec.~\protect\ref{sac.4} and Sec.~\protect\ref{sac.4c}). We have chosen $a=1$, $t = 1$ and arbitrarily equated $\mathfrak{m}$ with $t/1000$. Further, we have identified $u$ with the value corresponding to $K=1/2$, namely $\pi v_{\Sc f}/2$, Eq.~(\protect\ref{ec126}), where $v_{\Sc f} = 2at$, Eq.~(\protect\ref{ec40}). The overall magnitude of $\im[G(k;\omega+i 0^+)]$ diminishes rapidly for $k$ departing from $k_{\Sc f}$ (this is accompanied by an increasingly sharper rise of the right-most flank of the deep in $\re[G(k;\omega+i 0^+)]$ and a shift towards lower energies of the zero in this function). This signifies that the farther removed $k$ is from $k_{\Sc f}$, the more significant become the higher-order soliton/anti-soliton contributions to $G(k;\omega+i 0^+)$ that are neglected in the approximation of Eq.~(\protect\ref{ec249}). As expected (since the minimum value of $\chi$, Eq.~(\ref{ec279}), amounts to $2\mathfrak{m}$), the threshold at which $\im[G(k_{\Sc f};\omega+i 0^+)]\not\equiv 0$ satisfies $\vert\omega\vert/(\mathfrak{m} u) = 2$. Note that $G(k_{\Sc f};0) = 0$, which, in view of Eq.~(\protect\ref{ec242}), is an exact result. This outcome is significant in that in the present case $k_{\Sc l} = k_{\Sc f} \equiv \pi/2$ and $\mu=0$ (compare with Eq.~(\protect\ref{e9})).} \label{f2}
\end{figure}

In Fig.~\ref{f2} we present the Green function $G(k;\omega+i 0^+)$ for $k = k_{\Sc f}$ as calculated numerically on the basis of the expression Eq.~(\protect\ref{e33}), in which $\{\t{G}_{\h{\ul{\Sc a}},\h{\ul{\Sc b}}}(k\pm k_{\Sc f};z/u)\}$ are evaluated according to Eqs.~(\ref{ec239}) and (\protect\ref{ec272}). We point out that the nature of the approximations underlying our calculations (specifically, the approximation of Eq.~(\protect\ref{ec249})) restrict the validity of the calculated $G(k;\omega+i 0^+)$ to the regions of the $k$-$\omega$ plane close to $(\pm k_{\Sc f},\mu)$, where $\mu = 0$ in our calculations. The existence of gap in the single-particle spectrum of the system under consideration, as evidenced by $\im[G(k_{\Sc f};\omega +i 0^+)] \equiv 0$ for $\vert\omega\vert/(\mathfrak{m} u) <2$, trivially leads to the satisfaction of the Luttinger theorem, as discussed in Sec.~\ref{s4}. With reference to Eq.~(\ref{e22}), for the mass gap one has
\begin{equation}
\mathcal{M} = 2 u\, \mathfrak{m}, \label{e35}
\end{equation}
where $\mathfrak{m}$ is the soliton/antisoliton mass introduced in Eq.~(\ref{ec248}); from Eq.~(\ref{e35}) and Eq.~(\ref{e22}), one readily deduces the expression for $\mathfrak{m}$ specific to the case at hand. Fig.~\ref{f2} further reveals that our calculated Green function exactly satisfies Eq.~(\ref{e9}) for $k = k_{\Sc f}$ and $\mu =0$, whereby $k_{\Sc l}$ is to be identified with $k_{\Sc f} \equiv \pi/2$. This result is a direct consequence of the symmetry relation in Eq.~(\ref{ec241}), which, on the condition of continuity of $G(k_{\Sc f};z)$ at $z=0$, leads to Eq.~(\ref{ec242}).

For completeness, the contributions to $G(k_{\Sc f};\omega + i0^+)$, as depicted in Fig.~\ref{f2}, are essentially due to $\t{G}_{\h{\Sc r},\h{\Sc r}}(0;\omega + i0^+)$ and $\t{G}_{\h{\Sc r},\h{\Sc l}}(0;\omega + i0^+)$, Eq.~(\ref{e33}), which are moreover equal.

\subsection{The case of $V \gg t \approx t'$}
\label{s5b}

In the limit of infinitely large $V$, one of the two sites in the unit cell of the CDW GS is fully occupied and the other is fully empty. Thus for $V \gg t \approx t'$, the effective tunneling matrix element corresponding to nearest-neighbours is proportional to $t^2/V$, for which one has $t^2/V \ll t'$. It follows that in the case at hand, one can to a good approximation identify $t$ with zero and only consider next-nearest-neighbour hopping processes. Assuming that for $V=\infty$ the odd sites are fully occupied, and thus the even sites are fully empty, for $V \gg t \approx t'$ to a good approximation one has that electrons hop between even sites and holes hop between odd sites. One thus trivially obtains from Eq.~(\ref{e4}) that in the case at hand the energy dispersions of the single-particle excitations of the system under consideration are to leading order described by
\begin{equation}
E_k^{\pm} \sim -2 t'\cos(2k) + \left\{\begin{array}{c} V \\ \\ 0 \end{array} \right., \;\;\; k \in (-\frac{\pi}{2},\frac{\pi}{2}]. \label{e36}
\end{equation}
In the present case, the minimum spectral gap is an indirect one and its value amounts to $V - 4t'$.

\vspace{0.5cm}
\section{Concluding remarks}
\label{s6}

We have unequivocally demonstrated that the finite-size-scaling analysis as employed by KP \cite{KP08}, which is based on the explicitly-calculated values of the $G(k;\mu)$ pertaining to finite systems consisting of $N$ lattice sites, with $N$ taking the values presented in Eq.~(\ref{e8}), is \emph{not} capable of establishing breakdown of the Luttinger theorem for the insulating uniform GSs of the $t$-$t'$-$V$ model; using a simple model self-energy for which the corresponding $G(k;\mu)$ \emph{exactly} satisfies the Luttinger theorem, the approach by KP \cite{KP08} implies failure of the Luttinger model. Furthermore, we have shown that the statement by KP \cite{KP08}, that the numerical results corresponding to $N=26$ implied breakdown of the Luttinger theorem, is erroneous; through a simple \emph{counting} of the number of $k$ points for which $G(k;\mu) > 0$ (see Fig.~2 in Ref.~\cite{KP08}), one readily ascertains that for $N=26$, which according to Eq.~(\ref{e7}) corresponds to $N_{\rm e}=13$ at half-filling, the Luttinger theorem indeed applies: for exactly $13$ different $k$ points, out of the total of $26$ different $k$ points of which the underlying $\mathrm{1BZ}$ consists, Eq.~(\ref{e6}), one has $G(k;\mu) > 0$.

On the basis of their calculations corresponding to the insulating GSs of the Hubbard Hamiltonian on a two-dimensional triangular lattice, KP in an earlier paper \cite{KP07} arrived at a similar conclusion as the above concerning the Luttinger theorem, namely that in the absence of particle-hole symmetry the Luttinger theorem fails. In a separate Comment \cite{FT2D09}, one of us unequivocally demonstrates that this conclusion is also false and that the quantitative amount of the reported ``failure'' of the Luttinger theorem is fully attributable to an error arising from truncating by KP to \emph{leading} order in $\t{\mu} \doteq \mu -U/2$ the infinite series expansion for $G({\bm k};\mu)$ in powers of $\t{\mu}$ that has underlain the considerations by KP \cite{KP07}.

\section*{Acknowledgements}
We thank Grigori E. Volovik for his kind permission to present an account in this paper of the contents of his private communications with one of us (BF) on the subject matter of the Luttinger theorem. We further thank him for reading the final draft of this Comment. AMT acknowledges the support from US DOE under contract number DE-AC02-98 CH 10886.

\begin{appendix}

\begin{widetext}
\section{Some computational details}
\label{saa}

We have calculated the numerical results presented in Table~\ref{t1} with the aid of \emph{Mathematica} (version 6). The reader wishing to carry out calculations for the values of the parameters $\{\mu, t, A, B\}$ different from those to which the data in Table~\ref{t1} correspond, may find the following \emph{Mathematica} instructions helpful. The numerical results obtained for the specific values of $\mu$, $t$, $A$ and $B$ that we choose below, are presented in the fourth row of Table~\ref{t1}. As in Ref.~\cite{KP08}, for $N$ we have chosen the five values listed in Eq.~(\ref{e8}).

{\footnotesize
\begin{verbatim}
In[1]:= Ek[k_, t_, A_, B_] := -2 t Cos[k] + A + B/(k^2 - (Pi/2)^2)
In[2]:= G[k_, mu_, t_, A_, B_] := 1/(mu - Ek[k, t, A, B])
In[3]:= Gbar[mu_, t_, A_, B_, n_] := (G[Pi/2 + Pi/n, mu, t, A, B] + G[Pi/2 - Pi/n, mu,t,A, B])/2
In[4]:= DeltaG[mu_, t_, A_, B_, n_] := G[Pi/2 + Pi/n, mu, t, A, B] - G[Pi/2 - Pi/n, mu, t, A, B]
In[5]:= TGbar = Table[{n,Gbar[0.0, 1.0, 1.87, 1.33, n]}, {n, {14, 18, 22, 26, 30}}]
Out[5]:= {{14, 0.701368}, {18, 0.367454}, {22, 0.228655}, {26, 0.156993}, {30, 0.114867}}
In[6]:= InputForm[FindFit[TGbar, a1 + b1/n + c1/n^2, {a1, b1, c1}, n], NumberMarks -> False]
Out[6]//InputForm= {a1 -> 0.19006514037109792, b1 -> -10.33960282911629, c1 -> 244.78744001923275}
In[7]:= TDeltaG = Table[{n, DeltaG[0.0, 1.0, 1.87, 1.33, n]}, {n, {14, 18, 22, 26, 30}}]
Out[7]:= {{14, -1.89343}, {18, -1.17785}, {22, -0.85808}, {26, -0.678616}, {30, -0.563547}}
In[8]:= InputForm[FindFit[TDeltaG, b2/n + c2/n^2, {b2, c2}, n], NumberMarks -> False]
Out[8]//InputForm= {b2 -> -6.989008699335287, c2 -> -269.1504632514183}
In[9]:= N[1/2 - 2 (0.19006514037109792/-6.989008699335287), 7]
Out[9]:= 0.55439
\end{verbatim}
}
\end{widetext}

\section{The single-particle Green matrix of the CDW state}
\label{sab}

Here we determine the general form of the single-particle Green matrix $\mathbb{G}(k;z)$ pertaining to the CDW state of the one-dimensional $t$-$t'$-$V$ model that we consider in Sec.~\ref{s4}. Amongst others, we show that this matrix is a $2\times 2$ one. As we shall see, this result is a direct consequence of the fact that the model under consideration is defined on a lattice; had the model been a continuum model, consisting of $N/2$ unit cells (each containing two `atoms'), the relevant $\mathbb{G}(k,z)$ would be an $N/2 \times N/2$ matrix \cite{Note2}.

For the regular lattice under consideration, with lattice constant $a$, let $\{ \vert x_j\rangle\,\|\, j =1,2, \dots,N\}$ denote the complete set of the normalized eigenkets of the position operators $\h{x}$, and $\{ \vert k\rangle \,\|\, k \in \mathrm{1BZ}\}$ the complete set of the normalized eigenkets of the momentum operator $\h{p}$. For the following considerations it is more convenient to renumber the lattice sites in such a way that
\begin{equation}
j \in \Big\{-\frac{N}{2}+1,-\frac{N}{2}+2,\dots, 0, 1,\dots, \frac{N}{2}\Big\}, \label{eb1}
\end{equation}
where $N$ is even. Note that since the system under consideration is in the thermodynamic limit, $N$ is macroscopically large. One has
\begin{equation}
\langle x_j \vert k\rangle = \frac{1}{\sqrt{N}}\, \e^{i k (j a)}. \label{eb2}
\end{equation}
In the following, $\sum_j$ and $\sum_{j'}$ denote sums in which $j$ and $j'$ traverse the set in Eq.~(\ref{eb1}).

With $G(k,k';z)$ denoting the momentum representation of the Green operator $\wh{G}(z)$, of which $G_{j,j'}(z)$ is the site representation, by the completeness of the set $\{\vert x_j\rangle\}$, one has
\begin{equation}
G(k,k';z) = \sum_{j,j'} \langle k\vert x_j\rangle G_{j,j'}(z) \langle x_{j'}\vert k'\rangle. \label{eb3}
\end{equation}
On account of the fact that in the CDW state the unit cell consists of $2$ lattice points, one has the following symmetry relationship:
\begin{equation}
G_{j+2\nu,j'+2\nu}(z) = G_{j,j'}(z), \label{eb4}
\end{equation}
for any finite integer value of $\nu$ ($\nu$ may be negative, zero and positive). Using Eq.~(\ref{eb4}), from the expression in Eq.~(\ref{eb3}) one trivially obtains that
\begin{equation}
G(k,k';z) = \e^{-2\nu i (k-k') a}\, G(k,k';z), \label{eb5}
\end{equation}
from which one deduces that unless
\begin{equation}
k - k' = \frac{\pi}{a}\, \ell \doteq K_{\ell}',\;\;\; \ell = 0,\pm 1, \dots, \label{eb6}
\end{equation}
one has $G(k,k';z) \equiv 0$. One observes that $\{ K_{\ell}'\}$ is the set of the reciprocal-lattice vectors corresponding to the CDW GS of the system under consideration. Since $k, k' \in \mathrm{1BZ}$, it follows that the $\ell$ in Eq.~(\ref{eb6}) can take only the three values $-1$, $0$ and $+1$. With
\begin{equation}
\mathrm{1BZ}' \doteq (-\frac{\pi}{2 a}, \frac{\pi}{2 a}], \label{eb7}
\end{equation}
below we shall denote all non-trivial Fourier components of $G(k,k';z)$ by $G_{\ell,\ell'}(k;z)$, where
\begin{equation}
G_{\ell,\ell'}(k;z) \doteq G(k+K_{\ell}',k+K_{\ell'}';z), \;\;\; k \in \mathrm{1BZ}'. \label{eb8}
\end{equation}
Following Eq.~(\ref{eb3}), one thus has (see Eqs.~(\ref{eb2}) and (\ref{eb6})):
\begin{eqnarray}
&&\hspace{-0.8cm} G_{\ell,\ell'}(k;z) = \sum_{j,j'} \langle k + K_{\ell}'\vert x_j\rangle G_{j,j'}(z) \langle x_{j'} \vert k + K_{\ell'}'\rangle \nonumber\\
&&\hspace{0.8cm} \equiv \sum_{j,j'} (-1)^{j\ell + j' \ell'}\, \langle k \vert x_j\rangle G_{j,j'}(z) \langle x_{j'} \vert k \rangle. \label{eb9}
\end{eqnarray}

Using the closure relation for $\{ \vert k\rangle \,\|\, k \in \mathrm{1BZ}\}$, one can write
\begin{equation}
G_{j,j'}(z) = \sum_{k, k' \in \mathrm{1BZ}} \langle x_j\vert k\rangle G(k,k';z) \langle k'\vert x_{j'}\rangle. \label{eb10}
\end{equation}
Decomposing the $\mathrm{1BZ}$ as
\begin{equation}
\mathrm{1BZ} = (-\frac{\pi}{a},-\frac{\pi}{2 a}] \cup \mathrm{1BZ}' \cup  (\frac{\pi}{2 a},\frac{\pi}{a}], \label{eb11}
\end{equation}
and further
\begin{equation}
\mathrm{1BZ}' = (-\frac{\pi}{2 a},0] \cup (0,\frac{\pi}{2 a}] \equiv \mathrm{1BZ}_-' \cup \mathrm{1BZ}_+', \label{eb12}
\end{equation}
on account of the fact that the only non-trivial Fourier coefficients $G(k,k';z)$ are those presented in Eq.~(\ref{eb8}), after some algebra one obtains that
\begin{widetext}
\begin{eqnarray}
&&\hspace{0.0cm} G_{j,j'}(z) = \!\!\sum_{k \in \mathrm{1BZ}_-'} \!\!\Big\{ \langle x_j\vert k\rangle G_{0,0}(k;z) \langle k\vert x_{j'}\rangle + \langle x_j\vert k\rangle G_{0,1}(k;z) \langle k + K_1'\vert x_{j'}\rangle + \langle x_j\vert k + K_1'\rangle G_{1,0}(k;z) \langle k\vert x_{j'}\rangle  \nonumber\\
&&\hspace{3.0cm} + \langle x_j\vert k + K_1'\rangle G_{1,1}(k;z) \langle k + K_1'\vert x_{j'}\rangle \Big\} \nonumber\\
&&\hspace{1.3cm} + \!\!\sum_{k \in \mathrm{1BZ}_+'} \!\!\Big\{ \langle x_j\vert k\rangle G_{0,0}(k;z) \langle k\vert x_{j'}\rangle + \langle x_j\vert k\rangle G_{0,-1}(k;z) \langle k -K_1'\vert x_{j'}\rangle + \langle x_j\vert k - K_1'\rangle G_{-1,0}(k;z) \langle k\vert x_{j'}\rangle  \nonumber\\
&&\hspace{3.0cm} + \langle x_j\vert k - K_1'\rangle G_{-1,-1}(k;z) \langle k - K_1'\vert x_{j'}\rangle \Big\}. \label{eb13}
\end{eqnarray}
\end{widetext}
In arriving at this expression we have used the fact that e.g. the interval $(-\frac{\pi}{a}, -\frac{\pi}{2 a}]$ is translatable into $\mathrm{1BZ}_+'$ by $K_1'$. We note that the total number of terms contributing to $G_{j,j'}(z)$ is nine, of which three are identically vanishing (such as the one corresponding to $\sum_{k \in \mathrm{1BZ}_+'} \sum_{k' \in \mathrm{1BZ}_-'}$) on account of the fact that the pertinent $k$ and $k'$ do not satisfy the relationship in Eq.~(\ref{eb6}).

From Eq.~(\ref{eb2}) one obtains that
\begin{equation}
\langle x_j \vert k \pm K_1'\rangle = (-1)^j\, \langle x_j \vert k \rangle, \label{eb14}
\end{equation}
where the RHS is independent of whether on the left-hand side (LHS) one has $+K_1'$ or $-K_1'$. Using the result in Eq.~(\ref{eb14}), from Eq.~(\ref{eb3}) one immediately deduces that
\begin{equation}
G_{-\ell,-\ell'}(k;z) = G_{\ell,\ell'}(k;z),\;\; \ell, \ell' \in \{-1,0,1\}. \label{eb15}
\end{equation}
Using Eqs.~(\ref{eb14}) and (\ref{eb15}), Eq.~(\ref{eb13}) can thus be expressed as
\begin{equation}
G_{j,j'}(z) = \sum_{k \in \mathrm{1BZ}'} {\bm v}_j^{\dag}(k)\cdot \mathbb{G}(k;z) \cdot {\bm v}_{j'}(k), \label{eb16}
\end{equation}
where
\begin{equation}
\mathbb{G}(k;z) \doteq \left(\begin{array}{cc} G_{0,0}(k;z) & G_{0,1}(k;z) \\ \\
G_{1,0}(k;z) & G_{1,1}(k;z) \end{array} \right), \label{eb17}
\end{equation}
and (cf. Eqs.~(\ref{eb2}) and (\ref{eb14}))
\begin{equation}
{\bm v}_j(k) \doteq \left(\begin{array}{c} \langle k\vert x_j \rangle \\ \\
\langle k + K_1'\vert x_j \rangle \end{array} \right) \equiv \frac{1}{\sqrt{N}}\, \e^{i k (j a)} \left(\begin{array}{c} 1 \\ \\ (-1)^j \end{array} \right). \label{eb18}
\end{equation}
The vector ${\bm v}_j^{\dag}(k)$ is a row vector whose first and second components are $\langle x_j\vert k\rangle$ and $\langle x_j\vert k + K_1'\rangle$ respectively.

That $\mathbb{G}(k;z)$ is a finite-dimensional matrix, is related to the fact that the $\mathrm{1BZ}$ is finite, on account of the system under consideration being defined on a regular lattice of non-vanishing lattice constant; that it is a $2\times 2$ matrix, is a consequence of the fact that the extent of $\mathrm{1BZ}$ is \emph{twice} as large as that of $\mathrm{1BZ}'$. Note that for the homogeneous GS of the system under consideration (that is for $V < V_{\rm c}$), $G_{0,1}(k;z) = G_{1,0}(k;z) = 0$, $\forall k \in \mathrm{1BZ}'$. In such case, $G_{0,0}(k;z)$ and $G_{1,1}(k;z)$, with $k \in \mathrm{1BZ}'$, together describe the Green function $G(k;z)$ of the uniform GS, with $k \in \mathrm{1BZ}$. In this connection, on account of Eq.~(\ref{eb15}), $G_{1,1}(k;z)$ for $k \in \mathrm{1BZ}_+'$ is identical to $G(k;z)$ for $k \in (-\frac{\pi}{a},-\frac{\pi}{2 a}]$.

\section{The bosonization of the $t$-$t'$-$V$ model, the quantum sine-Gordon and the $XXZ$ spin-chain models}
\label{sac}

Here we determine the bosonized version of the one-dimensional $t$-$t'$-$V$ model in the continuum limit \cite{AL76}, where (below $a$ the lattice constant and $j$ a site index)
\begin{equation}
x \doteq j a \label{ec1}
\end{equation}
is considered as a continuous variable. In doing so, we formally assume that $a \to 0$ and $N \to \infty$, in such a way that
\begin{equation}
L = N a \label{ec2}
\end{equation}
is macroscopically large. In practice, the lattice constant $a$ is identified as the unit of length so that physically $a\to 0$ is to be understood as signifying that the correlation functions calculated in the continuum limit accurately describe those pertaining to the underlying lattice model over the length scales that are large in comparison with $a$.

Unless we indicate otherwise, in the following we shall explicitly deal with half-filled GSs, for which the Fermi points of the underlying non-interacting GS is located at
\begin{equation}
\pm k_{\Sc f} = \pm\frac{\pi}{2 a}. \label{ec3}
\end{equation}

\subsection{Preliminaries}
\label{sac.1}

Making use of the exact representation of $\h{c}_j$ in terms of its Fourier components $\{\h{a}_{k}\}$, that is (cf. Eq.~(\ref{eb2}))
\begin{equation}
\h{c}_j = \frac{1}{\sqrt{N}} \sum_{k \in \mathrm{1BZ}} \e^{i k\, (j a)}\, \h{a}_{k}, \label{ec4}
\end{equation}
disregarding the contributions corresponding to values of $k$ outside the interval
\begin{widetext}
\begin{equation}
[-k_{\Sc f} - \Lambda, -k_{\Sc f} + \Lambda] \cup [+k_{\Sc f} - \Lambda, +k_{\Sc f} + \Lambda] \subseteq \mathrm{1BZ}, \nonumber
\end{equation}
where $\Lambda > 0$, one obtains that
\begin{equation}
\h{c}_j \simeq \sqrt{a}\, \Big\{ \e^{+i k_{\Sc f} (j a)}\, \frac{1}{\sqrt{L}} \sum_{\vert k\vert \le \Lambda} \e^{i k (j a)}\, \h{a}_{k+k_{\Sc f}} + \e^{-i k_{\Sc f} (j a)}\, \frac{1}{\sqrt{L}}\sum_{\vert k\vert \le \Lambda} \e^{i k (j a)}\, \h{a}_{k-k_{\Sc f}} \Big\} \equiv \sqrt{a}\, \big\{(+i)^j\, \h{R}(x) + (-i)^j\, \h{L}(x)\big\}, \label{ec5}
\end{equation}
where the $(\pm i)^j$ are the values of $\e^{\pm i k_{\Sc f} (j a)}$ for the specific $k_{\Sc f}$ in Eq.~(\ref{ec3}).

The slowly-varying chiral field operators $\h{R}(x)$ and $\h{L}(x)$ can be expressed in terms of the bosonic field operators $\h{\Phi}(x)$ and $\h{\Theta}(x)$ as follows (cf. Eq.~(2.30) in Ref.~\cite{TG03}) \cite{Note3}:
\begin{eqnarray}
\left.\begin{array}{c} \h{R}(x)\\ \h{L}(x)\end{array} \right\} \times \e^{\pm i k_{\Sc f} x} &=&
\lim_{\alpha \to 0} \frac{\h{U}_{\pm}}{\sqrt{2\pi\alpha}} \, \e^{\pm i (k_{\Sc f} - \pi/L) x}\, \exp\!\big[\!\pm i\sqrt{\pi}\, \h{\Phi}(x) -i \sqrt{\pi}\, \h{\Theta}(x)\big] \nonumber\\
&=& \e^{\mp i\pi/4} \lim_{\alpha \to 0} \frac{\h{U}_{\pm}}{\sqrt{2\pi\alpha}} \, \e^{\pm i (k_{\Sc f} - \pi/L) x}\, \exp\!\big[\! -i \sqrt{\pi}\, \h{\Theta}(x)] \exp[\pm i\sqrt{\pi}\, \h{\Phi}(x + 0^+)\big], \label{ec6}
\end{eqnarray}
\end{widetext}
where $\h{U}_{\pm}$ is the Klein factor \cite{FDMH81,TG03} and $\alpha \approx 1/\Lambda$. Naturally, the operators $\h{\Phi}(x)$ and $\h{\Theta}(x)$ are functions of this cut-off parameter. Since the problem at hand is defined on a lattice, of lattice constant $a$, it is natural to identify $\alpha$ with $a$. \emph{This is the choice that we shall make in the following.}

The second equality in Eq.~(\ref{ec6}) is established through the application of the Baker-Campbell-Hausdorff formula \cite{RG74}
\begin{equation}
\e^{\h{A}} \e^{\h{B}} = \e^{\h{A} + \h{B}}\, \e^{\frac{1}{2} [\h{A},\h{B}]_{-}}, \label{ec7}
\end{equation}
which applies when $[\h{A},\h{B}]_-$ commutes with both $\h{A}$ and $\h{B}$, and making use of Eq.~(\ref{ec18}) below. With reference to the latter equation, it is evident that the phase factors $\e^{\mp i\pi/4}$ in the second expression in Eq.~(\ref{ec6}) are directly related to our use of $\h{\Phi}(x+0^+)$ (\emph{and} $L\to\infty$), instead of $\h{\Phi}(x)$. Use of $\h{\Phi}(x+0^+)$ is fundamental, as it prevents the possibility that $\partial_x \h{\Theta}(x)$ may occur alongside $\h{\Phi}(x)$ (or $\h{\Theta}(x)$ alongside $\partial_x\h{\Phi}(x)$) whose commutation is unbounded; see Eq.~(\ref{e24}) above. We note in passing that the first expression in Eq.~(\ref{ec6}) coincides with those in Refs.~\cite{GNT98}, \cite{AMT03} and \cite{TG03}. As we shall see later (Sec.~\ref{sac.6}), the second expression in Eq.~(\ref{ec6}) will he crucial in establishing a relationship between $\h{R}(x)$ and $\h{L}(x)$ and the soliton-generating fields pertaining to the quantum sine-Gordon Hamiltonian \cite{SC75} whose form factors have been determined by LZ \cite{LZ01}.

For $\h{\Phi}(x)$ and $\h{\Theta}(x)$ one has the explicit expressions (cf. Eq.~(2.23) in Ref.~\cite{TG03}; \cite{Note3})
\begin{widetext}
\begin{eqnarray}
\h{\Phi}(x) &=& +\frac{\sqrt{\pi} x}{L} \h{Q} +\frac{i\sqrt{\pi}}{L} \sum_{k\not=0} \big(\frac{L \vert k\vert}{2\pi}\big)^{1/2}\, \frac{1}{k}\, \e^{-\alpha \vert k\vert/2 - i k x} (\h{b}_{k}^{\dag} + \h{b}_{-k}), \label{ec8}\\
\h{\Theta}(x) &=& -\frac{\sqrt{\pi} x}{L} \h{J} -\frac{i\sqrt{\pi}}{L} \sum_{k\not=0} \big(\frac{L \vert k\vert}{2\pi}\big)^{1/2}\, \frac{1}{\vert k\vert}\, \e^{-\alpha \vert k\vert/2 - i k x} (\h{b}_{k}^{\dag} - \h{b}_{-k}), \label{ec9}
\end{eqnarray}
\end{widetext}
in which $\{\h{b}^{\dag}_k\}$ and $\{\h{b}_k\}$ are canonical bosonic creation and annihilation operators, respectively. Above
\begin{equation}
\h{Q} \doteq \h{Q}_{\Sc r} + \h{Q}_{\Sc l} \label{ec10}
\end{equation}
is the total-charge operator, and
\begin{equation}
\h{J} \doteq \h{Q}_{\Sc r} - \h{Q}_{\Sc l} \label{ec11}
\end{equation}
the total-current operator. Since $\h{Q}$ and $\h{J}$ commute with the Hamiltonian under consideration, the Hilbert space of the problem at hand can be classified in terms of the eigenvalues $Q$ and $J$ of these operators, which are \emph{integer valued} and satisfy the condition \cite{FDMH80,FDMH81} (see also p.~280 in Ref.~\cite{SS99})
\begin{equation}
(-1)^{J} = (-1)^{Q}. \label{ec12}
\end{equation}
For clarity, our $\h{Q}$ is in Ref.~\cite{FDMH81} (in contrast to Refs.~\cite{FDMH80} and \cite{FDMH81a}) denoted by $N-N_0$, where $N_0 = k_{\Sc f} L/\pi$ is assumed to be an \emph{odd} integer. This fact accounts for the absence of a minus sign on the RHS of Eq.~(\ref{ec12}) in comparison with Eq.~(2.3) in Ref.~\cite{FDMH81}.

For the explicit calculations to be carried out below, it is relevant to note that on the basis of the expressions in Eq.~(\ref{ec8}) and (\ref{ec9}) one can easily demonstrate that (below $x, x'$ are arbitrary)
\begin{itemize}
\label{items:i-v}
\item[(i)] $\h{\Phi}(x)$ and $\h{\Theta}(x)$ are Hermitian,
\item[(ii)] $\h{\Phi}(x)$ commutes with $\h{\Phi}(x')$, and
\item[(iii)] $\h{\Theta}(x)$ commutes with $\h{\Theta}(x')$.
\end{itemize}
Since $x$ and $x'$ are arbitrary, from (ii) and (iii) one immediately infers that
\begin{itemize}
\item[(iv)] $\partial_x \h{\Phi}(x)$ commutes with $\h{\Phi}(x')$, and
\item[(v)] $\partial_x \h{\Theta}(x)$ commutes with $\h{\Theta}(x')$.
\end{itemize}
On account of (ii) and (iv), in dealing with the functions of $\h{\Phi}(x)$ one can proceed as though $\h{\Phi}(x)$ were an ordinary function, and not an operator. Similarly for $\h{\Theta}(x)$, on account of (iii) and (v). In this connection, and in view of the exponential functions of $\h{\Phi}(x)$ and $\h{\Theta}(x)$ that we shall encounter below, we point out that for an arbitrary operator $\h{A}(x)$ one has \cite{RPF51}
\begin{equation}
\partial_x \e^{\h{A}(x)} = \int_0^1 \rd s\; \e^{(1-s) \h{A}(x)} [\partial_x \h{A}(x)] \e^{s \h{A}(x)}, \label{ec13}
\end{equation}
which, for $\h{A}(x)$ and $\partial_x\h{A}(x)$ not commuting, amounts to a non-trivial and generally a very complicated expression to work with.

It is important to specify the nature of the sums with respect to $k$ in Eqs.~(\ref{ec8}) and (\ref{ec9}). These sums are over \cite{FDMH81}
\begin{equation}
\Big\{k = \frac{2\pi n}{L}\,\|\, n = 0, \pm 1, \pm 2, \dots, \pm\infty \Big\}.
\label{ec14}
\end{equation}
This set is different from the discrete set
\begin{equation}
\Big\{ k = \frac{2\pi n}{L} \,\|\, -\frac{L}{2 a} < n \le \frac{L}{2 a} \Big\} \label{ec15}
\end{equation}
comprising the $\mathrm{1BZ}$ corresponding to the system under consideration, Eq.~(\ref{e6}). Thus, insofar as the sums in Eqs.~(\ref{ec8}) and (\ref{ec9}) (and those in the related expressions) are concerned, the usual \emph{sharp} cut-off of summations of the $k$-space, as imposed by the \emph{bounded} domain $\mathrm{1BZ}$, is replaced by the \emph{soft} cut-off $\e^{-\alpha \vert k\vert}$. This replacement is \emph{crucial} for the validity of some of the expressions that we shall encounter below.

From Eqs.~(\ref{ec8}) and (\ref{ec9}) one readily deduces that
\begin{eqnarray}
\h{\Phi}(x + L) &=& \h{\Phi}(x) + \sqrt{\pi}\, \h{Q}, \label{ec16}\\
\h{\Theta}(x+L) &=& \h{\Theta}(x) - \sqrt{\pi}\, \h{J}. \label{ec17}
\end{eqnarray}
For completeness, the $\h{Q}_{\Sc r}$ and $\h{Q}_{\Sc l}$ in $\h{Q} = \h{Q}_{\Sc r} + \h{Q}_{\Sc l}$ are in Ref.~\cite{TG03} denoted by $N_{\Sc r}$ and $N_{\Sc l}$ and the eigenvalues corresponding to these by respectively $\mathcal{N}_{\Sc r}$ and $\mathcal{N}_{\Sc l}$ (see Appendix B in Ref.~\cite{TG03} and in particular note that Haldane's above-mentioned $N_0$ is in this Appendix denoted by $2n_0+1$).

Making use of the explicit expressions in Eqs.~(\ref{ec8}) and (\ref{ec9}), and the canonical commutations relations for the bosonic operators $\{\h{b}_k^{\dag}\}$ and $\{\h{b}_k\}$, one trivially obtains that for $x, x' \in \mathds{R}$ (see Eq.~(2.24) in Ref.~\cite{TG03})
\begin{eqnarray}
&&\hspace{-1.0cm} \big[\h{\Phi}(x), \h{\Theta}(x')\big]_{-} = \frac{1}{L} \sum_{k\not=0} \frac{1}{k}\, \e^{i k (x' - x) - \alpha \vert k\vert}\nonumber\\
&&\hspace{0.0cm} \equiv \frac{i}{\pi} \tan^{-1}\Big(\frac{\e^{-2\pi\alpha/L} \sin(2\pi [x'-x]/L)}{1-\e^{-2\pi\alpha/L} \cos(2\pi [x'-x]/L)}\Big) \nonumber\\
&&\hspace{0.0cm} \sim \frac{i}{2}\, \sgn(x'-x)\;\; \mbox{\rm when}\;\; \alpha \downarrow 0,\; L \to \infty.  \label{ec18}
\end{eqnarray}
The last expression in Eq.~(\ref{ec18}) is the leading-order asymptotic contribution (corresponding to the asymptotic limit $L\to \infty$) to the function on the second line, subject to the condition that $\vert x'-x\vert \ll L$; the expression can be treated as an \emph{exact} one (i.e. the `$\sim$' can be replaced by `$=$') by taking the limit $L\to\infty$ \emph{prior to} allowing $x' - x$ to vary over the interval $(-\infty,\infty)$. We remark that on employing the set of $k$ points in Eq.~(\ref{ec15}), instead of the one in Eq.~(\ref{ec14}), one would obtain a function that for $\vert x'-x\vert \ll L$ is up to a finite multiplicative constant almost identical to $\mathrm{Si}(\pi [x'-x])$, where $\mathrm{Si}(z)$ is the sine-integral function (item 5.2.1 in Ref.~\cite{AS72}); importantly, the derivative of this function with respect to $x'$ is \emph{finite} at $x'=x$ for \emph{all} values of $L$ (compare with Eq.~(\ref{e24}) in the light of Eq.~(\ref{ec19})).

We emphasise that the commutator in Eq.~(\ref{ec18}) is a periodic function of $x'-x$, with period $L$. Since for $x, x' \in [0,L]$ one has $x'-x \in [-L,L]$, it follows that unless one takes the limit $L\to\infty$ \emph{prior to} allowing $x'-x$ to vary over the interval $(-\infty,\infty)$, use of the last expression in Eq.~(\ref{ec18}), which is not a periodic function of $x'-x$, is not justified. This fact has not be taken account of in Refs.~\cite{TG03,GNT98,AMT03}. In contrast, Haldane \cite{FDMH81} takes careful account of the periodicity of the commutator in Eq.~(\ref{ec18}), as well as other related commutators. In doing so, Haldane only considers the limit $\alpha = 0^+$, corresponding to the case where the cut-off function $\e^{- \alpha \vert k\vert}$ is infinitely soft.

Taking the limit $L\to\infty$, for
\begin{equation}
\h{\Pi}(x) \doteq \partial_x \h{\Theta}(x) \label{ec19}
\end{equation}
from Eq.~(\ref{ec18}) one obtains the commutation relation in Eq.~(\ref{e24}). It follows that for $L=\infty$, $\h{\Pi}(x)$ is the momentum operator \emph{canonically} conjugate to $\h{\Theta}(x)$. With reference to our earlier remarks, for finite but sufficiently large values of $L$, instead of Eq.~(\ref{e24}) one has the following asymptotic expression (cf. Eq.~(3.27) in Ref.~\cite{FDMH81})
\begin{equation}
\big[\h{\Phi}(x), \h{\Pi}(x')\big]_{-} \sim i\! \sum_{n=-\infty}^{\infty} \delta(x-x' + n L),\; \alpha=0^+,\; L\to \infty. \label{ec20}
\end{equation}

Within the framework of the bosonization technique, one defines the chiral density-fluctuation operators
\begin{equation}
\h{\rho}_{\Sc r}(x) \doteq \; :\!\h{R}^{\dag}(x) \h{R}(x)\!:,\;\;\; \h{\rho}_{\Sc l}(x) \doteq \; :\!\h{L}^{\dag}(x) \h{L}(x)\!:, \label{ec21}
\end{equation}
which are related to the bosonic fields $\h{\Phi}(x)$ and $\h{\Theta}(x)$ as follows (cf. Eq.~(\ref{ec19})):
\begin{eqnarray}
\partial_x \h{\Phi}(x) &=& +\sqrt{\pi}\, \big( \h{\rho}_{\Sc r}(x) + \h{\rho}_{\Sc l}(x)\big) \equiv \sqrt{\pi}\, \h{\rho}(x), \label{ec22} \\
\partial_x \h{\Theta}(x) &=& -\sqrt{\pi}\, \big( \h{\rho}_{\Sc r}(x) - \h{\rho}_{\Sc l}(x)\big) \equiv \h{\Pi}(x). \label{ec23}
\end{eqnarray}
In Eq.~(\ref{ec21}), $:\!\h{A} \h{B}\!:$ stands for the normal-ordered counterpart of the product $\h{A} \h{B}$. The expressions in Eqs.~(\ref{ec22}) and (\ref{ec23}) are readily obtained from Eqs.~(\ref{ec8}) and (\ref{ec9}) by employing the expression for the canonical boson operator $\h{b}_{k}$ in terms of the Fourier components of the chiral density operators $\h{\rho}_{\Sc r}(x)$ and $\h{\rho}_{\Sc l}(x)$ (see Eq.~(2.15) in Ref.~\cite{TG03}).

We note that $\h{\rho}_{\Sc r}(x) + \h{\rho}_{\Sc l}(x)$ is the \emph{slowly-varying} part of the total density-fluctuation operator \cite{Note4}. To appreciate this fact, one has to realise that, making use of Eq.~(\ref{ec5}),
\begin{eqnarray}
&&\hspace{-1.2cm}\frac{:\!\h{c}^{\dag}_j \h{c}_j\!:}{a} \approx\; :\!\h{R}^{\dag}(j a) \h{R}(j a)\!: + :\!\h{L}^{\dag}(j a) \h{L}(j a)\!: \nonumber\\
&&\hspace{-0.2cm}
+(-1)^{j}\, [ :\!\h{R}^{\dag}(j a) \h{L}(j a)\!: + :\!\h{L}^{\dag}(j a) \h{R}(j a)\!:]. \label{ec24}
\end{eqnarray}
One observes that $\h{\rho}_{\Sc r}(x) + \h{\rho}_{\Sc l}(x)$ does not take account of the last two terms on the RHS of Eq.~(\ref{ec24}). As we shall see later (see e.g. Eq.~(\ref{ec31}) below), the last-mentioned two terms do not contribute to the total number of particles in the system to leading order in the small parameter $a$. From Eqs.~(\ref{ec16}) - (\ref{ec22}) one has
\begin{equation}
\int_{0}^{L} \rd x\; [\h{\rho}_{\Sc r}(x) \pm\h{\rho}_{\Sc l}(x)] =
\left\{ \begin{array}{l} \h{Q},\\\\ \h{J}. \end{array} \right. \label{ec25}
\end{equation}
These expressions make explicit that $\h{Q}$ and $\h{J}$ are the $k=0$ Fourier components of $\h{\rho}_{\Sc r}(x) + \h{\rho}_{\Sc l}(x)$ and $\h{\rho}_{\Sc r}(x) - \h{\rho}_{\Sc l}(x)$ respectively.

To make contact with the work by LZ \cite{LZ01}, we now proceed with expressing $\h{\Theta}(x)$ in terms of $\h{\Pi}(x)$. To do so, it is required that $\h{\Theta}(x)$ be specified at one value of $x$; following Eq.~(\ref{ec19}), $\h{\Pi}(x)$ determines $\h{\Theta}(x)$ up to an additive constant. With reference to the Riemann-Lebesgue lemma (Sec.~9.41 in Ref.~\cite{WW62}), the second terms on the RHS of Eq.~(\ref{ec9}) becomes vanishingly small for $\vert x\vert \to\infty$, whereby
\begin{equation}
\h{\Theta}(-L) \sim \sqrt{\pi}\, \h{J}\;\;\; \mbox{\rm for}\;\;\; L \to \infty. \label{ec26}
\end{equation}
Thus, integrating both sides of Eq.~(\ref{ec19}) with respect to $x$, one obtains that \cite{Note5}
\begin{equation}
\h{\Theta}(x) \sim \sqrt{\pi}\, \h{J} + \int_{-L}^x \rd x'\; \h{\Pi}(x')\;\;\ \mbox{\rm for}\;\;\; L \to \infty, \label{ec27}
\end{equation}
where `$\sim$' is to be identified with `$=$' for $L=\infty$. In a Hilbert space corresponding to a fixed eigenvalue $J$ of $\h{J}$, the operator $\sqrt{\pi} \h{J}$ on the RHS of Eq.~(\ref{ec27}) gives rise to a constant phase factor in the expressions for $\h{R}(x)$ and $\h{L}(x)$, which, with reference to Eq.~(\ref{ec6}) and in view of the fact that $J$ takes integer values, is equal to $(-1)^J = \pm 1$, depending on whether $J$ is even or odd. This factor is of no consequence to $\mathscr{G}_{\h{\Sc a},\h{\Sc b}}(x,\tau)$, and the related functions, when $\h{A} = \h{B}$, where $\h{A}, \h{B} \in \{\h{L}, \h{R}\}$; it is however relevant when $\h{A} \not= \h{B}$. To clarify this, let $\vert\Psi\rangle$ denote an arbitrary eigenstate of $\h{Q}_{\Sc r}$ and $\h{Q}_{\Sc l}$ in the Fock space of the system under consideration. With reference to Eq.~(\ref{ec11}), and since the $J$ in $\h{J} \vert\Psi\rangle = J \vert\Psi\rangle$ is the difference in the numbers of right-moving and left-moving particles in $\vert\Psi\rangle$, in addition to the fact that the number of the left-moving (right-moving) particles in $\h{L}^{\dag} \vert\Psi\rangle$ (in $\h{R}^{\dag} \vert\Psi\rangle$) is by one unit more than that in $\vert\Psi\rangle$, one immediately infers that the expectation value of $\h{R} \h{L}^{\dag}$ (or of $\h{L} \h{R}^{\dag}$) with respect to $\vert\Psi\rangle$ differs by a minus sign from the one corresponding to the case in which the $\sqrt{\pi}\, \h{J}$ on the RHS of Eq.~(\ref{ec27}) has been suppressed altogether. The `$\pm$' pre-multiplying the expression on the RHS of Eq.~(\ref{e25}) (which we could equally have chosen to be `$\mp$') accounts for the last-mentioned minus sign. As we shall see later in this appendix, without this minus sign, the contributions of $\t{G}_{\h{\Sc r},\h{\Sc l}}(k-k_{\Sc f};z)$ and $\t{G}_{\h{\Sc l},\h{\Sc r}}(k+k_{\Sc f};z)$ to the calculated $G(k;z)$, Eq.~(\ref{e33}), would violate the causality principle (see in particular the second remark following Eq.~(\ref{ec236}) below).

\label{KleinF}
For $L \to \infty$, one generally identifies the $\e^{\pm i (-\pi/L)\, x}$ on the RHS of Eq.~(\ref{ec6}) with unity. Thus by equating $\alpha$ with the lattice constant $a$ (see above) and making use of Eq.~(\ref{ec27}), for $L \to\infty$ the expressions in Eq.~(\ref{ec6}) are transformed into the those presented in Eq.~(\ref{e25}). In our calculations we employ the expressions in Eq.~(\ref{e25}) by further suppressing the Klein factors $\h{U}_{\pm}$, this on account of the fact that $\h{U}_{\pm}$ do not affect the behaviours of correlation functions that we consider in this paper \cite{LP74} (for a case where Klein factors play a role, consult Sec.~4.3.2 in Ref.~\cite{TG03}). Consequently, the expressions in Eq.~(\ref{e25}) in which the $\h{U}_{\pm}$ are suppressed, are \emph{not} operator identities, however can be viewed as such in calculating the correlation functions $\{G_{\h{\Sc a},\h{\Sc b}}(x,i\omega_n)\}$ introduced in Eq.~(\ref{e31}). Barring the functions $\e^{\pm i k_{\Sc f} (j a)}$ that we have already separated from $\h{R}(x)$ and $\h{L}(x)$, the expression on the RHS of Eq.~(\ref{e25}) concerning $\h{R}(x)$ ($\h{L}(x)$) coincides with that of the operator $\h{O}_1$ ($\h{O}_2$) introduced by Luther and Peschel in Ref.~\cite{LP74}.

We note in passing that, with $\h{\Pi}(x)$ being associated with the quantum-mechanical momentum operator $\h{p}$, the contribution to $\h{R}(x)$ and $\h{L}(x)$ of the first term in the argument of the exponential function on the RHS of Eq.~(\ref{e25}) is the equivalent of the single-particle eigenfunction $\exp(i p\, x)$ of the one-dimensional Schr\"odinger equation describing a particle in a constant potential. Further, from Eq.~(\ref{ec22}) one observes that placing a particle at location $x=x_0$ gives rise to a `kink' (a discontinuity if the particle is infinitely localised), of magnitude $\sqrt{\pi}$, in $\h{\Phi}(x)$ at $x=x_0$. Therefore, physically the factor $\exp[\pm i\sqrt{\pi}\, \h{\Phi}(x)]$ on the RHS of Eq.~(\ref{e25}) introduces a phase jump equal to $+\pi$ ($-\pi$) in $\h{R}(x)$ ($\h{L}(x)$) for any $x$ at which a localised particle is encountered \cite{FDMH81,JV94} (see in particular Sec.~3.1, ``Phenomenological bosonization'', in Ref.~\cite{TG03}). The accumulation of these phase jumps amount to the Jordan-Wigner phase that ensures anti-commutation relations to be satisfied for Fermi operators expressed in terms of Bose operators.

The dynamics of the bosonic field operators $\h{\Phi}(x)$ and $\h{\Pi}(x)$ are determined by the quantum sine-Gordon Hamiltonian \cite{SC75}. Below we present the details underlying the derivation of this Hamiltonian for the one-dimensional $t$-$t'$-$V$ Hamiltonian under consideration. This derivation almost entirely coincides with that encountered in the treatments of the Heisenberg-Ising chain of the SU(2) spin operators, as described in e.g. Refs.~\cite{FDMH83,GNT98,AMT03,TG03}.

\begin{widetext}
\subsection{Bosonization and the sine-Gordon Hamiltonian}
\label{sac.2}

Using the expression in Eq.~(\ref{ec5}), one trivially obtains that
\begin{equation}
\h{c}_j^{\dag} \h{c}_{j+1} \simeq a i \big\{ \h{R}^{\dag}(x) \h{R}(x+a) - \h{L}^{\dag}(x) \h{L}(x+a) - (-1)^j\, \h{R}^{\dag}(x) \h{L}(x+a) + (-1)^j\, \h{L}^{\dag}(x) \h{R}(x+a) \big\}. \label{ec28}
\end{equation}
Making use of the Euler-Maclaurin summation formula (item 23.1.30 in Ref.~\cite{AS72}), whereby
\begin{equation}
\sum_{j=1}^{N} f(j a) = \frac{1}{a} \int_{a}^{N a} \rd x\; f(x) + \dots,
\label{ec29}
\end{equation}
for $a \to 0$ one obtains that
\begin{equation}
\sum_{j=1}^{N} \h{R}^{\dag}(j a) \h{R}(j a+a) \simeq \frac{1}{a} \int_{0}^{L} \rd x\; \h{R}^{\dag}(x) \h{R}(x+a) \approx \frac{1}{a} \int_{0}^{L} \rd x\; \h{R}^{\dag}(x) \h{R}(x) + \int_{0}^{L} \rd x\; \h{R}^{\dag}(x) \partial_x \h{R}(x). \label{ec30}
\end{equation}
For $\sum_{j=1}^{N} \h{L}^{\dag}(j a) \h{L}(j a+a)$ one obtains a similar expression.

For obtaining the sum with respect to $j$ of the last two terms on the RHS of Eq.~(\ref{ec28}), we assume that $N$ is \emph{even}, whereby one can write
\begin{eqnarray}
&&\hspace{-1.2cm}\sum_{j=1}^{N} (-1)^j\, \h{R}^{\dag}(x) \h{L}(x+a) = \sum_{j=1}^{N/2} \h{R}^{\dag}(j (2a)) \h{L}(j (2a) +a) - \sum_{j=1}^{N/2} \h{R}^{\dag}(j (2a)-a) \h{L}(j (2a))\nonumber\\
&&\hspace{-0.5cm} \approx \frac{1}{2}\! \int_{2 a}^{L} \rd x\; \h{R}^{\dag}(x) \partial_x \h{L}(x) + \frac{1}{2}\! \int_{2 a}^{L} \rd x\; \big(\partial_x \h{R}^{\dag}(x)\big) \h{L}(x) = \frac{1}{2} \big( \h{R}^{\dag}(L) \h{L}(L) - \h{R}^{\dag}(2a) \h{L}(2a)\big) = O(a), \label{ec31}
\end{eqnarray}
\end{widetext}
where the last equality follows from the periodic boundary condition
\begin{equation}
\h{R}^{\dag}(x+L) \h{L}(x+L) = \h{R}^{\dag}(x) \h{L}(x). \label{ec32}
\end{equation}
This condition can be readily deduced from Eq.~(\ref{ec68}) below in conjunction with Eq.~(\ref{ec16}) above, taking into account that the eigenvalues of $\h{Q}$ are integers.

For $\sum_{j=1}^{N} (-1)^j\, \h{L}^{\dag}(x) \h{R}(x+a)$ one obtains a similar result, implying that for $a \to 0$ the last two terms on the RHS of Eq.~(\ref{ec28}) do \emph{not} contribute to $\sum_{j=1}^{N} \h{c}_j^{\dag} \h{c}_{j+1}$. This is a direct consequence of half-filling for which $k_{\Sc f}$ has the value given in Eq.~(\ref{ec3}).

Combining the above results, for $a\to 0$ and $N\to\infty$ (see Eq.~(\ref{ec2})) one obtains that
\begin{widetext}
\begin{equation}
-t \big(\sum_{j=1}^{N} \h{c}_{j}^{\dag} \h{c}_{j+1} + \mathrm{H.c.}\big) = -(2a t)\, i \!\int_0^{L} \rd x\; \big( \h{R}^{\dag}(x) \partial_x \h{R}(x) - \h{L}^{\dag}(x) \partial_x \h{L}(x) \big). \label{ec33}
\end{equation}
In a similar fashion as above, one obtains that
\begin{equation}
-t' \big( \sum_{j=1}^{N} \h{c}_{j}^{\dag} \h{c}_{j+2} + \mathrm{H.c.}\big) = \frac{2 t'}{a} \int_0^{L} \rd x\; \big( \h{R}^{\dag}(x) \h{R}(x) + \h{L}^{\dag}(x) \h{L}(x) \big). \label{ec34}
\end{equation}
\end{widetext}

In view of Eqs.~(\ref{ec21}), (\ref{ec22}) and (\ref{ec16}) one has
\begin{eqnarray}
-t' \big( \sum_{j=1}^{N} :\!\h{c}_{j}^{\dag} \h{c}_{j+2}\!: + \mathrm{H.c.}\big)
&=& \frac{2t'}{a \sqrt{\pi}} \int_0^{L} \rd x\; \partial_x \h{\Phi}(x) \nonumber\\
&=& \frac{2 t'}{a}\, \h{Q}. \label{ec35}
\end{eqnarray}
This term can be absorbed in the contribution of the chemical potential in the thermodynamic Hamiltonian of the problem at hand (see in particular Eq.~(\ref{ec25}) above). It follows that \emph{at half-filling}, and for sufficiently small $V/t$, whereby the bosonization scheme as considered above is physically reasonable, the parameter $t'$ does \emph{not} enter into the bosonized Hamiltonian. As we shall discuss later, this Hamiltonian only to leading order in $V/t$ coincides with the bosonized Hamiltonian derived from the exact solution of the $t$-$V$ Hamiltonian (compare the $\wh{\mathcal{H}}$ in Eq.~(\ref{ec75}) with that in Eq.~(\ref{ec76})).

For completeness, we point out that the results in Eqs.~(\ref{ec33}) and (\ref{ec35}) are immediately obtained by employing the following exact representation (see Eq.~(\ref{ec4})):
\begin{equation}
-\sum_{j=1}^{N} (t\,\h{c}_{j}^{\dag} \h{c}_{j+1} + t'\, \h{c}_{j}^{\dag} \h{c}_{j+2})  + \mathrm{H.c.}
= \sum_{k \in \mathrm{1BZ}} \varepsilon(k)\, \h{a}_{k}^{\dag} \h{a}_k, \label{ec36}
\end{equation}
where for the energy dispersion $\varepsilon(k)$ one has
\begin{equation}
\varepsilon(k) = -2t \cos(k a) - 2t' \cos(2 k a). \label{ec37}
\end{equation}
On effecting (cf. Eq.~(\ref{ec5}))
\begin{equation}
\sum_{k \in \mathrm{1BZ}} \varepsilon(k)\, \h{a}_{k}^{\dag} \h{a}_k \rightharpoonup \sum_{\substack{-\Lambda < k + k_{\Sc f} < \Lambda\\ -\Lambda < k - k_{\Sc f} < \Lambda}} \!\!\varepsilon(k)\, \h{a}_{k}^{\dag} \h{a}_k, \label{ec38}
\end{equation}
where $\Lambda > 0$ is the cut-off in the momentum space, and linearising $\varepsilon(k)$ at $k =\pm k_{\Sc f}$, one has
\begin{equation}
\varepsilon(k) \approx -2 (t+t') \pm v_{\Sc f} (k \mp \pi/a), \label{ec39}
\end{equation}
where
\begin{equation}
v_{\Sc f} \doteq 2 a t, \label{ec40}
\end{equation}
one obtains that
\begin{eqnarray}
&&\hspace{-0.5cm} \sum_{k \in \mathrm{1BZ}} \varepsilon(k)\, \h{a}_{k}^{\dag} \h{a}_k \rightharpoonup -2(t+t') \!\!\sum_{\substack{-\Lambda < k + k_{\Sc f} < \Lambda\\ -\Lambda < k - k_{\Sc f} < \Lambda}} \h{a}_{k}^{\dag} \h{a}_k \nonumber\\
&&\hspace{-0.4cm} + v_{\Sc f} \Big\{\!\sum_{-\Lambda < k - k_{\Sc f} < \Lambda} \!\!\!\! (k-k_{\Sc f})\, \h{a}_{k}^{\dag} \h{a}_k - \!\!\sum_{-\Lambda < k + k_{\Sc f} < \Lambda} \!\!\!\! (k+k_{\Sc f})\, \h{a}_{k}^{\dag} \h{a}_k\Big\}.\nonumber\\ \label{ec41}
\end{eqnarray}
Evidently, $v_{\Sc f}$ is \emph{independent} of $t'$. The first term on the RHS of Eq.~(\ref{ec41}) can be absorbed in the contribution of the chemical potential in the thermodynamic Hamiltonian. With reference to Eq.~(\ref{ec40}), it should be evident that the normal-ordered form of the expression on the RHS of Eq.~(\ref{ec33}) coincides with the contribution proportional to $v_{\Sc f}$ in Eq.~(\ref{ec41}).

\label{zerot1}
We point out that the non-linear corrections to $\varepsilon(k)$, with $\varepsilon(k)$ as expanded around $k = \pm k_{\Sc f}$, depend non-trivially on $t'$. However, since the contributions arising from these corrections correspond at the lowest order to four-fermion interactions in the Hamiltonian, they are all \emph{irrelevant} operators, as can be established by simple power counting (for a precise treatment, one may consult Sec.~4.3 in Ref.~\cite{SS99}). It follows that for sufficiently small values of $t'$, at half-filling the consequences of $t' \not= 0$ can be accounted for by perturbation theory. \emph{In our explicit calculations in this paper, we assume that $t'$ is sufficiently small and thus neglect its consequences altogether. }

Following Eq.~(\ref{ec6}) and the remarks regarding e.g. $\h{U}_{\pm}$ in the paragraph subsequent to Eq.~(\ref{ec27}) on page~\pageref{KleinF}, we write (cf. Eq.~(14.31) in Ref.~\cite{SS99})
\begin{eqnarray}
&&\hspace{-0.8cm} \left. \begin{array}{r}\h{R}(x) \\ \h{L}(x) \end{array} \right\} \simeq \frac{1}{\sqrt{2\pi a}} \,\exp[-i \sqrt{\pi}\, ( \mp \h{\Phi}(x) +\h{\Theta}(x))] \nonumber\\
&&\hspace{-0.3cm} \equiv \frac{\e^{\mp\pi i/4}}{\sqrt{2\pi a}} \,\exp[-i \sqrt{\pi}\, \h{\Theta}(x)] \exp[\pm i \sqrt{\pi}\, \h{\Phi}(x+0^+)], \nonumber\\ \label{ec42}
\end{eqnarray}
from which one deduces that (see the remarks (i)-(v) on page~\pageref{items:i-v})
\begin{eqnarray}
\left. \begin{array}{r} \partial_x\h{R}(x)\\ \partial_x\h{L}(x) \end{array} \right\} \simeq -i\sqrt{\pi} (\partial_x \h{\Theta}(x) \mp \partial_x \h{\Phi}(x) ) \left\{ \begin{array}{l} \h{R}(x), \\ \h{L}(x). \end{array} \right. \label{ec43}
\end{eqnarray}
On account of the expressions in Eqs.~(\ref{ec21}), (\ref{ec22}) and (\ref{ec23}), one thus obtains that (see Eqs.~(\ref{ec33}), (\ref{ec19}) and (\ref{ec40}))
\begin{eqnarray}
&&\hspace{-0.5cm} -(2a t)\, i \!\int_0^{L} \rd x\; \big( \h{R}^{\dag}(x) \partial_x \h{R}(x) - \h{L}^{\dag}(x) \partial_x \h{L}(x) \big) \nonumber\\
&&\hspace{0.5cm} \simeq v_{\Sc f} \int_0^{L} \rd x\; \big( (\partial_x\h{\Theta}(x))^2 + (\partial_x\h{\Phi}(x))^2\big). \label{ec44}
\end{eqnarray}
\emph{The right-hand side of this expression, which coincides with the expression for the Hamiltonian in Eq.~(3.58) of Ref.~\cite{FDMH81}, is \emph{twice} as large as the correct non-interacting Luttinger Hamiltonian.} The truth of this statement is directly verified by substituting the expressions for $\h{\Phi}(x)$ and $\h{\Theta}(x)$ in Eqs.~(\ref{ec8}) and (\ref{ec9}) in the expression on the RHS of Eq.~(\ref{ec44}); following some algebra, one readily deduces that the resulting Hamiltonian is indeed twice the non-interacting Hamiltonian as expressed in terms of the bosonic operators $\{ \h{b}_{k}\}$ and $\{ \h{b}^{\dag}_k \}$, such as presented in e.g. Eq.~(2.22) of Ref.~\cite{TG03}.

To circumvent the above-mentioned problem, which has not been recognised earlier, following Refs.~\cite{GNT98} and \cite{AMT03} we proceed as follows. \emph{In order to keep the notation light, below we suppress the $0^+$ in $\h{\Phi}(x+0^+)$, with the understanding that when necessary the $x$ in $\h{\Phi}(x)$, and related operators, stands for $x+0^+$.}

With
\begin{equation}
\tau \doteq i t \label{ec45}
\end{equation}
the imaginary time, we define the complex `time' coordinates
\begin{equation}
z \doteq \tau + i x/v_{\Sc f},\;\;\; \b{z} \doteq \tau - i x/v_{\Sc f}, \label{ec46}
\end{equation}
and express the imaginary-time Heisenberg-picture counterparts of the operators $\h{\Phi}(x)$ and $\h{\Theta}(x)$, that is
\begin{equation}
\h{\Phi}(x,-i\tau) \doteq \h{\Phi}(z,\b{z}),\;\;\; \h{\Theta}(x,-i\tau) \doteq \h{\Theta}(z,\b{z}), \label{ec47}
\end{equation}
in terms of respectively the \emph{analytic} and \emph{anti-analytic} operators $\h{\varphi}(z)$ and $\h{\b\varphi}(\b{z})$, as follows (Eqs.~(22.11) and (22.13) in Ref.~\cite{AMT03}):
\begin{eqnarray}
\h{\Phi}(z,\b{z}) &=& \h{\varphi}(z) + \h{\b\varphi}(\b{z}), \label{ec48}\\
\h{\Theta}(z,\b{z}) &=& \h{\varphi}(z) - \h{\b\varphi}(\b{z}). \label{ec49}
\end{eqnarray}
From the first expression in Eq.~(\ref{ec42}), for the imaginary-time Heisenberg-picture counterparts of $\h{R}(x)$ and $\h{L}(x)$ one thus obtains that
\begin{eqnarray}
\h{R}(x,-i\tau) &\simeq& \frac{1}{\sqrt{2\pi a}} \exp[+i \sqrt{4\pi} \h{\b\varphi}(\b{z})], \label{ec50}\\
\h{L}(x,-i\tau) &\simeq& \frac{1}{\sqrt{2\pi a}} \exp[-i \sqrt{4\pi} \h{\varphi}(z)]. \label{ec51}
\end{eqnarray}
For $\h{R}^{\dag}(x,-i\tau)$ ($\h{L}^{\dag}(x,-i\tau)$) one has the same expression as $\h{R}(x,-i\tau)$ ($\h{L}(x,-i\tau)$) with $+i$ ($-i$) however replaced by $-i$ ($+i$). Compare with the relevant expressions in the bosonization tables 3.1 and 22.1 of Refs.~\cite{GNT98} and \cite{AMT03}, respectively.

With $f(z)$ and $g(\b{z})$ differentiable functions of $z$ and $\b{z}$, one has $\partial_{\tau} f(z) \equiv -i v_{\Sc f} \partial_x f(z)$ and $\partial_{\tau} g(\b{z}) \equiv +i v_{\Sc f} \partial_x g(\b{z})$, so that
\begin{equation}
\partial_{z} = \frac{1}{2} (\partial_{\tau} - i v_{\Sc f}\, \partial_x), \;\;\; \partial_{\b{z}} = \frac{1}{2} (\partial_{\tau} + i v_{\Sc f}\, \partial_x), \label{ec52}
\end{equation}
and consequently
\begin{equation}
\partial_x = \frac{i}{v_{\Sc f}}\, (\partial_z - \partial_{\b{z}}). \label{ec53}
\end{equation}
From Eqs.~(\ref{ec48}) and (\ref{ec49}) one trivially obtains that
\begin{eqnarray}
\partial_z \h{\Phi}(z,\b{z}) &\equiv& \frac{1}{2}\partial_z (\h{\Phi}(z,\b{z}) + \h{\Theta}(z,\b{z})), \;\;\;\;\;\label{ec54} \\
\partial_{\b{z}} \h{\Phi}(z,\b{z}) &\equiv& \frac{1}{2}\partial_{\b{z}} (\h{\Phi}(z,\b{z}) - \h{\Theta}(z,\b{z})). \;\;\;\;\; \label{ec55}
\end{eqnarray}
One can thus make the following identifications (see the bosonization tables 3.1 and 22.1 of Refs.~\cite{GNT98} and \cite{AMT03}, respectively):
\begin{eqnarray}
\h{R}^{\dag}(x,-i\tau) \h{R}(x,-i\tau) &=& \frac{-i}{\sqrt{\pi} v_{\Sc f}} \partial_{\b{z}} \h{\Phi}(z,\b{z}), \label{ec56}\\
\h{L}^{\dag}(x,-i\tau) \h{L}(x,-i\tau) &=& \frac{+i}{\sqrt{\pi} v_{\Sc f}} \partial_{z} \h{\Phi}(z,\b{z}). \label{ec57}
\end{eqnarray}
Making use of the equalities in Eqs.~(\ref{ec54}) and (\ref{ec55}), one readily verifies that for $\tau=0$ the expressions in Eqs.~(\ref{ec56}) and (\ref{ec57}) exactly reproduce $\h{\rho}_{\Sc r}(x)$ and $\h{\rho}_{\Sc l}(x)$ respectively, where $\h{\rho}_{\Sc r}(x)$ and $\h{\rho}_{\Sc l}(x)$ are defined in Eq.~(\ref{ec21}) (see Eqs.~(\ref{ec22}) and (\ref{ec23})).

In order to capture the correct pre-factor $\frac{1}{2}$ in the bosonized non-interacting Hamiltonian, with reference to Eq.~(\ref{ec53}) we propose the following \emph{substitutions} in going from the LHS to the RHS of Eq.~(\ref{ec44}):
\begin{eqnarray}
\h{R}^{\dag}(x) \partial_x \h{R}(x) &\rightharpoonup& -\frac{i}{v_{\Sc f}} \h{R}^{\dag}(x,0) \partial_{\b{z}}\h{R}(x,0), \nonumber\\
\h{L}^{\dag}(x) \partial_x \h{L}(x) &\rightharpoonup& +\frac{i}{v_{\Sc f}} \h{L}^{\dag}(x,0) \partial_{z}\h{L}(x,0). \label{ec58}
\end{eqnarray}
These substitutions are motivated by the fact that $\h{R}(x,-i\tau)$ depends only on $\b{z}$, Eq.~(\ref{ec50}), and $\h{L}(x,-i\tau)$ depends only on $z$, Eq.~(\ref{ec51}). Using the expressions in Eq.~(\ref{ec52}), one trivially obtains that
\begin{eqnarray}
-\frac{i}{v_{\Sc f}} \h{R}^{\dag}(x,0) \partial_{\b{z}}\h{R}(x,0) &\equiv& \frac{1}{2} \h{R}^{\dag}(x) \partial_x \h{R}(x), \nonumber\\
+\frac{i}{v_{\Sc f}} \h{L}^{\dag}(x,0) \partial_{z}\h{L}(x,0) &\equiv& \frac{1}{2} \h{L}^{\dag}(x) \partial_x \h{L}(x), \label{ec59}
\end{eqnarray}
where the $\partial_x \h{R}(x)$ and $\partial_x \h{L}(x)$ on the RHSs are to be expressed in terms of $\partial_x \h{\Theta}(x)$ and $\partial_x \h{\Phi}(x)$ on the basis of the expressions in Eq.~(\ref{ec43}). The pre-factors $\frac{1}{2}$ on the RHSs of the expressions in Eq.~(\ref{ec59}) account for the aforementioned missing $\frac{1}{2}$ on the RHS of Eq.~(\ref{ec44}). These $\frac{1}{2}$ pre-factors would turn into $1$ on replacing the $\partial_{\b{z}}\h{R}(x,0)$ and $\partial_{z}\h{L}(x,0)$ in Eq.~(\ref{ec58}) by respectively $\partial_{\b{z}}\h{R}(x,-i\tau)\vert_{\tau=0}$ and $\partial_{z}\h{L}(x,-i\tau)\vert_{\tau=0}$. Thus, the \emph{substitutions} in Eq.~(\ref{ec58}) prescribe a particular scheme, according to which the \emph{total} variations of $\h{R}(x,-i\tau)$ and $\h{L}(x,-i\tau)$ at $\tau=0$ are to be identified with their \emph{partial} variations along the $x$ axis.

We thus obtain that (cf. Eq.~(\ref{ec44}))
\begin{eqnarray}
&&\hspace{-0.5cm} -(2a t)\, i \!\int_0^{L} \rd x\; \big( \h{R}^{\dag}(x) \partial_x \h{R}(x) - \h{L}^{\dag}(x) \partial_x \h{L}(x) \big) \nonumber\\
&&\hspace{0.5cm} \simeq \frac{v_{\Sc f}}{2} \int_0^{L} \rd x\; \big( (\partial_x\h{\Theta}(x))^2 + (\partial_x\h{\Phi}(x))^2\big) \nonumber\\
&&\hspace{0.5cm} \equiv \frac{v_{\Sc f}}{2} \int_0^{L} \rd x\; \big( \h{\Pi}^2(x) + (\partial_x\h{\Phi}(x))^2\big), \label{ec60}
\end{eqnarray}
where in arriving at the last expression we have employed Eq.~(\ref{ec19}). Consequently, from Eqs.~(\ref{ec33}), (\ref{ec35}) and (\ref{ec40}) one obtains that
\begin{eqnarray}
&&\hspace{-1.0cm} -\sum_{j=1}^{N} (t\,\h{c}_{j}^{\dag} \h{c}_{j+1} + t'\, \h{c}_{j}^{\dag} \h{c}_{j+2})  + \mathrm{H.c.} \nonumber\\
&&\hspace{0.5cm} \rightleftharpoons \frac{v_{\Sc f}}{2} \int_0^{L} \rd x\; \big\{ \h{\Pi}^2(x) + \big(\partial_x \h{\Phi}(x)\big)^2\big\}, \label{ec61}
\end{eqnarray}
where `$\rightleftharpoons$' denotes the association between the original Hamiltonian on the left and its bosonized counterpart on the right.

Following Eq.~(\ref{ec5}), for $\h{n}_j \doteq \h{c}_j^{\dag} \h{c}_j$ one has
\begin{eqnarray}
&&\hspace{-1.0cm} \h{n}_j \simeq a \big\{ \h{R}^{\dag}(x) \h{R}(x) +\h{L}^{\dag}(x) \h{L}(x)\nonumber\\
&&\hspace{0.1cm} + (-1)^j \big[ \h{R}^{\dag}(x) \h{L}(x) + \h{L}^{\dag}(x) \h{R}(x) \big] \big\}, \label{ec62}
\end{eqnarray}
which on normal-ordering can be expressed as
\begin{equation}
\h{n}_j \simeq a \big\{ \h{\rho}(x) + (-1)^j\, \h{M}(x) \big\}, \label{ec63}
\end{equation}
where $\h{\rho}(x)$ is the total-density operator, Eq.~(\ref{ec22}), and
\begin{equation}
\h{M}(x) \doteq\; :\!\h{R}^{\dag}(x) \h{L}(x)\!: + :\!\h{L}^{\dag}(x) \h{R}(x)\!:, \label{ec64}
\end{equation}
an operator that measures the amount of chiral symmetry breaking. With reference to Eq.~(\ref{e20}), from Eq.~(\ref{ec63}) one observes that, at half-filling the GS expectation value of $a \h{\rho}(x)$ per unit length, i.e. $Q/N$, is equal to $1/2$, and that the GS expectation value of $a \h{M}(x)$ per unit length is equal to $\delta n$. In this connection, and with reference to Eqs.~(\ref{ec2}), (\ref{ec25}) and (\ref{ec29}), we note that \[\frac{1}{N} \sum_{j=1}^{N} \h{n}_j \simeq \frac{1}{L} \int_0^L \rd x\; a \h{\rho}(x) \equiv \frac{\h{Q}}{N}. \] As we shall indicate in Sec.~\ref{sac.4}, $\delta n$ is the counterpart of the staggered magnetization $\mathfrak{M}$ in the Ising anti-ferromagnetic GS of the $XXZ$ Heisenberg Hamiltonian for SU(2) spin operators.

Making use of the expression in Eq.~(\ref{ec63}), one obtains that
\begin{eqnarray}
&&\hspace{-1.2cm} \h{n}_j \h{n}_{j+1} \simeq a^2 \big\{ \h{\rho}(x) \h{\rho}(x+a) - \h{M}(x) \h{M}(x+a) \nonumber\\
&&\hspace{-0.4cm} - (-1)^j \big[ \h{\rho}(x) \h{M}(x+a) - \h{M}(x) \h{\rho}(x+a)\big] \big\}. \label{ec65}
\end{eqnarray}
By the same reasoning as leading to the result in Eq.~(\ref{ec31}), the contributions to $\sum_j \h{n}_j \h{n}_{j+1}$ of the two terms on the RHS pre-multiplied by $(-1)^j$ can be neglected in the limit $a\to 0$. In what follows we shall therefore only consider the contributions of $\h{\rho}(x) \h{\rho}(x+a)$ and $\h{M}(x) \h{M}(x+a)$ to the latter sum.

Following Eq.~(\ref{ec22}), for $a \to 0$ one has
\begin{equation}
\h{\rho}(x) \h{\rho}(x+a) \simeq \h{\rho}(x) \h{\rho}(x) \equiv \frac{1}{\pi} \big(\partial_x \h{\Phi}(x)\big)^2. \label{ec66}
\end{equation}

In order to determine $\h{M}(x) \h{M}(x+a)$, we make use of the expressions in Eqs.~(\ref{ec50}) and (\ref{ec51}). In doing so, we employ the Baker-Campbell-Hausdorff formula in Eq.~(\ref{ec7}) to express products of exponentials of $\h{\varphi}(z)$ and $\h{\b\varphi}(\b{z})$ as an exponential of $\h{\varphi}(z) +\h{\b\varphi}(\b{z}) \equiv \h{\Phi}(z,\b{z})$, Eq.~(\ref{ec48}). As we are dealing with $\tau=0$, we need only to determine $[\h{\varphi}(i x/v_{\Sc f}),\h{\b\varphi}(-i x'/v_{\Sc f})]_{-}$. Since $\h{\Phi}(z,\b{z})$ ($\h{\Theta}(z,\b{z})$) coincides with $\h{\Phi}(x+0^+)$ ($\h{\Theta}(x)$) for $\tau=0$, from Eqs.~(\ref{ec48}) and (\ref{ec49}) and the commutation relation in Eq.~(\ref{ec18}) (i.e. the one corresponding to $\alpha\downarrow 0$, $L\to\infty$) one readily obtains that
\begin{widetext}
\begin{equation}
\big[\h{\varphi}(\frac{i x}{v_{\Sc f}}),\h{\b\varphi}(-\frac{i x'}{v_{\Sc f}})\big]_{-}
\equiv  \frac{1}{4} \big[\h{\Phi}(x+0^+)+\h{\Theta}(x), \h{\Phi}(x'+0^+)-\h{\Theta}(x')\big]_{-} = \frac{i}{8} \big\{ \sgn(x-x'+0^+) - \sgn(x-x'-0^+)\big\}.  \label{ec67}
\end{equation}
\end{widetext}
Thus one has
\begin{eqnarray}
\h{R}^{\dag}(x,0) \h{L}(x,0) &\simeq& \frac{+i}{2\pi a} \exp[-i\sqrt{4\pi}\, \h{\Phi}(x)], \nonumber\\
\h{L}^{\dag}(x,0) \h{R}(x,0) &\simeq& \frac{-i}{2\pi a} \exp[+i\sqrt{4\pi}\, \h{\Phi}(x)]. \label{ec68}
\end{eqnarray}
These expressions can be directly obtained by making use of the first equality in Eq.~(\ref{ec42}). From Eqs.~(\ref{ec64}) and (\ref{ec68}) one deduces that
\begin{equation}
\h{M}(x) \simeq \frac{1}{\pi a} :\!\sin\!\big(\!\sqrt{4\pi}\, \h{\Phi}(x)\big)\!:. \label{ec69}
\end{equation}
We note that without the $0^+$ in $\h{\Phi}(x+0^+)$ in the expressions for $\h{R}(x)$ and $\h{L}(x)$ (here, those according to the second equality in Eq.~(\ref{ec42})), the pre-factors in the expressions in Eq.~(\ref{ec68}) would have involved $+1$ instead of $\pm i$, resulting in $\h{M}(x) \simeq -\frac{1}{\pi a} :\!\cos(\sqrt{4\pi}\,\h{\Phi}(x))\!:$ (compare with Eqs.~(6.23) and (6.24) in Ref.~\cite{TG03}; \cite{Note3}).

Since $\h{\Phi}(x)$ commutes with $\h{\Phi}(x+a)$ (see remark (ii) on page~\pageref{items:i-v}), in calculating $\h{M}(x) \h{M}(x+a)$ one can consider $\h{\Phi}(x)$ and $\h{\Phi}(x+a)$ as $c$ numbers. Making use of the identity
\begin{equation}
2 \sin(x) \sin(y) \equiv \cos(x-y) - \cos(x+y), \label{ec70}
\end{equation}
and of the expansion
\begin{equation}
\h{\Phi}(x + a) \simeq \h{\Phi}(x) + a\, \partial_x \h{\Phi}(x)\;\;\; \mbox{\rm for}\;\;\; a\to 0, \label{ec71}
\end{equation}
as well as the relationship
\begin{equation}
:\!\cos(\h{A})\!: \;\approx -\frac{1}{2} \h{A}^2, \label{ec72}
\end{equation}
one readily obtains that
\begin{widetext}
\begin{equation}
\h{M}(x) \h{M}(x+a) \approx \frac{-1}{2 (\pi a)^2}\, \cos\!\big(\!\sqrt{16\pi}\, \h{\Phi}(x)\big) - \frac{1}{\pi} \big(\partial_x \h{\Phi}(x)\big)^2. \label{ec73}
\end{equation}

From Eqs.~(\ref{ec65}), (\ref{ec66}) and (\ref{ec73}), one arrives at
\begin{equation}
\sum_{j=1}^{N} \h{n}_j \h{n}_{j+1} \approx \!\int_0^{L} \rd x\; \Big\{ \frac{2 a}{\pi} \big(\partial_x \h{\Phi}(x)\big)^2 +\frac{1}{2\pi^2 a} \cos\!\big(\!\sqrt{16\pi}\, \h{\Phi}(x)\big)\Big\}\;\;\; \mbox{\rm as}\;\;\; a\to 0. \label{ec74}
\end{equation}
Combining this result with that in Eq.~(\ref{ec61}), one obtains the following correspondence between $\wh{H}$ and the bosonic Hamiltonian $\wh{\mathcal{H}}$ in the continuum limit:
\begin{equation}
\wh{H} \rightleftharpoons \wh{\mathcal{H}} \doteq \int_0^{L} \rd x\; \Big\{ \frac{v_{\Sc f}}{2}\big[ \h{\Pi}^2(x) + \big(1 + \frac{2 V}{\pi t}\big) \big(\partial_x \h{\Phi}(x)\big)^2\big] + \frac{2 a V}{(2\pi a)^2} \cos\!\big(\!\sqrt{16\pi}\, \h{\Phi}(x)\big)\Big\}. \label{ec75}
\end{equation}
\end{widetext}
Provided that $V/t$ is sufficiently small, this sine-Gordon Hamiltonian \cite{SC75,GNT98,AMT03} governs the dynamics of the field operators $\h{\Pi}(x)$ and $\h{\Phi}(x)$. With reference to our remark following Eq.~(\ref{ec69}) above, we note that without the $0^+$ in $\h{\Phi}(x+0^+)$ in the expressions for $\h{R}(x)$ and $\h{L}(x)$ (i.e. the last expressions on the RHS of Eq.~(\ref{ec42})), the `$+$' pre-multiplying the last term on the RHS of Eq.~(\ref{ec75}) would have been `$-$' (compare with Eq.~(6.29) in Ref.~\cite{TG03}; see also the remarks centred around Eq.~(\ref{ec80}) below). This follows from the identity (cf. Eq.~(\ref{ec70})) \[ 2\cos(x) \cos(y) \equiv \cos(x-y) + \cos(x+y), \] which would have resulted in $+1$ instead of the $-1$ in the first term on the RHS of Eq.~(\ref{ec73}).

The Hamiltonian in Eq.~(\ref{ec75}) coincides with that in Eq.~(6.29) of Ref.~\cite{TG03} (taking into account that $J_z$ in the latter reference is the equivalent of the $V$ in our present considerations \cite{Note3}, that $g_3 \equiv a J_z$, and that, for the reason specified in the previous paragraph, the sign of the last term on the RHS of Eq.~(\ref{ec75}) is opposite to that of the corresponding term in Eq.~(6.29) of Ref.~\cite{TG03}). Further, the coefficient of the cosine term in Eq.~(29.27) of Ref.~\cite{AMT03}, and that in Eq.~(11.14) of Ref.~\cite{GNT98}, is twice too large; this is due to partially neglecting the $2$ on the LHS of the identity in Eq.~(\ref{ec70}) above. In verifying this statement, one should note that the ratio $V/(2t)$ is the equivalent of the dimensionless quantity $\Delta$, Eq.~(\ref{ec104}), the anisotropy parameter, in Eq.~(29.27) of Ref.~\cite{AMT03}, and in Eq.~(11.14) of Ref.~\cite{GNT98}.

\subsection{The (exact) bosonized Hamiltonian for small $\vert t'\vert$ (for $t'=0$) and arbitrary values of $t$ and $V$}
\label{sac.3}

As we have indicated earlier, the above analysis is valid for $V$ sufficiently small with respect to $t$ and $t'$. Here we consider the specific case corresponding to $t'=0$, for which the model under consideration is identically equivalent to the $XXZ$ Heisenberg Hamiltonian for SU(2) spin operators. Regarding the specific choice $t'=0$, we refer the reader to our pertinent remarks in the paragraph preceding Eq.~(\ref{ec42}) on page~\pageref{zerot1}. Since the transfer matrix of the zero-field eight-vertex model \cite{RJB72} (defined on a two-dimensional lattice) commutes with the Hamiltonian of the one-dimensional $XYZ$ model for SU(2) spin operators \cite{McCW68,BS70}, on the basis of the exact solution of the former model \cite{RJB72,RJB07}, one can deduce the low-energy properties of the latter model \cite{JKMC73,AL76}. This enables one to deduce the bosonized Hamiltonian of the $t$-$V$ model at half-filling for in principle arbitrary values of $V$.

\begin{widetext}
The bosonized Hamiltonian corresponding to the $t$-$V$ Hamiltonian can be brought into the standard sine-Gordon form
\begin{equation}
\wh{\mathcal{H}} = \int_0^{L}\!\! \rd x\, \Big\{\frac{u}{2} \big[K \h{\Pi}^2(x) + \frac{1}{K} \big(\partial_x \h{\Phi}(x)\big)^2\big] + g \cos\big(\sqrt{16\pi}\, \h{\Phi}(x)\big)\Big\}. \label{ec76}
\end{equation}
\end{widetext}
A comparison of this Hamiltonian with that in Eq.~(\ref{ec75}) reveals that for sufficiently small $V/t$ one has
\begin{eqnarray}
g &=& \frac{2 a V}{(2\pi a)^2}, \label{ec77}\\
\frac{1}{K} &\simeq& \big(1+ \frac{2V}{\pi t}\big)^{1/2}, \label{ec78}\\
u &\simeq& v_{\Sc f} \big(1+ \frac{2V}{\pi t}\big)^{1/2}.
\label{ec79}
\end{eqnarray}
Later we shall establish (see the discussions following Eqs.~(\ref{ec116}) and (\ref{ec123})) that the RHSs of the expressions in Eqs.~(\ref{ec78}) and (\ref{ec79}) are exact to leading order in $V/t$.

For completeness, we point out that the sign of the coupling constant $g$ is not crucial, as it can be reversed on effecting the shift transformation:
\begin{equation}
\h{\Phi}(x) \rightharpoonup \h{\Phi}(x) + \frac{2 l + 1}{4}\,\sqrt{\pi}, \label{ec80}
\end{equation}
where $l$ is an arbitrary integer (negative, zero and positive). Since the constant term on the RHS of Eq.~(\ref{ec80}) is a $c$ number, the transformed $\h{\Phi}(x)$ remains to satisfy the commutation relations in Eq.~(\ref{ec18}); consequently, the commutation relation in Eq.~(\ref{e24}) remains equally applicable to the transformed $\h{\Phi}(x)$ as the untransformed $\h{\Phi}(x)$. With reference to Eqs.~(\ref{ec6}) and (\ref{ec42}), the transformation in Eq.~(\ref{ec80}) gives rise to the multiplicative phase factors $\e^{+i (2 l +1)\pi/4}$ and $\e^{-i (2 l +1)\pi/4}$ in the expressions for $\h{R}(x)$ and $\h{L}(x)$ respectively.

In the following we shall have occasion to refer to the Euclidean action corresponding to the Hamiltonian in Eq.~(\ref{ec76}), which we determine now. With $\tau$ the imaginary time, Eq.~(\ref{ec45}), on the basis of the Hamilton equation of the classical mechanics of point particles, i.e. (Eq.~(40.4) in Ref.~\cite{LLV1})
\begin{equation}
\stackrel{\bm\cdot}{q}_i = \frac{\partial H}{\partial p_i}, \label{ec81}
\end{equation}
one has
\begin{equation}
\stackrel{\;\,\bm\cdot}{{\h\Phi}\!}(x,-i\tau) \equiv i \partial_{\tau} \h{\Phi}(x,-i\tau) = \frac{\delta \wh{\mathcal{H}}}{\delta \h{\Pi}(x,-i\tau)}, \label{ec82}
\end{equation}
where $\wh{\mathcal{H}}$ stands for the Hamiltonian as expressed in terms of the imaginary-time Heisenberg-picture counterparts of $\h{\Pi}(x)$ and $\h{\Phi}(x)$. From Eq.~(\ref{ec76}) one obtains that
\begin{equation}
\stackrel{\;\,\bm\cdot}{\h{\Phi}\!}(x,-i\tau) = u K\, \h{\Pi}(x,-i\tau). \label{ec83}
\end{equation}
For the Euclidean action one thus has (cf. Eqs.~(2.1) and (40.2) in Ref.~\cite{LLV1})
\begin{widetext}
\begin{eqnarray}
\wh{S} &\doteq& -\int_{0}^{\beta} \rd \tau\; \Big\{ \h{\Pi}(x,-i\tau) \stackrel{\;\,\bm\cdot}{\h{\Phi}\!}(x,-i\tau) - \wh{\mathcal{H}} \Big\} \nonumber\\ &\equiv& \int \rd x\; \rd \tau\; \Big\{ \frac{1}{2} \big[ \frac{1}{u K} \big(\partial_{\tau} \h{\Phi}(x,-i\tau)\big)^2 + \frac{u}{K} \big(\partial_x \h{\Phi}(x,-i\tau)\big)^2 \big] + g \cos\!\big(\sqrt{16\pi}\, \h{\Phi}(x,-i\tau)\big)\Big\}, \label{ec84}
\end{eqnarray}
\end{widetext}
where $\beta \doteq 1/T$ (with $\hbar= k_{\Sc b} =1$). In the second expression on the RHS of Eq.~(\ref{ec84}), $\int$ stands for $\int_0^{L} \int_{0}^{\beta}$. Introducing
\begin{eqnarray}
\ul{\h{\Phi}}(x,-i\ul{\tau}) &\equiv& \h{\Phi}(x,-i\tau), \label{ec85}\\
\ul{\h{\Pi}}(x,-i\ul{\tau}) &\equiv& \h{\Pi}(x,-i\tau), \label{ec86}
\end{eqnarray}
where
\begin{equation}
\ul{\tau} \doteq u \tau, \label{ec87}
\end{equation}
and subjecting the fields $\ul{\h{\Phi}}(x,-i\ul\tau)$ and $\ul{\h{\Pi}}(x,-i\ul\tau)$ to the canonical transformations (see Eq.~(\ref{e24}))
\begin{eqnarray}
\ul{\h{\Phi}}(x,-i\ul\tau) &\rightharpoonup& \sqrt{K}\, \ul{\h{\Phi}}(x,-i\ul\tau), \label{ec88} \\
\ul{\h{\Pi}}(x,-i\ul\tau) &\rightharpoonup& \frac{1}{\sqrt{K}}\, \ul{\h{\Pi}}(x,-i\ul\tau), \label{ec89}
\end{eqnarray}
the action in Eq.~(\ref{ec84}) can be expressed as
\begin{equation}
\wh{S} = \wh{S}_0 + \ul{g} \int {\rm d}^2\ul{r}\; \cos(\upbeta \ul{\h{\Phi}}(x,-i\ul\tau)), \label{ec90}
\end{equation}
where
\begin{equation}
\wh{S}_0 \doteq \frac{1}{2} \int {\rm d}^2\ul{r}\; \Big[ \big(\partial_{\ul\tau} \ul{\h{\Phi}}(x,-i\ul\tau)\big)^2 + \big(\partial_{x} \ul{\h{\Phi}}(x,-i\ul\tau)\big)^2\Big] \label{ec91}
\end{equation}
is the Gaussian action, and
\begin{equation}
\ul{g} \doteq \frac{g}{u},\;\;\;\;\; \upbeta \doteq \sqrt{16\pi K}. \label{ec92}
\end{equation}
The constant $\upbeta$ is not to be confused with $\beta \doteq 1/T$. Above,
\begin{equation}
{\rm d}^2\ul{r} \doteq \rd x\, \rd \ul\tau \equiv u\, \rd x\, \rd \tau \doteq u\, {\rm d}^2r. \label{ec93}
\end{equation}
The two-dimensional integrals in Eqs.~(\ref{ec90}) and (\ref{ec91}) are over the rectangle $[0,L] \times [0, \ul{\beta}]$, where
\begin{equation}
\ul{\beta} \doteq u \beta. \label{ec94}
\end{equation}
\emph{In the following, where no confusion can arise, we shall suppress the bars below the symbols $\ul{\tau}$, $\ul{g}$ and $\ul{\h{\Phi}}(x,-i\ul\tau)$.}

With $d$ denoting the \emph{scaling dimension} of the field $\cos(\upbeta \h{\Phi}(x,-i\tau))$ (see Eq.~(\ref{ec100})), for which one has (see Eq.~(\ref{ec102}))
\begin{equation}
d = \frac{\upbeta^2}{4\pi}, \label{ec95}
\end{equation}
and $D$ the dimension of the space over which the integral in Eq.~(\ref{ec90}) is carried out (here $D=2$), according to a well-known theorem one has that for $d > D$ the perturbation to the Gaussian action $\wh{S}_0$ in Eq.~(\ref{ec90}) is \emph{irrelevant}, for $d <D$ it is \emph{relevant}, and for $d = D$ \emph{marginal} (see pp.~206 and 207 in Ref.~\cite{AMT03}). In the case at hand, where $D=2$, the perturbation to the Gaussian action $\wh{S}_0$ is irrelevant when
\begin{equation}
\frac{\upbeta^2}{4\pi} > 2, \label{ec96}
\end{equation}
etc. Using the expression for $\upbeta$ in Eq.~(\ref{ec92}), one deduces that for
\begin{equation}
K > \frac{1}{2} \label{ec97}
\end{equation}
the perturbation to the Gaussian action is irrelevant, for $K <\frac{1}{2}$ it is relevant, and for $K = \frac{1}{2}$ marginal. Thus, for $K >\frac{1}{2}$, and sufficiently small $\ul{g}$, perturbation theory applies and the essential physics of the sine-Gordon action is determined by the Gaussian action $\wh{S}_0$. Amongst others, for $K >\frac{1}{2}$ the single-particle excitation spectrum of the system is gapless and thus the system is `critical' and the GS `disordered' (see later, in particular the remarks in the paragraph preceding Eq.~(\ref{ec145}) on page~\pageref{critical1}). With reference to Eq.~(\ref{ec78}), one observes that for $K \approx \frac{1}{2}$ one must have $V/t \approx 3\pi/2 \approx 4.71$, which is too large for Eq.~(\ref{ec78}) to be accurate. As we shall see later, $K =\frac{1}{2}$ exactly corresponds to $V/t = 2$. Thus, with $V_{\rm c}(t,t')$ denoting the critical value of $V$ for the one-dimensional $t$-$t'$-$V$ model on a regular lattice, one has (cf. Eq.~(\ref{ec119}) below)
\begin{equation}
V_{\rm c}(t,0) = 2 t. \label{ec98}
\end{equation}

With
\begin{equation}
z \doteq \tau + i x/u \;\;\;\; \mbox{\rm and} \;\;\;\; z' \doteq \tau' + i x'/u,
\label{ec99}
\end{equation}
the scaling dimension $d$ of the field $\cos(\upbeta \h{\Phi}(x,-i\tau))$ is defined according to the following expression:
\begin{widetext}
\begin{equation}
\langle \cos(\upbeta \h{\Phi}(x,-i\tau)) \cos(\upbeta \h{\Phi}(x',-i\tau')) \rangle_0 \sim \frac{C}{\vert z - z'\vert^{2d}} \;\;\; \mbox{\rm for}\;\;\; \vert z - z'\vert \to \infty, \label{ec100}
\end{equation}
where $C$ is a constant. The average $\langle\dots\rangle_0$ is with respect to the Gaussian action $\wh{S}_0$, whereby the $u$'s in Eq.~(\ref{ec99}) are to be momentarily identified with $v_{\Sc f}$, Eq.~(\ref{ec46}). To calculate $d$, one employs the identity $\cos(x) \equiv \frac{1}{2} (\e^{i x} + \e^{-i x})$. Using the left-most identification in Eq.~(\ref{ec47}), on the basis of the fact that  (cf. Eqs.~(22.8) and (22.9) in Ref.~\cite{AMT03}; see also Appendix C in Ref.~\cite{TG03})
\begin{equation}
\langle \e^{i \upbeta \h{\Phi}(z,\b{z})}\, \e^{i \upbeta \h{\Phi}(z',\b{z}')} \rangle_0 = 0\;\;\; \mbox{\rm for}\;\; \frac{L}{a} \to \infty,
\label{ec101}
\end{equation}
one obtains that
\begin{eqnarray}
\langle \cos(\upbeta \h{\Phi}(x,-i\tau)) \cos(\upbeta \h{\Phi}(x',-i\tau')) \rangle_0 &=& \frac{1}{4} \Big\{\langle \e^{i \upbeta \h{\Phi}(z,\b{z})}\, \e^{-i \upbeta \h{\Phi}(z',\b{z}')} \rangle_0 + \langle \e^{-i \upbeta \h{\Phi}(z,\b{z})}\, \e^{i \upbeta \h{\Phi}(z',\b{z}')} \rangle_0 \Big\}\nonumber\\
&=& \frac{1}{2} \left|\frac{a}{z - z'}\right|^{\upbeta^2/2\pi}\;\;\; \mbox{\rm for}\;\; \frac{L}{a} \gg 1, \; \vert z - z'\vert \gg a. \label{ec102}
\end{eqnarray}
Comparing the last expression in this equation with the RHS of Eq.~(\ref{ec100}), one arrives at the expression for $d$ in Eq.~(\ref{ec95}).
\end{widetext}

\subsection{Exact results for $K$, $u$ and $\mathcal{M}$}
\label{sac.4}

Following Eq.~(\ref{ec92}), $\upbeta$ is seen to be determined by $K$, Eq.~(\ref{ec76}). For determining $K$, one employs the correspondence between the $t$-$V$ Hamiltonian (or the $t$-$t'$-$V$ Hamiltonian at half-filling and for small values of $t'$ --- see our pertinent remarks in the paragraph preceding Eq.~(\ref{ec42}) on page~\pageref{zerot1}) and the $XXZ$ Heisenberg Hamiltonian for SU(2) spin operators (also known as the Heisenberg-Ising Hamiltonian), which is established through the following identifications (compare Eq.~(\ref{e4}) with Eq.~(6.10) in Ref.~\cite{TG03}):
\begin{equation}
J_{xy} \leftrightarrow 2 t,\;\;\; J_{z} \leftrightarrow V. \label{ec103}
\end{equation}
Here, $J_{xy}$ denotes the common value of $J_{x}$ and $J_{y}$. One defines
\begin{equation}
\Delta \doteq \frac{J_{z}}{J_{xy}} \equiv \frac{V}{2t}, \label{ec104}
\end{equation}
which measures the anisotropy of the exchange parameter in the $z$ direction in comparison with those in the $x$ and $y$ directions.

\label{chiralsb}
For the $XXZ$ Heisenberg Hamiltonian, in the thermodynamic limit, one can show that for $\vert\Delta\vert \le 1$ the system is `critical' and the GS `disordered' (see later, in particular the remarks in the paragraph preceding Eq.~(\ref{ec145}) on page~\pageref{critical1}); at $\Delta =-1$  the GS of the system is an isotropic Heisenberg ferromagnet, and at $\Delta = +1$ it is an Heisenberg anti-ferromagnet. For $\Delta < -1$ the GS is an Ising ferromagnet (FM), and for $\Delta> 1$ it is an Ising anti-ferromagnet (AFM). \emph{Whereas the Ising FMic GS is gapless, the Ising AFMic GS is gapped.} The Ising AFMic GS has a broken translational symmetry, characterised by a non-vanishing staggered magnetization $\mathfrak{M}$, the order parameter of the GS (for the relationship between $\mathfrak{M}$ and $\delta n$ as introduced in Eq.~(\ref{e20}), see the remarks following Eq.~(\ref{ec63}) above); the continuous symmetry of the GSs corresponding to $\Delta\le 1$ (i.e., the chiral symmetry of the underlying quantum sine-Gordon Hamiltonian) is for $\Delta>1$ spontaneously broken and reduced to a discrete $Z_2$ symmetry. Thus, the unit cell of this broken-symmetry GS is twice as large as that of the uniform GSs corresponding to $\Delta\le 1$ (Ch.~11 in Ref.~\cite{GNT98}, Ch. 6, in particular Fig.~6.1, in Ref.~\cite{TG03} and Ch. 29 in Ref.~\cite{AMT03}).

Below we consider in some detail the cases corresponding to $\vert\Delta\vert \le 1$, $\Delta <-1$ and $\Delta >1$.

\subsubsection{The case of $\vert \Delta\vert \le 1$\\ (including a discussion of the $XYZ$ model, with $J_x \not= J_y$)}
\label{sac.4a}

The value of $K$ is deduced from the leading-order asymptotic term of the correlation function
\begin{equation}
\langle \h{S}^z(x,-i\tau) \h{S}^z(0,0)\rangle_{\rm stag},\nonumber
\end{equation}
or of that of
\begin{equation}
\langle \h{S}^+(x,-i\tau) \h{S}^-(0,0)\rangle,\nonumber
\end{equation}
in the asymptotic region $\vert z\vert \equiv \vert \tau + i x/u\vert \to \infty$ (cf. Eq.~(\ref{ec99})). Here $\h{S}^z(x,-i\tau)$ is the $z$ component of the SU(2) spin operator in the continuum limit, and $\h{S}^{+}(x,-i\tau)$ and $\h{S}^-(x,-i\tau)$ are respectively the rasing and lowering SU(2) spin operators, also in the continuum limit. The latter three spin operators can be expressed in terms of the field operators $\h{\Phi}(x,-i\tau)$ and $\h{\Theta}(x,-i\tau)$, whereon the above-mentioned asymptotic terms are relatively straightforwardly calculated (see Ref.~\cite{LP75} and Ch.~6 of Ref.~\cite{TG03}; for a complementary calculation regarding the \emph{static} counterparts of these dynamic correlation functions, see Refs.~\cite{SL98,SL99}); on replacing $\langle\dots\rangle$ by $\langle\dots\rangle_0$, these asymptotic terms are readily calculated along the same lines as those leading to Eq.~(\ref{ec102}) above.

The subscript `stag', for `staggered', refers to that part of the correlation function $\langle \h{S}^z(x,-i\tau) \h{S}^z(0,0)\rangle$ whose amplitude is modulated by the phase factor $\e^{2 i k_{\Sc f} x}$ (at half-filling, for which $k_{\Sc f} = \frac{\pi}{2 a}$, Eq.~(\ref{ec3}), this phase factor gives rise to a sign alteration for the $j$ in $x \doteq j a$ increasing in steps of unity). We note that the asymptotic series expansion of $\langle \h{S}^z(x,-i\tau) \h{S}^z(0,0)\rangle$ corresponding to $\vert z\vert \to\infty$ contains a term that decays like $1/\vert z\vert^2$, so that, in view of Eq.~(\ref{ec106}) below, the leading-order term in the asymptotic series expansion of $\langle \h{S}^z(x,-i\tau) \h{S}^z(0,0)\rangle_{\rm stag}$, specific to $\vert z\vert \to\infty$, may not be the leading-order term of the asymptotic series expansion of $\langle \h{S}^z(x,-i\tau) \h{S}^z(0,0)\rangle$. This is in fact that case for $-1 \le \Delta < 0$, for which one has $1 < K \le \infty$; for $0 < \Delta \le 1$, in contrast, one has $\frac{1}{2} \le K < 1$ (see Eqs.~(\ref{ec108}) and (\ref{ec116})).

For the explicit asymptotic expressions of the first (second) of the above-indicated correlation functions, we refer the reader to Eqs.~(16) and (21) \cite{Note6} (Eqs.~(15) and $(15')$) of Ref.~\cite{LP75} (compare with Eqs.~(29.37) and (29.38) in Ref.~\cite{AMT03}, and with Eq.~(6.38) in Ref.~\cite{TG03}). In this connection, we note that
\begin{equation}
K \equiv \frac{1}{2\,\theta}, \label{ec105}
\end{equation}
where $\theta$ is the symbol employed in Ref.~\cite{LP75} ($\theta$ is denoted by $\eta$ in Refs.~\cite{SL98,SL99}). For $\vert z\vert \to \infty$ one has
\begin{equation}
|\langle \h{S}^z(x,-i\tau) \h{S}^z(0,0)\rangle_{\rm stag}| \sim  \frac{C_1}{\vert z\vert^{2K}}, \label{ec106}
\end{equation}
\begin{equation}
|\langle \h{S}^+(x,-i\tau) \h{S}^-(0,0)\rangle| \sim \frac{C_2}{\vert z\vert^{1/(2K)}}, \label{ec107}
\end{equation}
where $C_1$ and $C_2$ are some finite constants. These power-law decays are indicative of the mass-less spectrum of the uniform, or `disordered', GSs corresponding to $\vert\Delta\vert \le 1$, which, as we briefly indicated in the previous paragraph, correspond to $K \ge \frac{1}{2}$ (cf. Eq.~(\ref{ec97})). This is a peculiarity of the $XXZ$ model, since for the $XYZ$ model, the deviation of $J_x$ from $J_y$ gives rise to a massive spectrum, as we shall discuss later in this section (see Eqs.~(\ref{ec142}), (\ref{ec143}), (\ref{ec144}) and (\ref{ec163}) below).

For $\vert\Delta\vert \le 1$, $\Delta$ is expressed as
\begin{equation}
\Delta = + \cos(\upmu),\;\;\; 0\le \upmu \le \pi, \label{ec108}
\end{equation}
where $\upmu$ is one of the elliptic-function parameters that one encounters in the exact solution of the $8$-vertex model on a square lattice in the thermodynamic limit (see Ref.~\cite{JKMC73} and Ch.~10 in Ref.~\cite{RJB07}).

The plus sign on the RHS of Eq.~(\ref{ec108}), as opposed to the minus sign in the counterpart of this expression in Eq.~(10.16.8) of Ref.~\cite{RJB07}, calls for an explanation. The $8$-vertex model on a square lattice is solved in a so-called `fundamental region' (FR) of the parameter space (the parameters being conventionally denoted by $w_1$, $w_2$, $w_3$ and $w_4$ --- Eqs.~(10.2.16) and (10.1.2) in Ref.~\cite{RJB07}) and the solution for other regions of this space are deduced from the symmetry of the partition function of this model under some well-specified transformations of these parameters (Eq.~(10.2.17) in Ref.~\cite{RJB07}, and Appendix A in Ref.~\cite{JKMC73}). The essence of this symmetry property is encapsulated in a `rearrangement procedure' which for the specific case of the $XYZ$ model is very simple. For this model, the FR is defined as the region for which  (Sec. 10.15 in Ref.~\cite{RJB07})
\begin{equation}
\vert J_y \vert \le J_x \le -J_z \label{ec109}
\end{equation}
holds. For the cases where the parameters $\{ J_x, J_y, J_z\}$ do not satisfy the inequalities in Eq.~(\ref{ec109}), one defines the `rearranged' parameters $\{ J_x^{\rm r}, J_y^{\rm r}, J_z^{\rm r}\}$, satisfying these inequalities, obtained through a permutation of $\{ J_x, J_y, J_z\}$ in combination with possibly negation of a pair of the original parameters. The expressions presented in Eq.~(10.16.8) of Ref.~\cite{RJB07} correspond to
\begin{equation}
(J_x, J_y, J_z) = (1,1,\Delta), \label{ec110}
\end{equation}
for which in the case of $\vert \Delta\vert \le 1$ one has:
\begin{equation}
(J_x^{\rm r}, J_y^{\rm r}, J_z^{\rm r}) = (1,-\Delta,-1). \label{ec111}
\end{equation}
This in turn leads to $\Gamma_{\rm r} = -\Delta$ and $\Delta_{\rm r} = -1$ (for the definitions of $\Gamma_{\rm r}$ and $\Delta_{\rm r}$, see Eq.~(10.15.5) in Ref.~\cite{RJB07}). In our case, in contrast, one has (upon an appropriate normalisation of the parameters -- see Eqs.~(\ref{ec103}) and (\ref{ec104}); compare the Hamiltonian in Eq.~(19.1) of Ref.~\cite{AMT03}, or that in Eq.~(6.2) of Ref.~\cite{TG03}, with the Hamiltonian in Eq.~(10.14.1) of Ref.~\cite{RJB07} and in doing so note that $\h{S}_i^{\nu} \equiv \frac{1}{2} \h{\sigma}_i^{\nu}$, where $\nu = x, y, z$):
\begin{equation}
(J_x, J_y, J_z) = (-1,-1,-\Delta), \label{ec112}
\end{equation}
which in the case of $\vert \Delta\vert \le 1$ leads to
\begin{equation}
(J_x^{\rm r}, J_y^{\rm r}, J_z^{\rm r}) = (1,\Delta,-1). \label{ec113}
\end{equation}
In our case we thus have $\Gamma_{\rm r} = + \Delta$ and $\Delta_{\rm r} = -1$ (compare the latter $\Gamma_{\rm r}$ with the one presented following Eq.~(\ref{ec111}) above).

For completeness, in the case of $\vert \Delta\vert \le 1$, for the \emph{elliptic modulus} $\kappa$ (often denoted by $k$) one has $\kappa=1$ (Eq.~(10.15.6a) in Ref.~\cite{RJB07}). Due to the fact that \[\mathrm{snh}(\lambda,\kappa) \equiv -i\, \mathrm{sn}(i\lambda,\kappa),\] which for $\kappa=1$ is identical to $\tan(\lambda)$, from Eq.~(10.15.6b) in Ref.~\cite{RJB07} one immediately arrives at Eq.~(\ref{ec108}). In doing so one makes use of Eq.~(10.15.12) in Ref.~\cite{RJB07}, in which $I' \equiv K(\kappa')$, where $\kappa' \doteq (1-\kappa^2)^{1/2}$ is the \emph{elliptic conjugate modulus}; for $\kappa=1$, $\kappa' =0$, so that $I' = \pi/2$ (whereby $\lambda = \upmu/2$).

Above, $\mathrm{sn}(u,\kappa)$ is the Jacobian elliptic \emph{sinam} function (Sec.~22.11 in Ref.~\cite{WW62}) and $K$ is the \emph{complete} elliptic integral of the first kind (Ch.~15 in Ref.~\cite{RJB07}), or ``the constant $K$'' ($K \equiv K(\kappa)$) corresponding to $\kappa$ (Sec.~22.3 in Ref.~\cite{WW62}); by definition, $\mathrm{sn}(K,\kappa) =1$ (Sec.~22.3 in Ref.~\cite{WW62}). \emph{To avoid confusion with the $K$ as encountered in e.g. Eq.~(\ref{ec76}), in what follows we shall adopt the notation in Ref.~\cite{RJB07} and denote $K \equiv K(\kappa)$ by $I$ and $K' \equiv K(\kappa')$ by $I'$.}

As we shall have occasion to refer to some results in Ref.~\cite{AS72} regarding elliptic functions, we point out that whereas in our notation $\kappa^2$ is the parameter of elliptic functions and ${\kappa'}^2$ the complementary parameter, in Ref.~\cite{AS72} (in particular in Chaps. 16 and 17 herein), $m$ is the parameter of these functions and $m_1 \doteq 1 - m$ the complementary parameter (Sec.~16.1 in Ref.~\cite{AS72}).

For later use, we further remark that in the theory of elliptic integrals, one encounters the \emph{nome} $q \equiv q(m)$ and \emph{complementary nome} $q_1 \equiv q(m_1)$, defined according to (items 17.3.17 and 17.3.18 in Ref.~\cite{AS72}; see also Eq.~(15.7.6) in Ref.~\cite{RJB07})
\begin{equation}
q \equiv q(m) \doteq \exp[-\pi I'/I], \;\;\; q_1 \equiv q(m_1) \doteq \exp[-\pi I/I']. \label{ec114}
\end{equation}
For $q$ one has the following expansion (item 17.3.21 in Ref.~\cite{AS72}):
\begin{equation}
q(m) = \frac{m}{16} + 8 \left(\frac{m}{16}\right)^2 + 84 \left(\frac{m}{16}\right)^3 + \dots \;\;\; \mbox{\rm for}\;\;\; \vert m\vert <1. \label{ec115}
\end{equation}
This series is very useful in deducing the asymptotic series expansion of $q$ for $m \to 0$ and that of $q_1$ for $m\to 1$.

The elliptic-function parameter $\upmu$ in Eq.~(\ref{ec108}) is related to $K$ (as encountered in Eq.~(\ref{ec76})) as follows \cite{LP75} (cf. Eqs.~(\ref{ec105}) and (\ref{ec162})):
\begin{equation}
\upmu = \pi \big(1 - \frac{1}{2 K}\big) \iff K = \frac{\pi}{2 (\pi - \upmu)}. \label{ec116}
\end{equation}
Thus the $K$ corresponding to the cases where $\vert\Delta\vert \le 1$ is deduced by obtaining the value of $\upmu$ from Eq.~(\ref{ec108}) and substituting the resulting value into the right-most expression in Eq.~(\ref{ec116}). From Eq.~(\ref{ec116}) one observes that the non-interacting case, for which $K=1$, Eq.~(\ref{ec78}), corresponds to $\upmu=\frac{\pi}{2}$, and, in view of Eq.~(\ref{ec108}), to $\Delta = 0$. Following Eq.~(\ref{ec104}), indeed $\Delta =0$ corresponds to $V = 0$.

Making use of the series expansion \[\arccos(x) = \frac{\pi}{2} - x - \frac{1}{6} x^3 -\dots\;\;\; \mbox{\rm for}\;\;\; x\to 0, \] from Eqs.~(\ref{ec108}) and (\ref{ec116}) one deduces that
\begin{equation}
K = 1 - \frac{2\Delta}{\pi} + \frac{4 \Delta^2}{\pi^2} + \dots~, \label{ec117}
\end{equation}
to be compared with the result obtained from the expression in Eq.~(\ref{ec78}), i.e.
\begin{equation}
K = 1 - \frac{2 \Delta}{\pi} + \frac{6 \Delta^2}{\pi^2} + \dots~. \label{ec118}
\end{equation}
One observes that to the linear order in $\Delta \doteq V/(2 t)$ the two expressions coincide.

From Eqs.~(\ref{ec116}), (\ref{ec108}) and (\ref{ec104}) one deduces that
\begin{equation}
K = \frac{1}{2} \iff \upmu = 0 \iff \Delta = 1 \iff V = 2t. \label{ec119}
\end{equation}
In particular, the latter exact equality (cf. Eq.~(\ref{ec98})) deviates from the perturbative result $V \approx \frac{3\pi}{2} t$ (obtained from Eq.~(\ref{ec78})) to which we referred in the text following Eq.~(\ref{ec97}) above.

To make contact with Ref.~\cite{LP75}, we note that the expression in Eq.~(\ref{ec116}) is the equivalent of that in Eq.~$(21')$ of Ref.~\cite{LP75} where $J_{xy}$ is identified with unity and thus where $J_{z}$ is the equivalent of the $\Delta$ in our present considerations (cf. Eq.~(\ref{ec104})); for clarity, one should recall Eq.~(\ref{ec105}) above and note that $\arcsin(\Delta) + \arccos(\Delta) \equiv \pi/2$. To make contact with Ref.~\cite{TG03}, we note that the counterpart of the parameter $\upmu$ in our present considerations is the $\pi (1-\beta^2)$ (and \emph{not} $\pi\beta^2$) in Ref.~\cite{TG03} (see Eq.~(6.31) herein), where the latter $\beta$ is not to be confused with the $\beta$ in our present considerations, which stands for $1/T$; the $\beta^2$ of Ref.~\cite{TG03} is thus equivalent to the $\theta$ in Ref.~\cite{LP75}; this is easiest appreciated by comparing the expression $1/K = 2\beta^2$ in Eq.~(6.31) of Ref.~\cite{TG03} with Eq.~(\ref{ec105}) above.

Having determined the functional form of the exact $K$ corresponding to $\vert \Delta\vert \le 1$, we now proceed with doing the same for the exact $u$, for which for small values of $V/t$ one has the expression in Eq.~(\ref{ec79}).

For the excitation energy $\Delta E$ corresponding to the case $\vert \Delta \vert \le 1$ one has \cite{JKMC73} (see Eq.~(7.10b) herein)
\begin{equation}
\Delta E = t\, \frac{\pi \sin(\upmu)}{\upmu}\, \big(\vert \sin(a q_1)\vert + \vert \sin(a q_2)\vert\big), \label{ec120}
\end{equation}
where $\upmu$ is the parameter deduced from Eq.~(\ref{ec108}) and $a$ the lattice constant (identified with unity in Ref.~\cite{JKMC73}). We note that the last expression in Eq.~(7.10b) of Ref.~\cite{JKMC73} contains typing errors: instead of $\sin q_1$ and $\sin q_2$, one must have $\vert\sin q_1\vert$ and $\vert\sin q_2\vert$. This fact can be verified by realising that for $\vert \Delta \vert \le 1$, the $k_1$ and $k_2$ in Eq.~(7.8) of Ref.~\cite{JKMC73} are equal to $1$ whereby the terms involving $\cos^2 q_1$ and $\cos^2 q_2$ reduce to $\vert\sin q_1\vert$ and $\vert\sin q_2\vert$ respectively. Physically, this fact can be appreciated by the realisation that the excitation energy $\Delta E$ \emph{cannot} be negative (see the discussions concerning elementary excitations in metals in Sec.~1.1, p.~10, of Ref.~\cite{PN66}).

For clarity, comparing the Hamiltonian in Eq.~(19.1) of Ref.~\cite{AMT03} (or that in Eq.~(6.2) of Ref.~\cite{TG03}), with that in Eq.~(7.1) of Ref.~\cite{JKMC73}, and in view of $\h{S}_i^{\nu} = \frac{1}{2} \h{\sigma}_i^{\nu}$, $\nu = x, y, z$, we point out that the $-J_{z}$ in Eq.~(7.8) of Ref.~\cite{JKMC73} ($-J_z$ is identified with $1$ in Eq.~(7.10b) of this reference) is to be replaced by $t$ and \emph{not} $2 t$ (cf. Eq.~(\ref{ec103}) above); \emph{on the absolute scale, the magnitudes of energies in Ref.~\cite{JKMC73} are twice as large as those in our present considerations.} This clarifies the reason for the $t$, as opposed to $2 t$, on the RHS of Eq.~(\ref{ec120}) (compare with Eq.~(6.31) of Ref.~\cite{TG03} where one also encounters $\frac{1}{2} J_{xy}$, instead of $J_{xy}$, in the expression for $u$).

With
\begin{equation}
q \doteq q_1 + q_2 \label{ec121}
\end{equation}
denoting the wave number of the low-lying elementary excitations in the system, from Eq.~(\ref{ec120}) one obtains that
\begin{equation}
\frac{\partial}{\partial q} \Delta E = a t\, \frac{\pi\sin(\upmu)}{\upmu}\, \sgn\!\big(\!\sin(a [q-q_2])\big)\, \cos(a [q-q_2]). \label{ec122}
\end{equation}
One can readily verify that the minimum of $\Delta E$ corresponds to $q_1 = q_2 = m \pi$, where $m=0, \pm 1, \dots$~. For $q=2 m\pi + 0^+$ the expression in Eq.~(\ref{ec122}) thus yields the value for $u$. One thus obtains that (cf. Eq.~(29.31) in Ref.~\cite{AMT03} and Eq.~(6.31) in Ref.~\cite{TG03})
\begin{equation}
u = v_{\Sc f}\, \frac{\pi \sin(\upmu)}{2\upmu}, \label{ec123}
\end{equation}
where we have used Eq.~(\ref{ec40}). For small values of $\Delta \doteq V/(2 t)$, from Eqs.~(\ref{ec108}) and (\ref{ec123}) one obtains that
\begin{equation}
u = v_{\Sc f} \big( 1 + \frac{2\Delta}{\pi} - \frac{(\pi^2 -8) \Delta^2}{2\pi^2} + \dots\big), \label{ec124}
\end{equation}
which is to be compared with
\begin{equation}
u = v_{\Sc f} \big( 1 + \frac{2\Delta}{\pi} - \frac{2 \Delta^2}{\pi^2} + \dots\big), \label{ec125}
\end{equation}
which is obtained from the expansion of the expression in Eq.~(\ref{ec79}) in powers of $\Delta$. Thus indeed to linear order in $V/t$ the expression in Eq.~(\ref{ec79}) coincides with the exact expression in Eq.~(\ref{ec123}).

From Eqs.~(\ref{ec119}) and (\ref{ec123}) one deduces that
\begin{equation}
K = \frac{1}{2} \iff u = \frac{\pi}{2}\, v_{\Sc f}. \label{ec126}
\end{equation}
This result deviates from the perturbative result $u \approx 2 v_{\Sc f}$ that one obtains from Eq.~(\ref{ec79}) and $V \approx \frac{3\pi}{2} t$, the latter being deduced from Eq.~(\ref{ec78}) and the condition $K =\frac{1}{2}$.

\label{dynscal}
The value of the gap in the single-particle excitation spectrum corresponding to $\Delta > 1$ can be directly deduced either from the expression for the inverse correlation length $\xi^{-1}$, or from that for the singular part of the Gibbs, or Helmholtz, free energy near criticality (see Eqs.~(10.12.22a) -- (10.12.23b) and (10.14.39) -- (10.14.43) in \cite{RJB07}). In general, the energy gap is proportional to $\xi^{-{\sf z}}$, where ${\sf z}$ is the \emph{dynamic scaling exponent} (Eq.~(1.3) in Ref.~\cite{SS99}). In our case, the Lorentz invariance of the theory under consideration (see Eq.~(\ref{ec39})) implies that ${\sf z}$ is equal to unity (p. 88 in Ref.~\cite{SS99}) so that the energy gap is directly proportional to $\xi^{-1}$. For a general discussion of the inverse correlation length, the reader is referred to Sec. 16.2 in Ref.~\cite{KH87} and Sec. 5.2 in Ref.~\cite{CL00}.

In anticipation of the calculation of the mass gap $\mathcal{M}$ in the broken-symmetry GS corresponding to $\Delta >1$ in Sections \ref{sac.4c} and \ref{sac.5}, below we establish the critical nature of the GS of the system under consideration for $\vert\Delta \vert \le 1$ (a fact that is apparent from the power-law decays of the correlation functions in Eqs.~(\ref{ec106}) and (\ref{ec107}) above) by considering the case $J_{x} = J_{y}$ as the limit of $J_{x} \downarrow J_{y}$. To this end, we consider the case corresponding to
\begin{equation}
\vert J_{z}\vert  < J_y < J_x. \label{ec127}
\end{equation}
Following our earlier considerations (regarding the plus sign on the RHS of Eq.~(\ref{ec108})), one has
\begin{equation}
(J_{x}^{\rm r}, J_{y}^{\rm r}, J_{z}^{\rm r}) = (J_y, -J_z, -J_x), \label{ec128}
\end{equation}
so that (Eq.~(10.15.5) in Ref.~\cite{RJB07})
\begin{equation}
\Gamma_{\rm r} = -\frac{J_{z}}{J_{y}}, \;\;\; \Delta_{\rm r} = -\frac{J_{x}}{J_{y}}. \label{ec129}
\end{equation}
In view of Eq.~(\ref{ec127}), it will be convenient to introduce
\begin{equation}
\epsilon \doteq J_{x} - J_{y} \label{ec130}
\end{equation}
and consider the case $\epsilon\downarrow 0$. For convenience, in what follows we assume that
\begin{equation}
\eta \doteq \frac{\epsilon/J_y}{1-J_z^2/J_y^2} \ll 1. \label{ec131}
\end{equation}

Following Eq.~(10.15.6a) in Ref.~\cite{RJB07} and making use of the expressions in Eq.~(\ref{ec129}) for $\Gamma_{\rm r}$ and $\Delta_{\rm r}$, one has
\begin{equation}
\frac{2 \kappa^{1/2}}{1+\kappa} = 1 - \eta + O(\epsilon\eta) + O(\eta^2), \label{ec132}
\end{equation}
from which one obtains that ($\kappa$ is bound to satisfy $0\le \kappa \le 1$, Eq.~(10.15.7) in Ref.~\cite{RJB07})
\begin{equation}
\kappa = 1 -2\, (2\eta)^{1/2} + O(\eta). \label{ec133}
\end{equation}
Consequently,
\begin{equation}
\kappa' \doteq (1-\kappa^2)^{1/2} = 2\, (2\eta)^{1/4} + O\big(\eta^{3/4}\big). \label{ec134}
\end{equation}

Making use of
\begin{equation}
\mathrm{snh}(\lambda,\kappa) = \tan(\lambda) + \Big\{-\lambda \sec^2(\lambda) +\tan(\lambda)\Big\}\, (2\eta)^{1/2} + O\big(\eta\big),
\label{ec135}
\end{equation}
and expressing $\lambda$ as
\begin{equation}
\lambda = \lambda_0 + \delta\lambda, \label{ec136}
\end{equation}
where $\lambda_0$ is the solution of Eq.~(10.15.6b) in Ref.~\cite{RJB07} corresponding to $\epsilon=0$ (and thus to $\eta=0$) --- that is, $\lambda_0$ is the solution of
\begin{equation}
\tan(\lambda_0) = \Big(\frac{1+J_z/J_y}{1-J_z/J_y}\Big)^{1/2} \iff \cos(2\lambda_0) = -\frac{J_z}{J_y}, \label{ec137}
\end{equation}
after some algebra for $\delta\lambda$ one obtains that
\begin{equation}
\delta\lambda = \lambda_0\, (2\eta)^{1/2} + O\big(\eta\big). \label{ec138}
\end{equation}
We note that $0 \le \lambda \le I'$ (Eq.~(10.15.7) in Ref.~\cite{RJB07}), where, in view of Eq.~(\ref{ec134}), for $I'$ one has the expression in Eq.~(\ref{ec147}) below. Thus, $0 \le \lambda_0 \le \pi/2$, which is in conformity with the first equation in Eq.~(\ref{ec137}) of which the RHS is positive.

In order to establish the critical nature of the system under consideration for $J_x = J_y$ and $\vert\Delta\vert < 1$, we need to calculate the quantity ${\sf t}$, defined in Eq.~(10.12.2) of Ref.~\cite{RJB07}, which is vanishing on the critical surface; in general, this quantity vanishes \emph{linearly} with $T-T_{\rm c}$, where $T_{\rm c}$ is the critical temperature; $(T-T_{\rm s})/{\sf t}$ is positive (Sec.~10.12 in Ref.~\cite{RJB07}). In this connection, we note that the parameters $w_1$, $w_2$, $w_3$ and $w_4$, or $a$, $b$, $c$ and $d$, of the $8$-vertex model are functions of temperature $T$, Eqs.~(10.1.2) and (10.2.1) in Ref.~\cite{RJB07}. According to Eq.~(10.12.3) of Ref.~\cite{RJB07}, one has
\begin{equation}
\kappa + \kappa^{-1} = 2 - 4 {\sf t}. \label{ec139}
\end{equation}
Calculation of ${\sf t}$ from this expression is mediated by the calculation of the quantity $p$ for which one has (Eq.~(10.12.5) in Ref.~\cite{RJB07})
\begin{equation}
p \equiv (q_1)^2 = \frac{(1-\kappa^2)^2}{16^2} + \frac{(1-\kappa^2)^3}{16^2} + O\big((1-\kappa^2)^4\big), \label{ec140}
\end{equation}
where we have employed the series expansion in Eq.~(\ref{ec115}) (for $q_1$ see Eq.~(\ref{ec114})). Expressing $\kappa$ as $\kappa = \kappa_0 + \delta \kappa$, with $\kappa_0$ denoting the solution of the equation $(1-\kappa^2)^2 = 16^2 p$, i.e. \[\kappa_0 \equiv (1- 16 p^{1/2})^{1/2} = 1 - 8 p^{1/2} - 32 p + O(p^{3/2}),\] from Eq.~(\ref{ec140}) one obtains that
\begin{equation}
\kappa = 1 - 8 p^{1/2} + 32 p + O\big(p^{3/2}\big). \label{ec141}
\end{equation}
Substituting this expression into Eq.~(\ref{ec139}), for $\kappa\to 1$, and thus $p\to 0$, Eq.~(\ref{ec140}), one deduces that (cf. Eq.~(10.12.19) in Ref.~\cite{RJB07})
\begin{equation}
{\sf t} = -16\, p + O(p^2) \iff p = -\frac{1}{16} {\sf t} + O({\sf t}^2). \label{ec142}
\end{equation}
Making use of Eqs.~(\ref{ec140}), (\ref{ec133}) and (\ref{ec130}), one has (cf. Eq.~(10.14.37) in Ref.~\cite{RJB07}; see also Sec.~4.5 in Ref.~\cite{LZ01})
\begin{eqnarray}
p &=& \frac{1}{8}\,\eta + O\big(\eta^{3/2}\big) \nonumber\\
&=& \frac{1}{16} \frac{J_x^2 - J_y^2}{J_y^2 - J_z^2} + O\big(\eta^{3/2}\big) + O\big(\epsilon\eta/J_y\big).
\label{ec143}
\end{eqnarray}

For the inverse correlation length $\xi^{-1}$ near criticality, one has (Eq.~(10.12.23a) in Ref.~\cite{RJB07})
\begin{equation}
\xi^{-1} \sim \frac{C}{a} (-{\sf t})^{\pi/2\upmu}, \label{ec144}
\end{equation}
where $C >0$ is a finite dimensionless constant and $a$ the lattice constant. The following four remarks are in order:

First, since Eq.~(\ref{ec144}) applies for ${\sf t} <0$, and since $(T-T_{\rm c})/{\sf t}$ is positive (see the previous paragraph), it follows that the case at hand, which is characterised by $J_{x} > J_{y} > \vert J_z\vert$, corresponds to the `ordered region' ($T < T_{\rm c}$) of the $8$-vertex model.

Second, the mathematical details underlying the expression in Eq.~(\ref{ec144}) are strictly valid for $\lambda \le \frac{1}{2} I'$ (see pp. 241 and 242 in Ref.~\cite{RJB07}). Baxter however has shown that Eq.~(\ref{ec144}) itself is valid for $\lambda \le \frac{2}{3} I'$, i.e. for $\upmu \le \frac{2\pi}{3}$ (see pp. 240 and 253 in Ref.~\cite{RJB07}). Nonetheless, Johnson \emph{et al.} \cite{JKMC73} have demonstrated that Eq.~(\ref{ec144}) gives the correct critical behaviour even for $\upmu > \frac{2\pi}{3}$ (see p. 253 in Ref.~\cite{RJB07}).

Third, Eq.~(\ref{ec144}) is not valid for $\upmu=0$; in such case, $\xi^{-1}$ vanishes \emph{exponentially} as criticality is approached; the appropriate expression for $\xi^{-1}$ in such case is that presented in Eq.~(8.11.24) of Ref.~\cite{RJB07} (see p.~272 in Ref.~\cite{RJB07}). We shall encounter this behaviour in Eq.~(\ref{ec187}) below. The renormalization-group analysis in Sec.~\ref{sac.5} reveals that this behaviour is specific to the states inside the cross-over regime (CO), adjacent to the strong-coupling (SC) regime, to be discussed in Sec.~\ref{sac.5c} below. In contrast, the behaviour of $\xi^{-1}$ as described by Eq.~(\ref{ec144}) is specific to the states deep inside the SC regime, to be discussed in Sec.~\ref{sac.5a}.

Fourth, Eq.~(\ref{ec144}) applies strictly for $\pi/(2\upmu)$ different from an \emph{integer}; for $\pi/(2\upmu)$ an integer, the expression for $\xi^{-1}$ acquires a logarithmic correction (see p.~252 in Ref.~\cite{RJB07}; see also Sec.~8 in Ref.~\cite{RJB72}).

\label{critical1}
One observes that for $J_x \downarrow J_y$, one has $p \downarrow 0$ (see the inequalities in Eq.~(\ref{ec127})), whereby ${\sf t}\uparrow 0$, Eq.~(\ref{ec142}). In the limit $J_x = J_y$, while $\vert J_z\vert < J_x = J_y$ (cf. Eq.~(\ref{ec127})), or $\vert\Delta \vert < 1$ in the light of Eq.~(\ref{ec103}), ${\sf t}$ is vanishing, signifying the system under consideration as being \emph{critical}. For $J_x > J_y > \vert J_z\vert$, $p$ is non-vanishing and unless $\upmu = 0$ (see the summary in the paragraph following Eq.~(\ref{ec158}) on page~\pageref{critical2}), $\xi^{-1}$ is non-vanishing so that the spectrum of the system is gapped (see the remarks in the paragraph following Eq.~(\ref{ec126}) on page~\pageref{dynscal}).

We proceed now with the calculation of $\upmu$, for which one has (Eq.~(10.12.5) in Ref.~\cite{RJB07})
\begin{equation}
\upmu = \frac{\pi}{I'}\,\lambda. \label{ec145}
\end{equation}
For $J_{x} = J_{y}$ and $\vert\Delta\vert \le 1$ we should recover the expression in Eq.~(\ref{ec116}) above. Making use of (item 17.3.11 of Ref.~\cite{AS72})
\begin{equation}
I(m) = \frac{\pi}{2}\, \Big\{ 1 + \Big(\frac{1}{2}\Big)^2 m + \Big(\frac{1\cdot 3}{2\cdot 4}\Big)^2 m^2 +\dots \Big\},\;\;\; \vert m\vert <1, \label{ec146}
\end{equation}
from Eq.~(\ref{ec134}) (and $m_1 = {\kappa'}^2$) one obtains
\begin{equation}
I' = \frac{\pi}{2} + \frac{\pi}{2}\, (2\eta)^{1/2} + O\big(\eta\big). \label{ec147}
\end{equation}
Thus, with
\begin{equation}
\upmu_0 \doteq \frac{\pi}{I'}\, \lambda_0 = 2\lambda_0 -2 \lambda_0 (2\eta)^{1/2} + O\big(\eta\big), \label{ec148}
\end{equation}
and (cf. Eq.~(\ref{ec138}))
\begin{equation}
\delta \upmu \doteq \frac{\pi}{I'}\, \delta \lambda = 2\lambda_0\, (2\eta)^{1/2} + O\big(\eta\big), \label{ec149}
\end{equation}
one has
\begin{equation}
\upmu \equiv \upmu_0 + \delta\upmu = 2\lambda_0 + O\big(\eta\big), \label{ec150}
\end{equation}
so that on account of Eq.~(\ref{ec137}) (recall that $0\le \lambda_0 \le \pi/2$)
\begin{equation}
\frac{\pi}{2\upmu} = \frac{\pi}{4\lambda_0 + O(\eta)} \equiv \frac{\pi}{2\cos^{-1}(-J_z/J_y) + O(\eta)}. \label{ec151}
\end{equation}
For $J_{x} = J_{y}$ ($\equiv J_{xy}$, Eq.~(\ref{ec104})), i.e. for $\epsilon=0$, which, following Eq.~(\ref{ec131}), implies $\eta=0$, Eq.~(\ref{ec151}) yields the exact result $\Delta = -\cos(\upmu)$, Eq.~(10.16.8) in Ref.~\cite{RJB07}; the origin of the minus sign in this expression, which deviates from the plus sign in Eq.~(\ref{ec108}), lies in the fact that in arriving at Eq.~(\ref{ec151}) we have been dealing directly with $(J_x,J_y,J_z)$, ordered according to Eq.~(\ref{ec127}), leading to Eq.~(\ref{ec128}), where one has the \emph{signature} $(+,-,-)$. This is to be contrasted with the signature of the corresponding parameters in Eq.~(\ref{ec113}), which is $(+,+,-)$. For further relevant details, the reader is referred to the discussions in the paragraph following Eq.~(\ref{ec108}).

For the cases where
\begin{equation}
-\frac{J_z}{J_y} = 1 - \gamma \;\;\; \mbox{\rm with}\;\;\; 0 \le \gamma \ll 1, \label{ec152}
\end{equation}
one has
\begin{equation}
\lambda_0 = \Big(\frac{\gamma}{2}\Big)^{1/2} + O\big(\gamma^{3/2}\big). \label{ec153}
\end{equation}
It would therefore appear that for $\gamma \downarrow 0$, $\pi/(2\upmu)$ would approach a large value, of the order of $1/\eta$, which is independent of $\gamma$. This is however not the case, as $\eta$ is a non-trivial function of $\gamma$; from Eqs.~(\ref{ec131}) and (\ref{ec152}) one obtains that
\begin{equation}
\eta = \frac{1}{2 J_y}\, \frac{\epsilon}{\gamma} + O\big(\frac{\epsilon}{J_y}\big),
\label{ec154}
\end{equation}
which implies that as $\gamma$ approaches $\epsilon$ from above, the condition $\eta \ll 1$, Eq.~(\ref{ec131}), which underlies our above considerations, will eventually fail to be valid. It follows that the validity of Eq.~(\ref{ec151}) is conditioned on $\epsilon \ll \gamma$. Thus, although $\gamma$ can approach zero, it my not come closer to zero than $\epsilon \doteq J_x - J_y$ in order for Eq.~(\ref{ec151}) to be applicable.

For the cases where
\begin{equation}
-\frac{J_z}{J_y} = -1 + \gamma\;\;\; 0\le \gamma \ll 1, \label{ec155}
\end{equation}
one has
\begin{equation}
\lambda _0 = \frac{\pi}{2} - \Big(\frac{\gamma}{2}\Big)^{1/2} + O\big(\gamma^{3/2}\big), \label{ec156}
\end{equation}
so that
\begin{equation}
\upmu \equiv \upmu_0 + \delta\upmu = \pi - (2\gamma)^{1/2} + O(\eta),
\label{ec157}
\end{equation}
from which one obtains that
\begin{equation}
\frac{\pi}{2\upmu} = \frac{1}{2} + \frac{1}{2\pi} (2\gamma)^{1/2} + O(\eta). \label{ec158}
\end{equation}
It can be shown that Eq.~(\ref{ec154}) applies also here, so that the smallness of $\eta$ similarly requires $\epsilon \ll\gamma$.

\label{critical2}
Summarising, since the exponent $\pi/(2\upmu)$ as deduced from Eq.~(\ref{ec151}) is finite, with reference to our remarks in the paragraph preceding Eq.~(\ref{ec145}) on page~\pageref{critical1}, we conclude that indeed for $J_x > J_y$ the inverse correlation length is non-vanishing and thus the excitation spectrum of the system is gapped. In contrast, for $J_x = J_y$ (i.e. for $\epsilon=0$) the system is critical (${\sf t} = 0$) and the GS is `disordered' ($\xi^{-1}$ vanishing). This fact is reflected in the power-law decays of the GS correlation functions in Eqs.~(\ref{ec106}) and (\ref{ec107}).

For $J_{x} > J_{y}$ (generally, for $J_{x} \not= J_{y}$) and $\vert z\vert \to \infty$ one has \cite{LP75} (see p.~3912 herein)
\begin{equation}
\langle \h{S}^+(x,-i\tau) \h{S}^-(0,0)\rangle \sim \frac{C_3}{\mathcal{M}^{\theta}} \equiv \frac{C_3}{\mathcal{M}^{1/(2 K)}}, \label{ec159}
\end{equation}
instead of the expression in Eq.~(\ref{ec107}). Here $C_3$ is a finite constant and $\mathcal{M}$ the correlated mass gap, for which one has \cite{LP75}
\begin{equation}
\mathcal{M} \propto \mathcal{M}_0^{\nu}, \label{ec160}
\end{equation}
in which (cf. Eq.~(\ref{ec130}))
\begin{equation}
\mathcal{M}_0 \doteq \frac{1}{2} (J_{x} - J_{y}) \equiv \frac{\epsilon}{2}, \label{ec161}
\end{equation}
the basal-plane anisotropy, is the uncorrelated mass gap. One has the following scaling law \cite{LP75} (see Table I herein)
\begin{equation}
\nu = \frac{1}{2 (1-\theta)} \equiv \frac{\pi}{2\upmu} \equiv \frac{K}{2 K - 1}. \label{ec162}
\end{equation}
To appreciate the expressions in Eqs.~(\ref{ec160}) and (\ref{ec161}), as well as the equivalence of $\nu$ with $\pi/(2\upmu)$ (the second equality in Eq.~(\ref{ec162})), one should realise that on account of (as regards ${\sf z} =1$, see the remark in the paragraph following Eq.~(\ref{ec126}) on page~\pageref{dynscal})
\begin{equation}
\mathcal{M} \propto \xi^{-1}, \label{ec163}
\end{equation}
the expression in Eq.~(\ref{ec160}) represents the same fact as expressed by Eq.~(\ref{ec144}). In this connection, note that $-{\sf t} = 2\eta + O(\eta^{3/2})$, Eqs.~(\ref{ec142}), (\ref{ec143}), and that $\eta \propto \epsilon$, Eq.~(\ref{ec131}), so that $-{\sf t} \propto \mathcal{M}_0$ near criticality. With reference to Eq.~(\ref{ec158}), we note that since for $\epsilon\ll \gamma$ and $\gamma\downarrow 0$ one has $\nu \downarrow \frac{1}{2}$, it follows that for $\epsilon\ll \gamma$ and $\gamma\downarrow 0$ the constant $K$ approaches $+\infty$ from below; this is evident from $\nu \sim \frac{1}{2} + \frac{1}{4 K}$, Eq.~(\ref{ec162}), which applies for $\vert K\vert \to\infty$ and which implies $\nu > \frac{1}{2}$ for $K$ \emph{positive} and large.

In the case of the $XY$ model, for which $J_{z} = 0$, one has the mean-field result $\nu=1$ \cite{LSM61,BMC68,LP75} (see also page 270 in Ref.~\cite{RJB07}), which, following to Eq.~(\ref{ec162}), correctly corresponds to $K = 1$, $\upmu = \frac{\pi}{2}$ and $\theta = \frac{1}{2}$. The result $\nu = 1$ immediately follows from Eq.~(\ref{ec151}) for the \emph{isotropic} $XY$ model \cite{LSM61}, for which $\epsilon = 0$, and thus $\eta=0$. For the \emph{anisotropic} $XY$ model \cite{BMC68}, Eq.~(\ref{ec151}) would incorrectly suggest that $\nu$ deviated from $1$. One can explicitly demonstrate that for $J_z =0$, whereby $\Gamma_{\rm r} =0$ (see Eq.~(\ref{ec129})), Eq.~(10.15.6b) in Ref.~\cite{RJB07} yields the \emph{exact} result:
\begin{equation}
\lambda = \frac{1}{2} I', \label{ec164}
\end{equation}
from which and Eq.~(\ref{ec145}) one immediately obtains $\nu =1$. The result in Eq.~(\ref{ec164}) follows from $\mathrm{snh}(\lambda,\kappa) \equiv -i\,\mathrm{sn}(i\lambda,\kappa)$ and the exact result (\emph{Example}~2 in Sec.~22.41 of Ref.~\cite{WW62})
\begin{equation}
\mathrm{sn}\big(\frac{1}{2} i I',\kappa\big) = i \kappa^{-1/2}. \label{ec165}
\end{equation}
We note in passing that, on account of $\Gamma_{\rm r} =0$ for $J_z=0$, one has further the following exact solution of Eq.~(10.15.6a) in Ref.~\cite{RJB07} (cf. Eq.~(\ref{ec133})): \[\kappa = 2 \Delta_{\rm r}^2 - 2 \vert\Delta_{\rm r}\vert (\Delta_{\rm r}^2 -1)^{1/2} -1,\] where $\Delta_{\rm r}$ is specified in Eq.~(\ref{ec129}). Following the inequalities in Eq.~(\ref{ec127}), one has $\Delta_{\rm r} < -1$ so that indeed $0 \le \kappa \le 1$, Eq.~(10.15.7) in Ref.~\cite{RJB07}.

Below we consider the cases corresponding to $\vert\Delta\vert >1$. In doing so, we deal separately with the cases corresponding to $\Delta < -1$ and $\Delta > 1$. In what follows, we explicitly consider the case corresponding to the expression in Eq.~(\ref{ec112}) above. In view of the definition of the FR in Eq.~(\ref{ec109}), in the following we shall view the condition $J_x^r = \vert J_y^r\vert$ as representing
\begin{equation}
J_x^r = \vert J_y^r\vert + 0^+. \label{ec166}
\end{equation}

\subsubsection{The case of $\Delta < -1$}
\label{sac.4b}

Following Eq.~(\ref{ec112}), in this case one has
\begin{equation}
(J_x^r, J_y^r, J_z^r) = (1,-1,\Delta),\;\;\; \mbox{\rm so that}\;\;\; \Gamma_{\rm r} = -1,\;\; \Delta_{\rm r} = \Delta. \label{ec167}
\end{equation}
Since $\Gamma_{\rm r} =-1$, from Eq.~(10.15.6a) in Ref.~\cite{RJB07} one observes that in the present case
\begin{equation}
\kappa = 0. \label{ec168}
\end{equation}
With reference to Eq.~(10.15.6b) in Ref.~\cite{RJB07}, which involves $\kappa^{-1/2}$, we should consider $\kappa=0$ as the limit of $\kappa \downarrow 0$. In this limit, from Eq.~(10.15.6a) in Ref.~\cite{RJB07} one has
\begin{equation}
\kappa \sim \frac{1}{4}\, \frac{1-\Gamma_{\rm r}^2}{\Delta_{\rm r}^2 -\Gamma_{\rm r}^2}. \label{ec169}
\end{equation}
Making use of $\mathrm{snh}(\lambda,\kappa) \sim \sinh(\lambda)$ for $\kappa\downarrow 0$, from Eq.~(\ref{ec169}) and Eq.~(10.15.6b) in Ref.~\cite{RJB07} one obtains that
\begin{equation}
\sinh(\lambda) \sim \frac{2 (\Delta_{\rm r}^2 - \Gamma_{\rm r}^2)^{1/2}}{1+ \Gamma_{\rm r}}. \label{ec170}
\end{equation}
According to this expression, and in view of Eq.~(\ref{ec166}),
\begin{equation}
\lambda = +\infty. \label{ec171}
\end{equation}

Since $\kappa \downarrow 0$, for the \emph{nome} $q(m)$ ($m\equiv \kappa^2$) to leading order one has $q(m) \sim \kappa^2/16$, Eq.~(\ref{ec115}), on account of Eq.~(\ref{ec114}) one deduces that
\begin{equation}
\frac{I'}{I} \sim \frac{2}{\pi} \ln\Big(\frac{4}{\kappa}\Big). \label{ec172}
\end{equation}
From this, $I \sim \pi/2$ for $\kappa\downarrow 0$ (item 17.3.22 in Ref.~\cite{AS72}), and Eq.~(\ref{ec146}) one thus has
\begin{equation}
\upmu \equiv \frac{\pi}{I'} \lambda \sim \frac{\pi\lambda}{\ln(4/\kappa)}. \label{ec173}
\end{equation}
On account of $\sinh(\lambda) \sim \frac{1}{2} \e^{\lambda}$ for $\lambda \to \infty$, from Eq.~(\ref{ec170}) one deduces that
\begin{equation}
\lambda \sim \ln\Big(\frac{4 (\Delta_{\rm r}^2 - \Gamma_{\rm r}^2)^{1/2}}{1+\Gamma_{\rm r}}\Big). \label{ec174}
\end{equation}
Conform Eq.~(\ref{ec171}), one observes that $\lambda$ indeed diverges (logarithmically) for $\Gamma_{\rm r} \downarrow -1$. From the expression in Eq.~(\ref{ec174}) and that in Eq.~(\ref{ec169}) we thus arrive at
\begin{equation}
\lambda \sim \ln\Big(\frac{2}{\kappa}\Big), \label{ec175}
\end{equation}
according to which and Eq.~(\ref{ec173}) (see Eq.~(10.16.8) of Ref.~\cite{RJB07})
\begin{equation}
\lim_{\kappa \downarrow 0} \upmu = \pi. \label{ec176}
\end{equation}
Since for $\upmu=\pi$ the expression in Eq.~(\ref{ec108}), which is specific to all $\Delta \in [-1,1]$, yields $\Delta = -1$, it follows that $\upmu$ is a continuous function of $\Delta$ across the boundary $\Delta = -1$ whose derivative with respect to $\Delta$ is however finitely discontinuous at this boundary.

The phase transition at $\Delta =-1$ is a first-order one, discussed in Sec.~8.11 of Ref.~\cite{RJB07}. In Ch.~8 of this reference, ice-type models are considered and the phase relevant to $\Delta <-1$ (that is, $\Delta >1$ in the convention of Ref.~\cite{RJB07}; see our remarks concerning the plus sign on the RHS of Eq.~(\ref{ec108})) is the \emph{ferroelectric} one (i.e. the Ising FMic state in the magnetic language --- see the second paragraph of Sec.~\ref{sac.4}, on page~\pageref{chiralsb}), corresponding to $\varepsilon_1 < \varepsilon_3, \varepsilon_5$ (pp. 156 and 157 in Ref.~\cite{RJB07}); realising that the Boltzmann weight $d$ of the $8$-vertex model is vanishing in the case corresponding to $J_x=J_y$ (the Heisenberg-Ising chain; see pp.~271 and 272 in Ref.~\cite{RJB07}), one observes that indeed for $d=0$ the \emph{ferroelectric} phase of the $8$-vertex model (phase \emph{I} as specified on p.~246 of Ref.~\cite{RJB07})) coincides with that of ice-type models; in this phase, for the Boltzmann weights $a$, $b$ and $c$ one has $a > b+c$ so that the phase boundary of the ferroelectric phase is determined by the condition $a=b+c$. In this case, $T-T_{\rm c}$ is proportional to (Eq.~(8.11.4) in Ref.~\cite{RJB07})
\begin{equation}
{\sf t} \doteq \frac{b+c-a}{a}, \label{ec177}
\end{equation}
and for the free energy per site $f$ one has (p. 157 in Ref.~\cite{RJB07})
\begin{equation}
f = \left\{ \begin{array}{ll} \varepsilon_1 - \frac{1}{2} k_{\Sc b} T_{\rm c}\, {\sf t} + O\big({\sf t}^{3/2}\big), & {\sf t} > 0,\\ \\
\varepsilon_1, & {\sf t} < 0, \end{array} \right. \label{ec178}
\end{equation}
establishing that indeed $\partial f/\partial {\sf t}$ is \emph{discontinuous} at ${\sf t} =0$, or $\Delta=-1$. Approaching the phase boundary from the region ${\sf t} >0$, from Eq.~(\ref{ec178}) one further observes that the specific heat diverges like ${\sf t}^{-1/2}$.

Since in approaching the \emph{ferroelectric} phase (our Ising FMic phase) one has (Sec.~8.11 in Ref.~\cite{RJB07})
\begin{equation}
\upmu \uparrow \pi\;\; \mbox{\it and}\;\; w \downarrow 0 \;\;\; \mbox{\rm for}\;\;\; T \downarrow T_{\rm c}, \label{ec179}
\end{equation}
where $w$ is an elliptic-function parameter, on writing (Eq.~(8.11.2) in Ref.~\cite{RJB07})
\begin{equation}
\upmu = \pi -\delta,\;\;\; w = -\pi + \epsilon, \label{ec180}
\end{equation}
for the Boltzmann weights $a$, $b$ and $c$ one has
\begin{equation}
a : b : c = \sin(\frac{1}{2} [\epsilon + \delta]) : \sin(\frac{1}{2} [\epsilon-\delta]) : \sin(\delta), \label{ec181}
\end{equation}
resulting in (Eq.~(8.11.6) in Ref.~\cite{RJB07})
\begin{equation}
{\sf t} \sim \frac{1}{2} (\epsilon-\delta) \delta \;\;\; \mbox{\rm for}\;\;\; T \downarrow T_{\rm c}. \label{ec182}
\end{equation}
It follows that for ${\sf t}\downarrow 0$, both $\delta$ and $\epsilon$ vanish like ${\sf t}^{1/2}$. It should be noted that the approach $\upmu \uparrow \pi$ in Eq.~(\ref{ec179}) coincides with that required of $\upmu$ by the expression in Eq.~(\ref{ec108}) in order for $\Delta$ to \emph{decrease} towards $-1$.

In the \emph{ordered phase}, corresponding to ${\sf t} <0$, or $\Delta <-1$, the correlation length $\xi$ is vanishing, and in the ferroelectric phase, corresponding to ${\sf t} >0$, or $\vert\Delta \vert <1$, it is infinite (p.~157 in Ref.~\cite{RJB07}). The latter observation is in conformity with the one arrived at in the previous section (see the summary in the paragraph following Eq.~(\ref{ec158}) on page~\pageref{critical2}).

\subsubsection{The case of $\Delta > 1$}
\label{sac.4c}

Following Eq.~(\ref{ec112}), in this case one has
\begin{equation}
(J_x^r, J_y^r, J_z^r) = (1,1,-\Delta),\;\;\; \mbox{\rm so that}\;\;\; \Gamma_{\rm r} = 1,\;\; \Delta_{\rm r} = -\Delta. \label{ec183}
\end{equation}
As in the previous case, here also one has $\kappa=0$. Making use of Eq.~(\ref{ec170}), for the present case we thus deduce the following equivalent of Eq.~(10.15.6b) in Ref.~\cite{RJB07}:
\begin{equation}
\sinh(\lambda) = (\Delta^2 -1)^{1/2}. \label{ec184}
\end{equation}
From this and the fact that in the case under consideration $\Delta$ is positive, one arrives at
\begin{equation}
\Delta = +\cosh(\lambda). \label{ec185}
\end{equation}
This result deviates from its counterpart in Eq.~(10.16.8) of Ref.~\cite{RJB07} by a minus sign on the RHS, which deviation is due to the same mechanism as we discussed in connection with the plus sign on the RHS of Eq.~(\ref{ec108}) above.

The asymptotic expression in Eq.~(\ref{ec173}) applies equally for the present case. Since, however, Eq.~(\ref{ec185}) implies $\lambda$ to be finite for finite values of $\Delta$ (cf. Eq.~(\ref{ec171})), in the present case for any finite value of $\Delta$ one has
\begin{equation}
\lim_{\kappa\downarrow 0} \upmu = 0. \label{ec186}
\end{equation}
Since for $\upmu=0$ the expression in Eq.~(\ref{ec108}), which is specific to all $\Delta \in [-1,1]$, yields $\Delta = +1$, it follows that $\upmu$ is a continuous function of $\Delta$ across the boundary $\Delta = +1$ whose derivative with respect to $\Delta$ is however finitely discontinuous at this boundary.

As we have indicated following Eq.~(\ref{ec144}) above, the expression for $\xi^{-1}$ in this equation is \emph{invalid} for $\upmu = 0$. For $\upmu=0$, and $T$ near $T_{\rm c}$, one instead has (Eq.~(8.11.24) in Ref.~\cite{RJB07}; see also Eq.~(6.12) in Ref.~\cite{JKMC73})
\begin{equation}
\xi^{-1} \sim \frac{4}{a}\, \exp\!\big(\!-\frac{\pi^2}{2\lambda}\big). \label{ec187}
\end{equation}
As before, the mass gap is proportional to this expression, Eq.~(\ref{ec163}). The expression in Eq.~(\ref{ec187}) can be simplified in the case of $\Delta - 1 \ll 1$. From Eq.~(\ref{ec185}) one observes that $\lambda$ approaches zero for $\Delta\downarrow 1$. Further, since the RHS of Eq.~(\ref{ec184}) is positive for $\Delta >1$, one infers that for the case at hand, where $\Delta >1$, $\lambda$ is positive. Using the expansion
\begin{equation}
\cosh(\lambda) \sim 1 + \frac{1}{2} \lambda^2 \;\;\; \mbox{\rm for}\;\;\; \lambda\to 0, \label{ec188}
\end{equation}
from Eq.~(\ref{ec185}) one thus obtains that
\begin{equation}
\lambda \sim +\sqrt{2 (\Delta -1)}\;\;\; \mbox{\rm for}\;\;\; \Delta \downarrow 1, \label{ec189}
\end{equation}
whereby, following Eqs.~(\ref{ec163}) and (\ref{ec187}),
\begin{equation}
\mathcal{M} \propto \xi^{-1} \sim \frac{4}{a} \exp\!\Big(\!-\frac{\pi^2}{2 \sqrt{2 (\Delta-1)}}\Big)\;\;\; \mbox{\rm for}\;\;\; \Delta \downarrow 1. \label{ec190}
\end{equation}
This expression is to be compared with that presented in Eq.~(\ref{ec226}) below.

\subsection{The mass gap $\mathcal{M}$: a perturbative Renormalization-Group approach}
\label{sac.5}

Comparing the action $\wh{S}$ in Eq.~(\ref{ec90}) above with that in Eq.~(10.1) in Ref.~\cite{GNT98}), for the following considerations we introduce the dimensionless coupling constant
\begin{equation}
\mathrm{z} \doteq \ul{g}\, a^2 \equiv \frac{a V}{2\pi^2 u}, \label{ec191}
\end{equation}
where $\ul{g}$, the scaled coupling constant, is defined in Eq.~(\ref{ec92}) above. Assuming that $\mathrm{z}$ is small, one can analyse the consequences of the cosine term in $\wh{\mathcal{S}}$ by means of the perturbative renormalization-group (RG) analysis. This analysis is standard and its details can be found in e.g. Refs.~\cite{GNT98,AMT03,TG03}. The considerations in this section follow closely those in Ch.~10 of Ref.~\cite{GNT98}, and Sec.~2.3.2 of Ref.~\cite{TG03}.

As we indicated in Sec.~\ref{sac.3}, the value of the scaling dimension $d$, Eq.~(\ref{ec95}), of the cosine field in Eq.~(\ref{ec90}) determines the significance or otherwise of the contribution to $\wh{\mathcal{S}}$ of the term pre-multiplied by $\ul{g}$ in Eq.~(\ref{ec90}): for $d < 2$ the latter term is relevant, for $d > 2$ it is irrelevant, and for $d=2$ it is marginal. The boundary value $d=2$ is more generally expressed as $d=D$, where $D$ is the dimension of the space-time domain over which the $\wh{S}$ under consideration is defined.

With $\Lambda \sim 1/a$ the actual cut-off of the problem in the momentum space (see Eq.~(\ref{ec38}) \emph{et seq.}), one writes
\begin{eqnarray}
\h{\Phi}(x) \equiv \h{\Phi}_{\Lambda}(x) &=& \frac{1}{\sqrt{L}} \sum_{\vert k\vert \le \Lambda'} \!\!\e^{i k x}\, \h{\phi}_{k} + \frac{1}{\sqrt{L}} \sum_{\Lambda' <\vert k\vert \le \Lambda} \!\!\!\e^{i k x}\, \h{\phi}_{k}\nonumber\\
&\equiv& \h{\Phi}_{\Lambda'}(x) + \h{h}(x), \label{ec192}
\end{eqnarray}
where $\Lambda' <\Lambda$ so that $\h{h}(x)$ amounts to the spatially `fast' mode of $\h{\Phi}_{\Lambda}(x)$. Integrating out the contribution of $\h{h}(x)$ to the action $\wh{S}$, Eq.~(\ref{ec90}), and expanding the resulting effective action to second order in $\ul{g}$, with
\begin{equation}
\rd l \doteq \frac{\Lambda - \Lambda'}{\Lambda} \label{ec193}
\end{equation}
for
\begin{equation}
\mathrm{z}_{\parallel}(l) \doteq d(l)-2, \;\;\; \mathrm{z}_{\perp}(l) \doteq \sqrt{8 A}\, \mathrm{z}(l) \label{ec194}
\end{equation}
one obtains the following coupled Kosterlitz-Thouless-type RG differential equations:
\begin{eqnarray}
\frac{\rd \mathrm{z}_{\parallel}(l)}{\rd l} &=& - \mathrm{z}_{\perp}^2(l), \label{ec195}\\ \frac{\rd \mathrm{z}_{\perp}(l)}{\rd l} &=& - \mathrm{z}_{\parallel}(l)\, \mathrm{z}_{\perp}(l). \label{ec196}
\end{eqnarray}
In Eq.~(\ref{ec194}), $A$ is a well-defined constant whose value is of no immediate relevance to our present considerations (see however the last paragraph of this section, on page~\pageref{valueofA}), and $d(l)$ is a function of $l$ which is constrained by the `initial' condition
\begin{equation}
d(0) = d, \label{ec197}
\end{equation}
where $d$ is the scaling dimension of the cosine field under consideration, Eq.~(\ref{ec95}). For the `initial' conditions of Eqs.~(\ref{ec195}) and (\ref{ec196}) one has
\begin{equation}
\mathrm{z}_{\parallel}(0) \doteq \mathrm{z}_{\parallel}^0 = d - 2, \;\;\; \mathrm{z}_{\perp}(0) \doteq \mathrm{z}_{\perp}^0 = \sqrt{8 A}\, \mathrm{z}. \label{ec198}
\end{equation}
Note that $\mathrm{z}(0) = \mathrm{z}$, Eq.~(\ref{ec191}).

Multiplying both sides of Eq.~(\ref{ec195}) by $\mathrm{z}_{\parallel}(l)$, and those of Eq.~(\ref{ec196}) by $\mathrm{z}_{\perp}(l)$, subtracting the resulting equations, one deduces that
\begin{equation}
{\bm\upmu}^2 \doteq \mathrm{z}_{\parallel}^2(l) - \mathrm{z}_{\perp}^2(l) \label{ec199}
\end{equation}
is independent of $l$; the quantity ${\bm\upmu}^2$ is therefore commonly referred to as the `constant of motion' or the `scaling invariant' (this ${\bm\upmu}$ is not to be confused with $\mu$ and $\upmu$ that we encounter elsewhere in this paper). One therefore has
\begin{equation}
{\bm\upmu}^2 \equiv (\mathrm{z}_{\parallel}^0)^2 - (\mathrm{z}_{\perp}^0)^2. \label{ec200}
\end{equation}
The constancy of ${\bm\upmu}^2$, for given values of $\mathrm{z}_{\parallel}^0$ and $\mathrm{z}_{\perp}^0$, implies that under the RG transformation the locus of $\mathrm{z}_{\parallel}(l)\, \h{\bm e}_{\parallel} + \mathrm{z}_{\perp}(l)\, \h{\bm e}_{\perp}$ on the $\mathrm{z}_{\parallel}$-$\mathrm{z}_{\perp}$ plane, with $\{ \h{\bm e}_{\parallel}, \h{\bm e}_{\perp}\}$ a Cartesian basis of this plane, is a parabola. Owing to this fact, under the RG transformation, the trajectory $\mathrm{z}_{\parallel}(l)\, \h{\bm e}_{\parallel} + \mathrm{z}_{\perp}(l)\, \h{\bm e}_{\perp}$ associated with $(\mathrm{z}_{\parallel}^0, +\mathrm{z}_{\perp}^0)$ is the exact mirror image of that associated with $(\mathrm{z}_{\parallel}^0, -\mathrm{z}_{\perp}^0)$. For convenience one therefore only considers the trajectories in the upper-half of the $\mathrm{z}_{\parallel}$-$\mathrm{z}_{\perp}$ plane. \emph{Consequently, unless we indicate otherwise, in the following $\mathrm{z}_{\perp}(l) \ge 0$, $\forall l$} (with reference to Eqs.~(\ref{ec198}) and (\ref{ec191}), in the following we are therefore tacitly assuming that $V\ge 0$). In this upper half-plane, the region $\mathrm{z}_{\perp} \le \mathrm{z}_{\parallel}$ corresponds to the weak-coupling (WC) regime, the region $\mathrm{z}_{\perp} > \vert\mathrm{z}_{\parallel}\vert$ to cross-over (CO) regime, and $\mathrm{z}_{\perp} \le -\mathrm{z}_{\parallel}$ to the strong-coupling (SC) regime \cite{GNT98,TG03}. The WC and SC sectors are characterised by ${\bm\upmu}^2 > 0$, that is by real ${\bm\upmu}$, and the CO sector by $-{\bm\upmu}^2 \doteq {\bm\upmu}_1^2 >0$, that is by purely imaginary ${\bm\upmu}$.

One readily verifies that the following expressions \emph{exactly} satisfy the equations in Eqs.~(\ref{ec195}) and (\ref{ec196}):
\begin{eqnarray}
\mathrm{z}_{\parallel}(l) &=& {\bm\upmu} \coth({\bm\upmu} l + \phi), \label{ec201}\\
\mathrm{z}_{\perp}(l) &=& {\bm\upmu}\, \mathrm{csch}({\bm\upmu} l + \phi), \label{ec202}
\end{eqnarray}
where $\phi$ is a constant, independent of $l$. One trivially verifies that these expressions indeed yield $\mathrm{z}_{\parallel}^2(l) - \mathrm{z}_{\perp}^2(l) = {\bm\upmu}^2$ for all $l$ (cf. Eq.~(\ref{ec199})). Dividing the first of the above expressions by the second, identifying $l$ with $0$, and making use of the initial conditions in Eq.~(\ref{ec198}), one obtains that
\begin{equation}
\phi= \cosh^{-1}\!\Big(\frac{\mathrm{z}_{\parallel}^0}{\mathrm{z}_{\perp}^0}\Big) \equiv \cosh^{-1}\!\Big(\frac{d-2}{\sqrt{8 A}\,\mathrm{z}}\Big). \label{ec203}
\end{equation}

The gap $\mathcal{E}_{\Sc g}$ in the single-particle excitation spectrum of $\wh{\mathcal{H}}$ close to the CDW transition is obtained as follows. Let the gap energy corresponding to $l \not=0$ be denoted by $\mathcal{E}_{\Sc g}(l)$ so that for the actual gap one has $\mathcal{E}_{\Sc g} = \mathcal{E}_{\Sc g}(0)$. Since $\mathcal{E}_{\Sc g}(l)$ has the dimension of energy (or inverse-time, whereby, and by the Lorentz invariance of our continuum theory, it has the \emph{scaling dimension} ${\sf z} = 1$), under the RG transformation it transforms according to
\begin{equation}
\mathcal{E}_{\Sc g}(l) = \mathcal{E}_{\Sc g}(0)\, \e^{l} \iff \mathcal{E}_{\Sc g} = \mathcal{E}_{\Sc g}(l) \,\e^{-l}. \label{ec204}
\end{equation}

\label{RGlist}
Following the common practice, we consider the following three cases corresponding to gapped states (below, as elsewhere, $\mathrm{z}_{\perp}\ge 0$):
\begin{itemize}
\item[(1)] $\mathrm{z}_{\perp} \ll -\mathrm{z}_{\parallel}$, specific to states deep inside the SC regime,
\item[(2)] $\mathrm{z}_{\perp} = - \mathrm{z}_{\parallel}$, corresponding to states on the boundary separating the CO and SC regimes, and
\item[(3)] $\mathrm{z}_{\perp} \gtrsim -\mathrm{z}_{\parallel}$, signifying states inside the CO regime adjacent to the SC regime.
\end{itemize}

Before dealing with the above three cases, we determine the value for the $\mathcal{E}_{\Sc g}(l)$ specific to the strong-coupling regime, where $\ul{g}$, the coupling-constant of the non-Gaussian contribution to the $\wh{S}$ in Eq.~(\ref{ec90}), is large and $\upbeta$, Eq.~(\ref{ec92}), is finite. In this connection, we note that $\ul{g} \propto 1/u$, Eq.~(\ref{ec92}), so that large values of $\ul{g}$ can be achieved not only by large values of $V$, but also by small values of $u$. In view of Eq.~(\ref{ec123}), $u$ approaches zero according to $u \sim \frac{1}{2} v_{\Sc f} (\pi-\mu)$ for $\upmu \uparrow \pi$. However, following Eqs.~(\ref{ec92}) and (\ref{ec116}), $\upbeta \sim 2\pi (v_{\Sc f}/u)^{1/2}$ for $\upmu \uparrow \pi$, so that for $u$ approaching zero, the magnitude of $\ul{g} u^2 \upbeta^2 \equiv g u\upbeta^2$ (see Eq.~(\ref{ec210}) below) only weakly depends on $u$ and is wholly determined by that of $V$, Eq.~(\ref{ec77}). It is interesting to note that, following Eq.~(\ref{ec108}), $\upmu \uparrow \pi$ corresponds to $\Delta \downarrow -1$, and in the light of the considerations in Sec.~\ref{sac.4a}, the GS in the region $\vert\Delta\vert <1$ is disordered and the spectrum of the system is gapless.

In the GS of the system corresponding to large values of $g$, $\ul{\h{\Phi}}(x,-i\ul\tau)$ is locked into a minimum of $\ul{g} \cos(\upbeta \ul{\h{\Phi}}(x,-i\ul\tau))$ and the low-lying excitation spectrum of the system is gapped. This spectrum is readily obtained by expanding to quadratic order the $\cos\big(\upbeta \ul{\h{\Phi}}(x,-i\ul\tau)\big)$ in the action in Eq.~(\ref{ec90}) around the above-mentioned minimum. The position $x_0$ of the minimum of $\ul{g} \cos(x)$ depends on the sign of $\ul{g}$; for $\ul{g} < 0$ one has $x_0 = 0$ (modulo $2\pi$) and for $\ul{g} >0$, $x_0 = \pi$ (modulo $2\pi$). Assuming that $\ul{g} > 0$, on neglecting the $-1$ in the expansion of $\cos(x)$ around $x=\pi$ (by normal ordering) and effecting the transformation
\begin{equation}
\ul{\h{\Phi}}(x,-i\ul\tau) \rightharpoonup \ul{\h{\Phi}}(x,-i\ul\tau) + \frac{\pi}{\upbeta}, \label{ec205}
\end{equation}
the action in Eq.~(\ref{ec90}) reduces into
\begin{equation}
\wh{S} \approx \wh{S}_0 + \frac{\upbeta^2}{2}\, \vert\ul{g}\vert\! \int {\rm d}^2\ul{r}\; \ul{\h{\Phi}}^2(x,-i\ul\tau), \label{ec206}
\end{equation}
where we have used $\vert\ul{g}\vert$ instead of $\ul{g}$. This we have done on account of the fact that $\cos(x) = 1 - x^2/2 + O(x^4)$ for $x \to 0$ (note the minus sign), whereby Eq.~(\ref{ec206}) indeed applies also for $\ul{g} <0$. For completeness, the transformation in Eq.~(\ref{ec205}) does not apply for the cases where $\ul{g} <0$.

Expressing $\ul{\h{\Phi}}(x,-i\ul\tau)$ in the double Fourier series
\begin{equation}
\ul{\h{\Phi}}(x,-i\ul\tau) = \frac{1}{\ul{\beta} \Omega} \sum_{\b{k},\ul{\omega}_n} \e^{i (\b{k} x - \ul{\omega}_n \ul{\tau})}\, \ul{\h{\t\Phi}}(\b{k},i\ul\omega_n), \label{ec207}
\end{equation}
where (cf. Eq.~(\ref{ec94}))
\begin{equation}
\ul{\omega}_n \doteq \frac{1}{u}\, \omega_n, \label{ec208}
\end{equation}
with $\{\omega_n\}$ Matsubara frequencies, and (cf. Eq.~(\ref{e33}) and Eq.~(\ref{ec3}))
\begin{equation}
\b{k} \doteq k \pm k_{\Sc f}, \label{ec209}
\end{equation}
for the $\wh{S}$ in Eq.~(\ref{ec206}) one obtains that
\begin{equation}
\wh{S} \approx \frac{1}{2 u^2} \frac{1}{\ul{\beta} \Omega}\! \sum_{\b{k},\ul{\omega}_n} \!\!\Big[ \omega_n^2 + (u \b{k})^2 + \vert g\vert u \upbeta^2 \Big] \ul{\h{\t\Phi}}^*(\b{k},i\ul\omega_n)\, \ul{\h{\t\Phi}}(\b{k},i\ul\omega_n), \label{ec210}
\end{equation}
where \[\ul{\h{\t\Phi}}^*(\b{k},i\ul\omega_n)\equiv \ul{\h{\t\Phi}}(-\b{k},-i\ul\omega_n).\] The action in Eq.~(\ref{ec210}) being quadratic in the modes $\{\ul{\h{\t\Phi}}(\b{k},i\ul\omega_n)\}$, the zeros of the expression enclosed by the square brackets in Eq.~(\ref{ec210}), subsequent to the application of the analytic continuation \[\ul\omega_n \to -i \ul\omega \pm 0^+,\;\;\; \ul\omega \in \mathds{R},\] coincide with the energy dispersions of these modes. One thus obtains that
\begin{equation}
E_k^{\pm} = \pm \big( (u \b{k})^2 + \sf{M}^2\big)^{1/2}, \label{ec211}
\end{equation}
where
\begin{equation}
{\sf M} \doteq \sqrt{\vert g\vert u}\, \upbeta \equiv 2 \sqrt{\frac{2 \vert V\vert K u}{\pi a}} \label{ec212}
\end{equation}
is the mass gap at $k = \pm k_{\Sc f}$. Since in this section we are tacitly assuming that $V \ge 0$ (see our pertinent remark following Eq.~(\ref{ec200}) above), in what follows we shall denote $\vert V\vert$ by $V$.

In the case at hand, for the spectral gap $\mathcal{E}_{\Sc g}$ at $k = \pm k_{\Sc f}$ one has $\mathcal{E}_{\Sc g} = 2 {\sf M}$. As the \emph{physical} spectral gap is small for $\Delta \downarrow 1$ (see Eqs.~(\ref{e22}) and (\ref{ec190})), for $0< \Delta -1 \ll 1$ one can identify the $u$ in Eq.~(\ref{ec212}) with that specific to $\Delta =1$, that is $u = \pi v_{\Sc f}/2 = \pi a t$ (Eqs.~(\ref{ec126}) and (\ref{ec40})). Similarly for $K$, which for $0< \Delta - 1 \ll 1$ takes a value satisfying $0 < \frac{1}{2} - K \ll 1$, Eq.~(\ref{ec119}); for $K < \frac{1}{2}$ the cosine term of the sine-Gordon Hamiltonian is relevant (see Eqs.~(\ref{ec95}), (\ref{ec97}) and the subsequent remarks). Thus for $0< \Delta - 1\ll 1$ from Eq.~(\ref{ec212}) one obtains that
\begin{equation}
{\sf M} \approx 2\sqrt{2}\, t\;\;\; \mbox{\rm for}\;\;\; 0 <\Delta -1 \ll 1. \label{ec213}
\end{equation}
Consequently,
\begin{equation}
\mathcal{E}_{\Sc g}(l_{\star}) = 2 {\sf M} \approx 4\sqrt{2}\, t \approx 5.7\, t.\label{ec214}
\end{equation}
With $2 t$ the bandwidth (for $t'=0$), one observes that $\mathcal{E}_{\Sc g}(l_{\star})$ is of the order of the bandwidth.

We should emphasise that, in contrast to $\mathcal{E}_{\Sc g}(l_{\star}) \e^{-l_{\star}}$, Eq.~(\ref{ec204}), the quantity $\mathcal{E}_{\Sc g}(l_{\star})$ is \emph{physically} meaningless. This follows from the fact that in expanding the action as in Eq.~(\ref{ec206}), we have taken no account of the reality that deep inside the massive phase (and indeed ${\sf M} \approx 2\sqrt{2}\, t$ is a very large mass), the parameter $u$, which is well-defined only in the region $-1 <\Delta \le 1$, carries no physically relevant information about the system under consideration.

We are now in a position to consider the cases (1), (2) and (3) indicated above, on page \pageref{RGlist}.

\subsubsection{Case 1: $\mathrm{z}_{\perp} \ll -\mathrm{z}_{\parallel}$}
\label{sac.5a}

In this case, one may identify the RHS of Eq.~(\ref{ec195}) with zero (recall our convention, whereby $\mathrm{z}_{\perp}\ge 0$), implying that $\mathrm{z}_{\parallel}(l)$ is independent of $l$ and thus equal to $\mathrm{z}_{\parallel}^0$. Using this result in the expression on the RHS of Eq.~(\ref{ec196}), one immediately obtains that
\begin{equation}
\mathrm{z}_{\perp}(l) \approx \mathrm{z}_{\perp}^0\, \e^{-\mathrm{z}_{\parallel}^0\, l} \iff \e^{-l} \approx \Big(\frac{\mathrm{z}_{\perp}(l)}{\mathrm{z}_{\perp}^0}\Big)^{1/\mathrm{z}_{\parallel}^0}. \label{ec215}
\end{equation}
By defining $l_{\star}$ as the value of $l$ for which $\mathrm{z}_{\perp}(l) = 1$ (or, more generally, $\mathrm{z}_{\perp}(l) = O(1)$), we deduce that
\begin{equation}
\e^{-l_{\star}} \approx \left| \mathrm{z}_{\perp}^0\right|^{-1/\mathrm{z}_{\parallel}^0}, \label{ec216}
\end{equation}
and thus, following Eqs.~(\ref{ec204}) and (\ref{ec212}),
\begin{equation}
\mathcal{E}_{\Sc g} \doteq 2 \mathcal{M} \approx 4 \sqrt{\frac{2 V K u}{\pi a}}\, \left|\mathrm{z}_{\perp}^0\right|^{-1/\mathrm{z}_{\parallel}^0}. \label{ec217}
\end{equation}

\subsubsection{Case 2: $\mathrm{z}_{\perp} = - \mathrm{z}_{\parallel}$}
\label{sac.5b}

In this case, the differential equations in Eqs.~(\ref{ec195}) and (\ref{ec196}) can be expressed as
\begin{equation}
\frac{\rd \mathrm{z}_{\alpha}(l)}{\rd l} = \pm \mathrm{z}_{\alpha}^2(l),\;\;\; \alpha = \left\{ \begin{array}{l} \perp \\ \parallel \end{array} \right.. \label{ec218}
\end{equation}
One readily verifies that
\begin{equation}
\mathrm{z}_{\alpha}(l) = \frac{\mathrm{z}_{\alpha}^0}{1 \mp \mathrm{z}_{\alpha}^0\, l}, \;\;\; \alpha = \left\{ \begin{array}{l} \perp \\ \parallel \end{array} \right.,\label{ec219}
\end{equation}
satisfies both the differential equation in Eq.~(\ref{ec218}) and the appropriate boundary condition. Denoting, as above, the value of $l$ for which $\mathrm{z}_{\perp}(l) = 1$ by $l_{\star}$, from Eq.~(\ref{ec219}) one obtains that
\begin{equation}
1 = \frac{\mathrm{z}_{\perp}^0}{1-\mathrm{z}_{\perp}^0\, l_{\star}} \iff l_{\star} = \frac{1}{\mathrm{z}_{\perp}^0} -1.\label{ec220}
\end{equation}
Hence, by Eqs.~(\ref{ec204}) and (\ref{ec212}) one arrives at
\begin{equation}
\mathcal{E}_{\Sc g} \doteq 2 \mathcal{M} = 4 \sqrt{\frac{2 V K u}{\pi a}}\, \exp\!\Big[1-\frac{1}{\mathrm{z}_{\perp}^0}\Big]. \label{ec221}
\end{equation}

\subsubsection{Case 3: $\mathrm{z}_{\perp} \gtrsim -\mathrm{z}_{\parallel}$}
\label{sac.5c}

In this case ${\bm\upmu}_1 \doteq  \sqrt{-{\bm\upmu}} > 0$, which however close to transition it is nearly equal to zero. Consequently, on account of Eq.~(\ref{ec200}), in the case at hand to a good approximation one has $\mathrm{z}_{\perp}^0/\mathrm{z}_{\parallel}^0 = -1$. Following Eq.~(\ref{ec203}), one therefore to a good approximation has
\begin{equation}
\phi = \pi i, \label{ec222}
\end{equation}
as a result of which Eqs.~(\ref{ec201}) and (\ref{ec202}) reduce into
\begin{eqnarray}
\mathrm{z}_{\parallel}(l) &=& {\bm\upmu}_1 \cot({\bm\upmu}_1 l), \label{ec223}\\
\mathrm{z}_{\perp}(l) &=& {\bm\upmu}_1 \csc({\bm\upmu}_1 l). \label{ec224}
\end{eqnarray}
Inspired by the expression in Eq.~(\ref{ec190}), here we define $l_{\star}$ by the condition \[{\bm\upmu}_1 l_{\star} = \frac{1}{2} \pi^2 \vert \mathrm{z}_{\parallel}^0\vert \equiv \pi^2 (1-d(0)/2) \equiv \pi^2 (1-2 K),\] where $d(0) = d \equiv \upbeta^2/(4\pi) \equiv 4 K$, Eqs.~(\ref{ec197}), (\ref{ec95}) and (\ref{ec92}). One thus has
\begin{equation}
l_{\star} = \frac{\pi^2}{2 {\bm\upmu}_1} \equiv \frac{\pi^2\, \vert \mathrm{z}_{\parallel}^0\vert}{2 \sqrt{(\mathrm{z}_{\perp}^0)^2 - (\mathrm{z}_{\parallel}^0)^2}} \approx \frac{\pi^2}{2 \sqrt{2 (\mathrm{z}_{\perp}^0/\vert \mathrm{z}_{\parallel}^0\vert - 1)}}, \label{ec225}
\end{equation}
where the last approximate expression is based on the consideration that in the case at hand $\mathrm{z}_{\perp}^0/\vert \mathrm{z}_{\parallel}^0\vert \gtrapprox 1$ (recall the above reasoning leading to Eq.~(\ref{ec222})). Consequently, following Eqs.~(\ref{ec204}) and (\ref{ec212}), one obtains that
\begin{equation}
\mathcal{E}_{\Sc g} \doteq 2\mathcal{M} \approx 4 \sqrt{\frac{2 V K u}{\pi a}}\, \exp\!\Big[\!-\frac{\pi^2}{2 \sqrt{2\, (\mathrm{z}_{\perp}^0/\vert \mathrm{z}_{\parallel}^0\vert - 1)}} \Big]. \label{ec226}
\end{equation}
This is the RG-based counterpart of the expression in Eq.~(\ref{ec190}). The similarity between the two expressions clarifies our above choice for the condition ${\bm\upmu}_1 l_{\star} = \frac{1}{2} \pi^2 \vert \mathrm{z}_{\parallel}^0\vert$.

\label{valueofA}
We note in passing that on identifying $\mathrm{z}_{\perp}^0/\vert \mathrm{z}_{\parallel}^0\vert$ with $\Delta  \doteq V/(2t)$ for $\Delta \downarrow 1$, one can deduce the explicit expression of the constant $A$, introduced in Eq.~(\ref{ec194}), in terms of $K$; for $\Delta \downarrow 1$, one obtains $A \sim \frac{1}{2} \pi^6 (1-2K)^2$. Since for $\Delta \downarrow 1$ one has $K \uparrow \frac{1}{2}$, it follows that $A$ vanishes in the limit of $\Delta = 1$.

\subsection{The single-particle Green functions $\t{G}_{\h{\Sc A},\h{\Sc B}}(k;i \omega_n)$, $\h{A}, \h{B} \in \{ \h{L},\h{R}\}$}
\label{sac.6}

Here we determine the single-particle Green functions $\t{G}_{\h{\Sc a},\h{\Sc b}}(k;i \omega_n)$, $\h{A}, \h{B} \in \{ \h{L},\h{R}\}$, as defined through Eqs.~(\ref{e31}) and (\ref{e32}), in terms of the form-factors of the soliton-generating field operators pertaining to the quantum sine-Gordon Hamiltonian as determined by Lukyanov and Zamolodchikov (LZ) \cite{LZ01}.

\subsubsection{Preliminaries}
\label{sac.6a}

We first introduce (cf. Eqs.~(\ref{ec92}) and (\ref{ec77}))
\begin{eqnarray}
\uul{\h{\upvarphi}}(x,-i\ul\tau) &\doteq& -\sqrt{8\pi}\, \ul{\h{\Phi}}(x,-i\ul\tau), \label{ec227} \\
{\bm\upbeta} &\doteq& \frac{\upbeta}{\sqrt{8\pi}} \equiv \sqrt{2 K},
\label{ec228} \\
{\bm\mu} &\doteq& -\frac{1}{2}\, \ul{g} \equiv -\frac{a V}{(2\pi a)^2\, u}, \label{ec229}
\end{eqnarray}
whereby the action in Eq.~(\ref{ec90}) can be written as
\begin{widetext}
\begin{equation}
\wh{S} = \int {\rm d}^2\ul{r}\; \Big\{\frac{1}{16\pi} \big[\big(\partial_{\ul\tau} \uul{\h{\upvarphi}}(x,-i\ul\tau)\big)^2 + \big(\partial_{x} \uul{\h{\upvarphi}}(x,-i\ul\tau)\big)^2 \big] - 2{\bm\mu} \cos({\bm\upbeta} \uul{\h{\upvarphi}}(x,-i\ul\tau))\Big\}. \label{ec230}
\end{equation}
This Euclidean action coincides with that underlying the considerations by LZ in Ref.~\cite{LZ01}. The minus sign on the RHS of Eq.~(\ref{ec227}) is not significant and serves only to bring our subsequent notation in conformity with that of LZ \cite{LZ01}.

With reference to Eqs.~(\ref{ec82}) and (\ref{ec83}), and in view of the fact that through the scaling transformations in Eq.~(\ref{ec88}) and (\ref{ec89}) the $K$ in the action in Eq.~(\ref{ec230}) has been identified with unity, one has
\begin{equation}
\h{\ul{\Pi}}(x,-i\ul\tau) = i\, \partial_{\ul\tau} \h{\ul{\Phi}}(x,-i\ul\tau) \equiv \frac{-i}{\sqrt{8\pi}}\, \partial_{\ul\tau} \uul{\h{\upvarphi}}(x,-i\ul\tau). \label{ec231}
\end{equation}
Making use of this expression and that in Eq.~(\ref{ec227}), taking into account the transformations in Eqs.~(\ref{ec88}) and (\ref{ec89}), from the second expression in Eq.~(\ref{ec6}) (in which we suppress the Klein factors $\h{U}_{\pm}$ --- see the remark in the paragraph following Eq.~(\ref{ec27}), on page~\pageref{KleinF}), combined with Eq.~(\ref{ec27}) (cf. Eq.~(\ref{e25})), one readily obtains that
\begin{equation}
\left.\begin{array}{c} \h{\ul{R}}(x,-i\ul\tau)\\ \h{\ul{L}}(x,-i\ul\tau)\end{array} \right\} = \pm\frac{\e^{\mp i\pi/4}}{\sqrt{2\pi a}} \exp\!\big[-\frac{1}{2 {\bm\upbeta}} \int_{-\infty}^x \rd x'\; \partial_{\ul\tau} \uul{\h{\upvarphi}}(x',-i\ul\tau)\big]\, \exp\!\big[\mp \frac{i {\bm\upbeta}}{4}\, \uul{\h{\upvarphi}}(x+0^+,-i\ul\tau)\big]. \label{ec232}
\end{equation}
\end{widetext}
These expressions are to be compared with that of the non-local field $\h{\mathcal{O}}_{s}^n(x)$ considered by LZ \cite{LZ01} and presented in Eq.~(2.4) of Ref.~\cite{LZ01} (here $n$ is an index and is not to be confused with a power). From this comparison one deduces that in employing the form factors as calculated by LZ \cite{LZ01} in the context of our present considerations, one has the following relationships:
\begin{equation}
\left.\begin{array}{c} \h{\ul{R}}(x,-i\ul\tau)\\ \h{\ul{L}}(x,-i\ul\tau)\end{array} \right\} \iff \pm\frac{\e^{\mp i\pi/4}}{\sqrt{2\pi a}}\, \h{\mathcal{O}}_{s}^n(x),\;\, {s} = \mp \frac{1}{4} {\bm\upbeta},\; n=2. \label{ec233}
\end{equation}
We note that in Ref.~\cite{LZ01} $x \equiv (\mathrm{x},\mathrm{y})$, which is to be identified with the Cartesian vector $\ul{\bm r} \equiv (x,\ul{\tau})$ in our notation. In the following we shall therefore denote the $\h{\mathcal{O}}_{s}^n(x)$ of LZ by $\h{\mathcal{O}}_{s}^n(x,-i\ul\tau)$. With reference to Eqs.~(\ref{ec85}), (\ref{ec86}) and (\ref{ec87}), we point out that
\begin{equation}
\h{L}(x,-i\tau) \equiv \h{\ul{L}}(x,-i \ul\tau),\;\;\; \h{R}(x,-i\tau) \equiv \h{\ul{R}}(x,-i \ul\tau), \label{ec234}
\end{equation}
so that, in view of Eqs.~(\ref{e31}) and (\ref{e32}) and the second equality in Eq.~(\ref{e33}) (for $L\to\infty$), one has
\begin{equation}
\t{G}_{\h{\Sc a},\h{\Sc b}}(k;i\omega_n \to z) = \frac{1}{u}\, \t{G}_{\h{\ul{\Sc a}},\h{\ul{\Sc b}}}(k;i\ul\omega_n \to \frac{1}{u}\, z), \label{ec235}
\end{equation}
where $\ul\omega_n \doteq \omega_n/u$ (see Eqs.~(\ref{ec208}), (\ref{ec87}) and (\ref{e26})). The result in Eq.~(\ref{ec235}) is a direct consequence of the Lorentz invariance of the continuum theory under consideration, so that it does \emph{not} strictly apply for the CDW GS (see Eq.~(\ref{ec273}) below and the subsequent remarks). Since the energy gap $\mathcal{E}_{\Sc g}$ in the single-particle excitation spectrum of the system under consideration diminishes exponentially for $V \downarrow V_{\rm c}$, Eq.~(\ref{e22}), for $V$ sufficiently close to $V_{\rm c}$ the deviation of the function on the LHS of Eq.~(\ref{ec235}) from its exact counterpart is however exponentially small.

\label{LorentzI}
To determine the functional forms of the operators $\h{\ul{L}}(x,-i\ul\tau)$ and $\h{\ul{R}}(x,-i\ul\tau)$, Eq.~(\ref{ec234}), one should realise that for $V=0$, $\h{R}(x,-i\tau)$ is a function of $\b{z}$ and $\h{L}(x,-i\tau)$ a function of $z$, Eqs.~(\ref{ec46}), (\ref{ec50}) and (\ref{ec51}); this remains the case for $V\not=0$, and $\vert \Delta\vert \le 1$, barring the fact that for $V\not=0$, and $\vert \Delta\vert \le 1$, the $v_{\Sc f}$ in Eq.~(\ref{ec46}) is transformed into the $u$ in Eq.~(\ref{ec99}). Similarly as regards $\h{\Phi}(x,-i\tau)$ and $\h{\Theta}(x,-i\tau)$, in the light of the expressions in Eq.~(\ref{ec47}). Using the identities $z \equiv \frac{1}{u} (\ul\tau + i x)$ and $\b{z} \equiv \frac{1}{u} (\ul\tau - i x)$, one immediately infers that the expression for $\h{\ul{L}}(x,-i\ul\tau)$ is deduced from that of $\h{L}(x,-i\tau)$ by substituting, in the explicit expression for the latter operator, $x/u$ by $x$ and $\tau$ by $\ul\tau$. Similarly for $\h{\ul{R}}(x,-i\ul\tau)$. Conversely, the expression for $\h{L}(x,-i\tau)$ is deduced from that of $\h{\ul{L}}(x,-i\ul\tau)$ by substituting, in the explicit expression for the latter operator, $x$ by $x/u$ and $\ul\tau$ by $\tau$. Similarly for $\h{R}(x,-i\ul\tau)$ (see Eq.~(\ref{ec270})).

With $\vert\Psi_{N_{\rm e}}\rangle$ denoting the $N_{\rm e}$-particle GS of $\wh{\mathcal{H}}$, we are to evaluate the following quantities:
\begin{equation}
\langle \Psi_{N_{\rm e}}\vert \h{\mathcal{O}}_{s}^2(x,-i\ul\tau) \, \h{\mathcal{O}}_{s'}^{2\,\dag}(0,0) \vert \Psi_{N_{\rm e}} \rangle, \;\, {s}, {s'} \in \{ -\frac{{\bm\upbeta}}{4}, +\frac{{\bm\upbeta}}{4} \}, \nonumber
\end{equation}
this on account of the fact that for $\ul\tau > 0$ one has (cf. Eq.~(\ref{e32}))
\begin{widetext}
\begin{eqnarray}
\lim_{\ul\beta\to\infty}\mathscr{G}_{\h{\ul{\Sc a}},\h{\ul{\Sc b}}}(x,\ul\tau) &=& -\frac{\e^{-\pi i(\varsigma-\varsigma')/4}}{2\pi a}\, \langle \Psi_{N_{\rm e}}\vert \h{\mathcal{O}}_{\varsigma {\bm\upbeta}/4}^2(x,-i\ul\tau) \, \h{\mathcal{O}}_{\varsigma' {\bm\upbeta}/4}^{2\,\dag}(0,0) \vert \Psi_{N_{\rm e}} \rangle, \;\;\varsigma,\varsigma' = \left\{ \begin{array}{ll} -1, & \h{\ul{A}}, \h{\ul{B}} = \h{\ul{R}}, \\ \\ +1, & \h{\ul{A}}, \h{\ul{B}} = \h{\ul{L}}. \end{array} \right. \label{ec236}
\end{eqnarray}
\end{widetext}
Following Eqs.~(\ref{e31}) and (\ref{e33}), from these quantities one calculates $G(k;z)$ (see the following two paragraphs). We note that, without the `$\pm$' with which the expression on the RHS of Eq.~(\ref{ec232}) (and similarly Eq.~(\ref{e25})) is pre-multiplied (see our remarks following Eq.~(\ref{ec27}) above), the phase factor on the RHS of Eq.~(\ref{ec236}) would have been $\e^{+\pi i(\varsigma-\varsigma')/4}$.

The expression in Eq.~(\ref{ec236}) is in general \emph{not} sufficient to calculate the zero-temperature Green function $G_{\h{\Sc a},\h{\Sc b}}(x,i\omega_n)$, for the limit $\beta\to\infty$ in Eq.~(\ref{e31}) is to be effected \emph{after} evaluating the integral with respect to $\tau$. Similarly for $\t{G}_{\h{\Sc a},\h{\Sc b}}(k;z)$. To be explicit, by effecting the limit $\beta\to\infty$ \emph{before} evaluating the last-mentioned integral, one can demonstrate that the resulting Green function satisfies $\im[\t{G}_{\h{\Sc a},\h{\Sc b}}(k;\omega +i\eta)] \equiv 0$ for \emph{all} $\omega < 0$ when $\eta = 0^{\pm}$. One can convince oneself of this fact by considering Eq.~(25.18) in Ref.~\cite{FW03}, which reveals that evaluating the integral with respect to $\tau$ over $[0,\infty)$ of the equivalent of the expression in Eq.~(\ref{ec236}) is tantamount to identifying the $(1-n_j^0)$ in Eq.~(25.18) of Ref.~\cite{FW03} with its zero-temperature limit, which is identically vanishing for all $j$ corresponding to $\epsilon_j^0 > \mu$. In contrast, use of the expression in Eq.~(\ref{e31}) has the effect of totally compensating $(1-n_j^0)$ by a factor $-1/(1-n_j^0)$ which in the limit $\beta=\infty$ is only operative for $\epsilon_j^0 > \mu$ (in Ref.~\cite{FW03}, $n_j^0$ denotes the non-interacting Fermi distribution function).

In our case, the expression in Eq.~(\ref{ec236}) contains sufficient information for calculating the zero-temperature Green function $G_{\h{\ul{\Sc a}},\h{\ul{\Sc b}}}(x,i\ul\omega_n)$, or $\t{G}_{\h{\ul{\Sc a}},\h{\ul{\Sc b}}}(k;z)$. This follows from the Lorentz invariance of the theory (which applies only approximately for $V > V_{\rm c}$) whereby for $\h{A} \not= \hat{B}$ one has
\begin{eqnarray}
\t{G}_{\h{\ul{\Sc a}},\h{\ul{\Sc a}}}(-k;+z) &\equiv& +\t{G}_{\h{\ul{\Sc b}},\h{\ul{\Sc b}}}(k;z), \nonumber\\
\t{G}_{\h{\ul{\Sc a}},\h{\ul{\Sc a}}}(+k;-z) &\equiv& -\t{G}_{\h{\ul{\Sc b}},\h{\ul{\Sc b}}}(k;z), \nonumber\\
\t{G}_{\h{\ul{\Sc a}},\h{\ul{\Sc b}}}(-k;-z) &\equiv& -\t{G}_{\h{\ul{\Sc b}},\h{\ul{\Sc a}}}(k;z). \label{ec237}
\end{eqnarray}
These results are obtained from the spectral representation of the Green functions $\t{G}_{\h{\ul{\Sc a}},\h{\ul{\Sc b}}}(k;z)$, $\h{A}, \hat{B} \in \{ \h{L},\h{R}\}$, in terms of the single-particle spectral function $A_{\h{\ul{\Sc a}},\h{\ul{\Sc b}}}(k;\omega)$, Eq.~(\ref{ec287}), and the symmetry properties (for $\h{A} \not= \hat{B}$): \[ A_{\h{\ul{\Sc a}},\h{\ul{\Sc a}}}(k;\omega) \equiv A_{\h{\ul{\Sc b}},\h{\ul{\Sc b}}}(k;-\omega) \equiv A_{\h{\ul{\Sc b}},\h{\ul{\Sc b}}}(-k;\omega),\] and \[A_{\h{\ul{\Sc a}},\h{\ul{\Sc b}}}(k;\omega) \equiv A_{\h{\ul{\Sc b}},\h{\ul{\Sc a}}}(-k;-\omega).\] In what follows, we shall denote the functions $\t{G}_{\h{\ul{\Sc a}},\h{\ul{\Sc b}}}(x,i\ul\omega_n)$ and $\t{G}_{\h{\ul{\Sc a}},\h{\ul{\Sc b}}}(k;z)$ as deduced from the expression in Eq.~(\ref{ec236}) by $\t{G}_{\h{\ul{\Sc a}},\h{\ul{\Sc b}}}^{+}(x,i\ul\omega_n)$ and $\t{G}_{\h{\ul{\Sc a}},\h{\ul{\Sc b}}}^{+}(k;z)$ respectively. It can be shown that for $\h{A} \not=\h{B}$ (see Eq.~(\ref{ec282}) below) \begin{eqnarray}
\t{G}_{\h{\ul{\Sc a}},\h{\ul{\Sc a}}}^{+}(-k;z) &\equiv& \t{G}_{\h{\ul{\Sc b}},\h{\ul{\Sc b}}}^{+}(k;z), \nonumber\\
\t{G}_{\h{\ul{\Sc a}},\h{\ul{\Sc b}}}^{+}(-k;z) &\equiv& \t{G}_{\h{\ul{\Sc b}},\h{\ul{\Sc a}}}^{+}(k;z), \nonumber\\
\t{G}_{\h{\ul{\Sc a}},\h{\ul{\Sc b}}}^{+}(-k;z) &\equiv& \t{G}_{\h{\ul{\Sc a}},\h{\ul{\Sc b}}}^{+}(k;z). \label{ec238}
\end{eqnarray}

Making use the identities in Eq.~(\ref{ec238}), one readily verifies that, for  $\h{A} \not=\h{B}$, the following Green functions satisfy the identities in Eq.~(\ref{ec237}):
\begin{eqnarray}
\t{G}_{\h{\ul{\Sc a}},\h{\ul{\Sc a}}}(k;z) &\doteq& \t{G}_{\h{\ul{\Sc a}},\h{\ul{\Sc a}}}^{+}(k;z) - \t{G}_{\h{\ul{\Sc b}},\h{\ul{\Sc b}}}^{+}(-k;-z), \nonumber\\
\t{G}_{\h{\ul{\Sc a}},\h{\ul{\Sc b}}}(k;z) &\doteq& \t{G}_{\h{\ul{\Sc a}},\h{\ul{\Sc b}}}^{+}(k;z) - \t{G}_{\h{\ul{\Sc b}},\h{\ul{\Sc a}}}^{+}(-k;-z).
\label{ec239}
\end{eqnarray}
One can gain confidence that the Green functions as defined here are indeed the correct representations of $\t{G}_{\h{\ul{\Sc a}},\h{\ul{\Sc a}}}(k;z)$ and $\t{G}_{\h{\ul{\Sc a}},\h{\ul{\Sc b}}}(k;z)$ by identifying $z$ with $\omega + i\eta$, $\eta\to 0$, and realising that $\im[\t{G}_{\h{\ul{\Sc a}},\h{\ul{\Sc b}}}^+(k;\omega + i\eta)] \equiv 0$ for $\omega <0$ and $\im[\t{G}_{\h{\ul{\Sc a}},\h{\ul{\Sc b}}}^+(k;\omega + i\eta)] \equiv \im[\t{G}_{\h{\ul{\Sc a}},\h{\ul{\Sc b}}}(k;\omega + i\eta)]$ for $\omega>0$, $\h{A}, \h{B} \in \{ \h{L}, \h{R}\}$.

We note that, following Eq.~(\ref{ec237}), from Eq.~(\ref{e33}) one obtains that
\begin{equation}
G(-k;-z) \equiv -G(k;z). \label{ec240}
\end{equation}
From Eqs.~(\ref{ec237}) and (\ref{e33}) one further deduces that
\begin{equation}
G(\pm k_{\Sc f};-z) \equiv -G(\pm k_{\Sc f};z). \label{ec241}
\end{equation}
This result is remarkable in that it establishes that for $G(\pm k_{\Sc f};z)$ \emph{continuous} at $z=0$, one has
\begin{equation}
G(\pm k_{\Sc f};0) = 0. \label{ec242}
\end{equation}
This equality is to be compared with that in Eq.~(\ref{e9}). One should also note the numerical results in Fig.~\ref{f2}.

\subsubsection{Technical details}
\label{sac.6b}

Let $\h{Z}_{\epsilon}(\theta)$ and $\h{Z}_{\epsilon}^{\dag}(\theta)$ denote the annihilation and creation operators for the single-particle state characterized by isotopic index $\epsilon$, $\epsilon \in \{-,+\}$, and rapidity $\theta$. These operators are not canonical, however satisfy the Zamolodchikov-Fadeev algebra \cite{SL95} (also Ch. IV, Sec.~34 in Ref.~\cite{AMT03}). Here $\epsilon = -$ refers to solitons (kinks) and $\epsilon= +$ to anti-solitons (anti-kinks). Since $\vert\Psi_{N_{\rm e}}\rangle$ is a state free from solitons and anti-solitons, it is a vacuum state of the theory. Below we shall therefore employ the notation
\begin{equation}
\vert 0\rangle \equiv \vert\Psi_{N_{\rm e}}\rangle. \label{ec243}
\end{equation}
One has
\begin{equation}
\h{Z}_{\epsilon}(\theta) \vert 0\rangle = 0. \label{ec244}
\end{equation}

A multi-particle state, consisting of a total of $n$ solitons and anti-solitons is determined according to
\begin{equation}
\vert \theta_1,\epsilon_1;\dots;\theta_n,\epsilon_n\rangle = \h{Z}_{\epsilon_1}^{\dag}(\theta_1)\dots \h{Z}_{\epsilon_n}^{\dag}(\theta_n) \vert 0\rangle. \label{ec245}
\end{equation}
The identity operator $\h{I}$ of the Fock space under consideration can therefore be decomposed as follows \cite{BFKZ99}:
\begin{eqnarray}
&&\hspace{-1.4cm} \h{I} = \vert 0\rangle \langle 0\vert + \sum_{n=1}^{\infty} \sum_{\{ \epsilon_1, \dots,\epsilon_n\}} \int_{-\infty}^{\infty} \frac{\rd \theta_1 \dots \rd \theta_n}{(2\pi)^n\, n!}\; \nonumber \\
&&\hspace{0.15cm} \times\vert \theta_1,\epsilon_1;\dots;\theta_n,\epsilon_n\rangle \langle \theta_1,\epsilon_1;\dots;\theta_n,\epsilon_n\vert. \label{ec246}
\end{eqnarray}
\begin{widetext}
We note that in the notation of LZ \cite{LZ01} the vacuum state $\vert 0\rangle$ is denoted by $\vert\mathrm{vac}\rangle$, and a basis state containing $n+N$ solitons, with rapidities $\theta_1$,\dots, $\theta_{n+N}$, and $N$ anti-solitons, with rapidities $\theta_1'$,\dots, $\theta_N'$, by
\begin{equation}
\vert A_-(\theta_1)\dots A_-(\theta_{n+N})\, A_+(\theta_1')\dots A_+(\theta_N')\rangle. \nonumber
\end{equation}

With $E(\theta)$ and $P(\theta)$ denoting respectively the energy and momentum of a soliton/antisoliton corresponding to rapidity $\theta$, one obtains that
\begin{eqnarray}
&&\hspace{-1.0cm} \langle \Psi_{N_{\rm e}}\vert \h{\mathcal{O}}_{s}^2(x,-i\ul\tau) \, \h{\mathcal{O}}_{s'}^{2\,\dag}(0,0) \vert \Psi_{N_{\rm e}} \rangle = \sum_{n=1}^{\infty} \sum_{\{\epsilon_1,\dots,\epsilon_n\}} \int_{-\infty}^{\infty} \frac{\rd \theta_1 \dots \rd \theta_n}{(2\pi)^n\, n!}\;\exp\!\Big[\sum_{j=1}^n \big(i P(\theta_j) x - E(\theta_j) \ul\tau\big)\Big]\, \nonumber\\
&&\hspace{6.5cm} \times \langle 0 \vert \h{\mathcal{O}}_{s}^2\vert \theta_1,\epsilon_1;\dots;\theta_n,\epsilon_n\rangle \langle 0 \vert \h{\mathcal{O}}_{s'}^2\vert \theta_1,\epsilon_1;\dots;\theta_n,\epsilon_n\rangle^*. \label{ec247}
\end{eqnarray}
\end{widetext}
For $E(\theta)$ and $P(\theta)$ one has (Eq.~(2.5) in Ref.~\cite{SL95}, and Eq.~(34.1) in Ref.~\cite{AMT03})
\begin{equation}
E(\theta) = \mathfrak{m}\, \cosh(\theta),\;\;\;
P(\theta) = \mathfrak{m}\, \sinh(\theta), \label{ec248}
\end{equation}
where $\mathfrak{m}$ is the soliton/antisoliton mass (see the remark following Eq.~(\ref{e35})). The expression in Eq.~(\ref{ec247}) makes explicit the way in which the Green functions $G_{\h{\ul{\Sc a}},\h{\ul{\Sc b}}}^+(x,i\ul\omega_n)$, $\h{\ul{A}}, \h{\ul{B}} \in \{\h{\ul{L}}, \h{\ul{R}} \}$, are determined in terms of \emph{form factors} $\langle 0 \vert \h{\mathcal{O}}_{s}^2\vert \theta_1,\epsilon_1;\dots;\theta_n,\epsilon_n\rangle$, ${s} \in \{-\frac{{\bm\upbeta}}{4},+\frac{{\bm\upbeta}}{4}\}$. Conservation of topological charge implies that in the case at hand only those form factors are non-vanishing for which number of solitons is by two units in excess of number of anti-solitons; in general, only those form factors corresponding to  $\h{\mathcal{O}}_{s}^p$ are in principle non-vanishing for which number of solitons is by $p$ units in excess of number of anti-solitons. Consequently, the first non-vanishing term on the RHS of Eq.~(\ref{ec247}) corresponds to $n=2$. By the same reasoning, we have already suppressed the contribution of $\vert 0\rangle\langle 0\vert$ in $\h{I}$ to the expectation value in Eq.~(\ref{ec247}).

Close to insulator-metal transition (i.e. for $\Delta\downarrow 1$), the first sum on the RHS of Eq.~(\ref{ec247}) is dominated by the terms corresponding to $n=2$. By the reasoning of the last paragraph, in this region only the exited states corresponding to two solitons, i.e. to $\epsilon_1 = -$ and $\epsilon_2=-$, contribute to the sum $\sum_{\{\epsilon_1,\epsilon_2\}}$ on the RHS of Eq.~(\ref{ec247}). For $\Delta\downarrow 1$ we therefore employ the following approximation:
\begin{widetext}
\begin{equation}
\langle \Psi_{N_{\rm e}}\vert \h{\mathcal{O}}_{s}^2(x,-i\ul\tau) \, \h{\mathcal{O}}_{s'}^{2\,\dag}(0,0) \vert \Psi_{N_{\rm e}} \rangle \approx \frac{1}{2}\! \int_{-\infty}^{\infty}\!\! \frac{\rd \theta_1 \rd \theta_2}{(2\pi)^2} \exp\!\Big[\sum_{j=1}^2 \big(i P(\theta_j) x - E(\theta_j) \ul\tau\big)\Big] \langle 0 \vert \h{\mathcal{O}}_{s}^2\vert \theta_1,-;\theta_2,-\rangle \langle 0 \vert \h{\mathcal{O}}_{s'}^2\vert \theta_1,-;\theta_2,-\rangle^*. \label{ec249}
\end{equation}
One has (Eq.~(2.12) in Ref.~\cite{LZ01})
\begin{eqnarray}
\langle 0 \vert \h{\mathcal{O}}_{s}^2\vert \theta_1,-;\theta_2,-\rangle &=&
\sqrt{Z_2({s})}\, \exp\!\big[\frac{\pi i {s}}{{\bm\upbeta}}\big]\, \exp\!\big[\frac{s}{{\bm\upbeta}} (\theta_1 + \theta_2)\big]\, \mathcal{G}(\theta_1-\theta_2) \nonumber\\
&\equiv& \sqrt{Z_2'(\varsigma)}\, \exp\!\big[\frac{\varsigma \pi i}{4}\big]\, \exp\!\big[\frac{\varsigma}{4} (\theta_1 + \theta_2)\big]\, \mathcal{G}(\theta_1-\theta_2),\;\;\; {s} \equiv \varsigma\,\frac{{\bm\upbeta}}{4}, \; \varsigma \in \{-1,+1\},
\label{ec250}
\end{eqnarray}
where $Z_2({s}) \equiv Z_2'(\varsigma)$ is a normalization constant, and (Eq.~(2.13) in Ref.~\cite{LZ01} --- see appendix \ref{sad})
\begin{equation}
\mathcal{G}(\theta) \doteq i \,\mathcal{C}_1\, \sinh(\theta/2)\, \exp\!\Big[\int_0^{\infty} \frac{\rd t}{t}\, \frac{\sinh^2\!\big((1 - i\theta/\pi) t\big) \sinh\!\big((\upxi-1) t\big)}{\sinh(2t) \cosh(t) \sinh(\upxi t)}\Big], \label{ec251}
\end{equation}
in which (Eq.~(2.14) of Ref.~\cite{LZ01})
\begin{equation}
\mathcal{C}_1 \doteq \exp\!\Big[\!-\!\int_0^{\infty} \frac{\rd t}{t}\, \frac{\sinh^2(t/2) \sinh\!\big((1-\upxi) t\big)}{\sinh(2t) \cosh(t) \sinh(\upxi t)}\Big].
\label{ec252}
\end{equation}
\end{widetext}
and (Eq.~(2.15) in Ref.~\cite{LZ01})
\begin{equation}
\upxi \doteq \frac{{\bm\upbeta}^2}{1-{\bm\upbeta}^2} \equiv \frac{2 K}{1-2K}, \label{ec253}
\end{equation}
where we have used Eq.~(\ref{ec228}) above. The quantity $\upxi$ is not to be confused with the correlation length $\xi$. The coefficient $\mathcal{C}_1$ in Eq.~(\ref{ec251}) is a function of $\upxi$. Since $K \uparrow \frac{1}{2}$ for $\Delta \downarrow 1$ (Eq.~(\ref{ec97}) and the surrounding text, as well as Sec.~\ref{sac.4c}), it follows that
\begin{equation}
\upxi \uparrow +\infty \;\;\; \mbox{\rm for}\;\;\; \Delta \downarrow 1.
\label{ec254}
\end{equation}
Since our interest is restricted to the region where $\Delta \downarrow 1$, in what follows we shall employ the following simplifying result:
\begin{equation}
\lim_{\upxi\to \infty} \frac{\sinh\big((\upxi -1) t\big)}{\sinh(\upxi t)} = \e^{-t},\;\; t \ge 0. \label{ec255}
\end{equation}
For $t >0$ and $\upxi$ finite, the leading-order correction to the RHS of this expression is equal to $-\e^{-2\upxi t}$.

\begin{widetext}
LZ \cite{LZ01} have \emph{conjectured} an expression for $Z_{n}(s)$ which has been presented in Eq.~(3.1) of Ref.~\cite{LZ01}. For the specific case of $n=2$, one has (appendix \ref{sae})
\begin{equation}
\sqrt{Z_2(s)} = \frac{2}{\mathcal{C}_1} \sqrt{\frac{2}{\upxi}}\, \Big[\frac{\sqrt{\pi}\, \mathfrak{m}\, \Gamma\big(\frac{3}{2} + \frac{\upxi}{2}\big)}{\Gamma\big(\frac{\upxi}{2}\big)}\Big]^{d(s,2)}\!\! \exp\!\Big[\!\int_0^{\infty}\! \frac{\rd t}{t}\, \Big\{\frac{1}{4\sinh(\upxi t)} \Big(\frac{\cosh(\upxi t)\, \e^{-2 (\upxi+1) t} -1}{\sinh\big( (\upxi+1) t\big) \cosh(t)} + 2\Big) - d(s,2)\,\e^{-2 t}\Big\} \Big],\, s = \pm \frac{{\bm\upbeta}}{4}, \label{ec256}
\end{equation}
\end{widetext}
where (Eq.~(2.8) in Ref.~\cite{LZ01})
\begin{equation}
d(s,2) \doteq 2 s^2 + \frac{1}{2 {\bm\upbeta}^2} \equiv \frac{{\bm\upbeta}^2}{8} + \frac{1}{2 {\bm\upbeta}^2}\;\;\; \mbox{\rm for}\;\;\; s = \pm\frac{{\bm\upbeta}}{4}. \label{ec257}
\end{equation}
One has
\begin{equation}
d\big(\!\pm\frac{{\bm\upbeta}}{4},2\big) \sim \frac{5}{8}\;\;\; \mbox{\rm as}\;\;\; \upxi\to\infty. \label{ec258}
\end{equation}
From Eqs.~(\ref{ec256}) and (\ref{ec257}) one observes that \emph{for the case at hand, $Z_2(s)$ takes the same value for $s=+{\bm\upbeta}/4$ and $s=-{\bm\upbeta}/4$.} In appendix \ref{sae} we consider the behaviour of $\sqrt{Z_2(s)}$ and in particular show that for a constant value of the soliton mass $\mathfrak{m}$, this function decays like $1/\upxi^{3/16}$ as $\upxi\to\infty$.

As is evident from the expressions in Eqs.~(\ref{ec249}) and (\ref{ec250}), for our considerations we need to calculate $\vert\mathcal{G}(\theta)\vert$, and not $\mathcal{G}(\theta)$. Consequently, we can simplify the calculations to be performed by separating the pure phase factor that may contribute to $\mathcal{G}(\theta)$. To this end, we employ the identity
\begin{widetext}
\begin{equation}
\sinh^2\!\big((1 - i\theta/\pi) t\big) \equiv -\sin^2(\theta t/\pi)\, \cosh(2 t) + \sinh^2(t) - i \sin(\theta t/\pi) \cos(\theta t/\pi) \sinh(2 t). \label{ec259}
\end{equation}
One observes that for $\theta, t \in \mathbb{R}$, the last term on the RHS of this expression gives rise to a pure phase in the expression for $\mathcal{G}(\theta)$. Since the contribution to $\mathcal{G}(\theta)$ of the second term on the RHS of Eq.~(\ref{ec259}) is independent of $\theta$, we absorb the contribution to $\mathcal{G}(\theta)$ as arising from this term into $\mathcal{C}_1$ and thus write
\begin{eqnarray}
\mathcal{G}(\theta) &=& i\, \t{\mathcal{C}}_1\, \mathcal{U}(\theta) \sinh(\theta/2) \exp\!\Big[\!-\!\int_0^{\infty} \frac{\rd t}{t}\, \frac{\sin^2(\theta t/\pi) \sinh\big((\upxi-1) t\big)}{\tanh(2t) \cosh(t) \sinh(\upxi t)} \Big] \nonumber\\
&\sim& i\, \t{\mathcal{C}}_1\, \mathcal{U}(\theta) \sinh(\theta/2) \exp\!\Big[\!-\!\int_0^{\infty} \frac{\rd t}{t}\, \frac{\sin^2(\theta t/\pi)\, \e^{-t}}{\tanh(2t) \cosh(t)} \Big] \;\;\; \mbox{\rm for}\;\;\; \upxi \to\infty, \label{ec260}
\end{eqnarray}
where
\begin{equation}
\t{\mathcal{C}}_1 \doteq \mathcal{C}_1 \exp\!\Big[\int_0^{\infty}\! \frac{\rd t}{t}\, \frac{\sinh^2(t) \sinh\!\big((\upxi -1) t\big)}{\sinh(2t) \cosh(t) \sinh(\upxi t)}\Big] \sim\frac{\mathfrak{G}^6}{2^{1/6} \sqrt{\e \pi}}\, \mathcal{C}_1 \equiv 1.356\,133\,165\,\dots \times \mathcal{C}_1\;\;\; \mbox{\rm for}\;\;\; \upxi\to\infty,
\label{ec261}
\end{equation}
\begin{equation}
\mathcal{U}(\theta) \doteq \exp\!\Big[\! -i\!\!\int_0^{\infty}\!\frac{\rd t}{t}\, \frac{\sin(\theta t/\pi) \cos(\theta t/\pi) \sinh\!\big((\upxi-1) t\big)}{\cosh(t) \sinh(\upxi t)}\Big], \label{ec262}
\end{equation}
in which $\mathfrak{G} \doteq \exp(1/12 -\zeta'(-1)) \equiv 1.282\,427\,129\,\dots$ is the Glaisher number; here $\zeta'(z)$ is the derivative of the Riemann zeta function (Ch.~13 in Ref.~\cite{WW62}). Clearly, $\vert\mathcal{U}(\theta)\vert = 1$, $\forall \theta \in \mathds{R}$. For $\upxi \to\infty$ one has
\begin{equation}
\mathcal{C}_1 \sim \exp\!\Big[\!-\!\int_0^{\infty}\! \frac{\rd t}{t}\, \frac{\sinh^2(t/2) \,\e^{-t}}{\sinh(2t) \cosh(t)}\Big] = \frac{\e^{1/4}}{2^{1/6} \mathfrak{G}^3}\, \sqrt{\frac{\Gamma\big(\frac{1}{4}\big)}{\Gamma\big(\frac{3}{4}\big)}} \equiv 0.932\,937\,401\,\dots~. \label{ec263}
\end{equation}
Following Eq.~(\ref{ec261}) and (\ref{ec263}), one thus has $\t{\mathcal{C}}_1 \sim 1.265\,187\,350\,\dots$ for $\upxi \to\infty$.

Making use of the expressions in Eqs.~(\ref{ec249}) and (\ref{ec250}), one obtains that
\begin{eqnarray}
&&\hspace{-2.0cm}\langle\Psi_{N_{\rm e}}\vert \h{\mathcal{O}}_{s}^2(x,-i\ul\tau) \, \h{\mathcal{O}}_{s'}^{2\,\dag}(0,0) \vert\Psi_{N_{\rm e}} \rangle \approx \mathcal{A}_{{s},{s'}}\! \int_{-\infty}^{\infty}\! \rd \theta_1 \rd \theta_2\; \vert\mathcal{G}(\theta_1-\theta_2)\vert^2 \nonumber\\
&&\hspace{1.0cm} \times\exp\!\Big[\mathfrak{m} \Big( i \big[\sinh(\theta_1) + \sinh(\theta_2)\big] x - \big[\cosh(\theta_1) + \cosh(\theta_2)\big] \ul\tau \Big) + \frac{\varsigma+\varsigma'}{4} (\theta_1+\theta_2)\Big], \label{ec264}
\end{eqnarray}
where
\begin{equation}
\mathcal{A}_{{s},{s'}} \doteq \frac{\sqrt{Z_2({s}) Z_2({s'})}}{8 \pi^2}\, \exp\!\big[\frac{\pi i}{4} (\varsigma-\varsigma')\big] = \frac{Z_2(s)}{8\pi^2} \times \left\{\begin{array}{ll} +1, & s=s', \\ \\
\pm i, & s = - s' = \pm \frac{{\bm\upbeta}}{4}. \end{array} \right. \label{ec265}
\end{equation}
In writing the last expression, we have made explicit the fact that $\sqrt{Z_2(s)}$ takes the same values for $s=+{\bm\upbeta}/4$ and $s=-{\bm\upbeta}/4$ (see our remark subsequent to Eq.~(\ref{ec258}) above). In appendix \ref{sad} we present the closed expression for $\vert\mathcal{G}(\theta)\vert$ specific to large values of $\upxi$.

Introducing the auxiliary variable $\theta \doteq \theta_1 - \theta_2$, and using the transformation $\exp(\theta_1) \rightharpoonup y$, the expression in Eq.~(\ref{ec264}) can be expressed as (with $s = \varsigma\,{\bm\upbeta}/4$, $s' = \varsigma'\,{\bm\upbeta}/4$)
\begin{equation}
\langle\Psi_{N_{\rm e}}\vert \h{\mathcal{O}}_{s}^2(x,-i\ul\tau) \, \h{\mathcal{O}}_{s'}^{2\,\dag}(0,0) \vert\Psi_{N_{\rm e}} \rangle \approx \mathcal{A}_{{s},{s'}}\! \int_{-\infty}^{\infty}\! \rd \theta\; \vert\mathcal{G}(\theta)\vert^2\; \ul{\mathcal{I}}_{\varsigma,\varsigma'}^{(\mathfrak{m})}(\theta;x,-i\ul\tau), \label{ec266}
\end{equation}
where
\begin{eqnarray}
\ul{\mathcal{I}}_{\varsigma,\varsigma'}^{(\mathfrak{m})}(\theta;x,-i\ul\tau)\! &\doteq& \!\e^{ -(\varsigma+\varsigma') \theta/4}\!\! \int_0^{\infty}\!\!\! \rd y\, \exp\!\Big[ \mathfrak{m} \Big( i \big[ \frac{y^2-1}{2y} + \frac{y^2 \exp(-2\theta)-1}{2y \exp(-\theta)}\big] x - \big[\frac{y^2 +1}{2y} + \frac{y^2 \exp(-2\theta) +1}{2y \exp(-\theta)}\big]\ul\tau\Big) \Big]\, y^{\frac{1}{2}(\varsigma+\varsigma') -1}\nonumber\\
&=& \! 2\, \Big(\frac{\ul\tau + i x}{\ul\tau - ix}\Big)^{(\varsigma+\varsigma')/4}\, K_{(\varsigma+\varsigma')/2}\big(2 \mathfrak{m} \cosh(\theta/2) \sqrt{{\ul\tau}^2 + x^2}\big),\;\; \ul\tau \ge 0, \label{ec267}
\end{eqnarray}
\end{widetext}
in which $K_{\nu}(z)$ is the modified Bessel function of the second kind (Ch.~9 in Ref.~\cite{AS72}). With
\begin{equation}
\nu \doteq \frac{\varsigma+\varsigma'}{2}, \label{ec268}
\end{equation}
on account of $\varsigma, \varsigma' \in \{-1,+1\}$, one has
\begin{equation}
\nu \in \{-1,0,+1\}. \label{ec269}
\end{equation}
Since $K_{-\nu}(z) \equiv K_{\nu}(z)$ (item 9.6.2 in Ref.~\cite{AS72}), it follows that the difference between the $\ul{\mathcal{I}}_{\varsigma,\varsigma'}^{(\mathfrak{m})}(\theta;x,-i\ul\tau)$ corresponding to $\nu = -1$ and that corresponding to $\nu = +1$ is entirely due to the function pre-multiplying the Bessel function on the RHS of Eq.~(\ref{ec267}). From the last expression in Eq.~(\ref{ec267}) one trivially deduces that
\begin{equation}
\ul{\mathcal{I}}_{\varsigma,\varsigma'}^{(\mathfrak{m})}(\theta;x,-i\ul\tau) \equiv \ul{\mathcal{I}}_{\varsigma,\varsigma'}^{(u \mathfrak{m})}(\theta;x/u,-i\tau) \doteq \mathcal{I}_{\varsigma,\varsigma'}^{(u \mathfrak{m})}(\theta;x,-i\tau). \label{ec270}
\end{equation}

By introducing the function (see Eqs.~(\ref{e31}), (\ref{e33}), (\ref{ec94}) above as well as Eq.~(25.14) in Ref.~\cite{FW03})
\begin{widetext}
\begin{eqnarray}
\t{\ul{\mathcal{I}}}_{\varsigma,\varsigma'}^{(\mathfrak{m})}(\theta;k,i\ul\omega_n) \doteq \lim_{\substack{L\to\infty\\ \ul{\beta}\to \infty}} \int_{0}^{\ul{\beta}} \rd \ul\tau \int_{-L/2}^{L/2} \rd x\; \e^{i (\ul\omega_n \ul\tau - k x)} \, \ul{\mathcal{I}}_{\varsigma,\varsigma'}^{(\mathfrak{m})}(\theta;x,-i\ul\tau), \label{ec271}
\end{eqnarray}
from Eqs.~(\ref{ec236}) and (\ref{ec266}) one obtains that
\begin{equation}
\t{G}_{\h{\ul{\Sc a}},\h{\ul{\Sc b}}}^{+}(k;i\ul\omega_n) = -\frac{Z_2}{(2\pi)^3 a}\, \int_0^{\infty} \rd \theta\; \vert\mathcal{G}(\theta)\vert^2\;
\t{\ul{\mathcal{I}}}_{\varsigma,\varsigma'}^{(\mathfrak{m})}(\theta;k,i\ul\omega_n),
\;\;\varsigma,\varsigma' = \left\{ \begin{array}{ll} -1, & \h{\ul{A}}, \h{\ul{B}} = \h{\ul{R}}, \\ \\ +1, & \h{\ul{A}}, \h{\ul{B}} = \h{\ul{L}}, \end{array} \right.  \label{ec272}
\end{equation}
\end{widetext}
where $Z_2$ is the short-hand notation for $Z_2(s)$, which, as we have indicated following Eq.~(\ref{ec265}) above, takes the same value for $s=+{\bm\upbeta}/4$ and $s=-{\bm\upbeta}/4$. In arriving at Eq.~(\ref{ec272}), we have made use of the fact that $\vert\mathcal{G}(\theta)\vert$ and $\t{\ul{\mathcal{I}}}_{\varsigma,\varsigma'}^{(\mathfrak{m})}(\theta;k,i\ul\omega_n)$ are both even functions of $\theta$.

Since $\ul{\omega}_n \ul{\tau} \equiv \omega_n \tau$, Eqs.~(\ref{ec87}), (\ref{ec208}), from Eq.~(\ref{ec270}) one deduces that
\begin{equation}
\t{\mathcal{I}}_{\varsigma,\varsigma'}^{(u \mathfrak{m})}(\theta;k,i\omega_n) \equiv \frac{1}{u}\, \t{\ul{\mathcal{I}}}_{\varsigma,\varsigma'}^{(\mathfrak{m})}(\theta;k,i\ul\omega_n),
\label{ec273}
\end{equation}
from which and the expressions in Eq.~(\ref{ec239}) one obtains the equality in Eq.~(\ref{ec235}). The result in Eq.~(\ref{ec273}) is the manifestation of the Lorentz invariance of the continuum theory under consideration. Although this invariance breaks down in the CDW GS, where $\mathfrak{m}\not=0$, we have not taken explicit account of this breakdown by the consideration that, for $V > V_{\rm c}$ and $V$ sufficiently close
$V_{\rm c}$, $\mathfrak{m}$ is exponentially small, a fact that we have emphasized in our remarks preceding Eqs.~(\ref{e30}) and (\ref{e31}). As regards the relationships between \[\t{\mathcal{I}}_{\varsigma,\varsigma'}^{(u \mathfrak{m})}(\theta;k,i\omega_n) \;\;\; \mbox{\rm and}\;\;\; \t{\ul{\mathcal{I}}}_{\varsigma,\varsigma'}^{(\mathfrak{m})}(\theta;k,i\ul\omega_n)\] with respectively \[\t{G}_{\h{\Sc a},\h{\Sc b}}(k;i\omega_n) \;\;\; \mbox{\rm and}\;\;\; \t{G}_{\h{\ul{\Sc a}},\h{\ul{\Sc b}}}(k;i\ul\omega_n),\] we refer the reader to the remarks in the paragraph following Eq.~(\ref{ec235}), on page~\pageref{LorentzI}, which underlie the defining expression in Eq.~(\ref{ec270}).

We evaluate the double-integral in Eq.~(\ref{ec271}) by employing the circular coordinates $(\varrho,\varphi)$:
\begin{equation}
\ul\tau = \varrho\, \cos(\varphi),\;\;\; x = \varrho\, \sin(\varphi), \;\;\; \varphi \in [-\frac{\pi}{2},\frac{\pi}{2}], \label{ec274}
\end{equation}
whereby from Eq.~(\ref{ec271}) one obtains that
\begin{equation}
\t{\ul{\mathcal{I}}}_{\varsigma,\varsigma'}^{(\mathfrak{m})}(\theta;k,i\ul\omega_n) =\!\left. \frac{-4}{\chi^2} \frac{\partial}{\partial\lambda}\! \int_{-\pi/2}^{\pi/2}\!\!\! \rd \varphi\, \e^{i\nu\varphi}\! f_{\nu}^{(\lambda)}(\theta;k,i\ul\omega_n;\varphi)\right|_{\lambda=1}\!\!\!, \label{ec275}
\end{equation}
where
\begin{equation}
f_0^{(\lambda)}(\theta;k,i\ul\omega_n;\varphi) \doteq \displaystyle\frac{\sqrt{1 + (\frac{\alpha}{\chi})^2} + \frac{\alpha}{\chi} \big(\frac{i\pi}{2} - \sinh^{-1}(\frac{\alpha}{\chi})\big)}{\sqrt{\lambda + (\frac{\alpha}{\chi})^2}}, \label{ec276}
\end{equation}
\begin{equation}
f_{\pm 1}^{(\lambda)}(\theta;k,i\ul\omega_n;\varphi) \doteq \displaystyle \frac{\frac{i\alpha}{\chi} \sqrt{1 + (\frac{\alpha}{\chi})^2} + \frac{\pi}{2} + i\sinh^{-1}(\frac{\alpha}{\chi})}{\sqrt{\lambda + (\frac{\alpha}{\chi})^2}}, \label{ec277}
\end{equation}
\begin{equation}
\alpha \equiv \alpha(i\ul\omega_n,k;\varphi) \doteq \ul\omega_n \cos(\varphi) - k \sin(\varphi), \label{ec278}
\end{equation}
\begin{equation}
\chi \equiv \chi(\theta) \doteq 2\mathfrak{m}\, \cosh(\theta/2). \label{ec279}
\end{equation}
In Eqs.~(\ref{ec276}) and (\ref{ec277}), $\sinh^{-1}(z)$ stands for the inverse of $\sinh(z)$, for which one has (item 4.6.20 in Ref.~\cite{AS72})
\begin{equation}
\sinh^{-1}(z) = \ln(z + \sqrt{1+z^2}). \label{ec280}
\end{equation}
The property $f_{-1}^{(\lambda)} \equiv f_{+1}^{(\lambda)}$ is a consequence of $K_{-\nu}(z) \equiv K_{+\nu}(z)$, to which we referred following Eq.~(\ref{ec269}) above.

From Eqs.~(\ref{ec275}), (\ref{ec276}) and (\ref{ec277}) one trivially deduces that for $\nu \in \{-1,0,+1\}$, and any finite value of $\vert i\ul\omega_n\vert$, one has
\begin{equation}
\t{\ul{\mathcal{I}}}_{\varsigma,\varsigma'}^{(\mathfrak{m})}(\theta;k,i\ul\omega_n) \sim \frac{2\pi}{\chi^2} \sim \frac{2\pi}{\mathfrak{m}^2}\, \e^{-\vert\theta\vert}\;\; \mbox{\rm as}\;\;\ \vert\theta\vert\to\infty. \label{ec281}
\end{equation}
In view of the leading-order asymptotic behaviour of $\vert \mathcal{G}(\theta)\vert$ corresponding to $\vert\theta\vert \to\infty$ (see Eq.~(\ref{ed19}) below), one infers that for $\vert\theta\vert \to\infty$ the integrand of the integral on the RHS of Eq.~(\ref{ec266}) decays like $\vert\theta\vert^{1/2}\, \e^{-\vert\theta\vert/2}$.

The order of differentiation with respect to $\lambda$ and integration with respect to $\varphi$ in Eq.~(\ref{ec275}) is significant; exchanging this order, the integral with respect to $\varphi$ may not exist. In our calculations, we have evaluated both the integral with respect to $\varphi$ and the derivative with respect to $\lambda$ in Eq.~(\ref{ec275}) numerically. For the latter, we have employed the central finite difference method as presented in item 25.3.21 of Ref.~\cite{AS72} (using $2h \le 10^{-4}$, for which reason we have evaluated the integral with respect to $\varphi$ to a relative accuracy better than $(2h)^2$). In Fig.~\ref{f2} we present some numerical results of our calculations based on the details presented in this section.

In closing this section, we remark that through the variable transformation $\varphi \rightharpoonup -\varphi$ in Eq.~(\ref{ec275}), one immediately deduces that, with $\b\varsigma \doteq -\varsigma$,
\begin{eqnarray}
\t{\ul{\mathcal{I}}}_{\varsigma,\varsigma}^{(\mathfrak{m})}(\theta;k,i\ul\omega_n) &\equiv& \t{\ul{\mathcal{I}}}_{\b\varsigma,\b\varsigma}^{(\mathfrak{m})}(\theta;-k,i\ul\omega_n), \nonumber\\
\t{\ul{\mathcal{I}}}_{\varsigma,\b\varsigma}^{(\mathfrak{m})}(\theta;k,i\ul\omega_n) &\equiv& \t{\ul{\mathcal{I}}}_{\b\varsigma,\varsigma}^{(\mathfrak{m})}(\theta;-k,i\ul\omega_n), \nonumber\\
\t{\ul{\mathcal{I}}}_{\varsigma,\b\varsigma}^{(\mathfrak{m})}(\theta;k,i\ul\omega_n) &\equiv& \t{\ul{\mathcal{I}}}_{\varsigma,\b\varsigma}^{(\mathfrak{m})}(\theta;-k,i\ul\omega_n),
\label{ec282}
\end{eqnarray}
which, following Eq.~(\ref{ec272}), lead to the symmetry relations presented in Eq.~(\ref{ec238}).

\subsubsection{Some general remarks}
\label{sac.6c}

For $\vert z\vert \to\infty$ and $\vert \arg z\vert < \frac{3}{2} \pi$, one has the asymptotic expression (item 9.7.2 in Ref.~\cite{AS72})
\begin{equation}
K_{\nu}(z) \sim \sqrt{\frac{\pi}{2 z}}\, \e^{-z} \Big\{ 1 + \frac{\gamma - 1}{8 z} + \frac{(\gamma -1) (\gamma-9)}{2! (8 z)^2} + \dots \Big\}, \label{ec283}
\end{equation}
where $\gamma \doteq 4\nu^2$. This asymptotic expression reveals that the infinite series in powers of $1/(8 z)$, enclosed by curly brackets, is terminating for \[\gamma=1 \iff \nu = \pm \frac{1}{2} \] (in which case it only consists of the single term $1$), \[\gamma = 9 \iff \nu = \pm \frac{3}{2}\] (in which case it consists of two terms, $1$ and $(\gamma-1)/(8 z)$), etc. In our case, where $\nu$ takes the tree values indicated in Eq.~(\ref{ec269}), the function $K_{(\varsigma+\varsigma')/2}(z) \equiv K_{\nu}(z)$ (as encountered on the RHS of Eq.~(\ref{ec267})) is \emph{not} expressible as a product of $z^{-1/2} \e^{-z}$ and a finite series in powers of $1/z$.

If one were to evaluate $\t{G}_{\h{\ul{\Sc a}},\h{\ul{\Sc b}}}(k;z)$ for small values of $\vert k\vert$ and $\vert z\vert$, it would be \emph{reasonable} to approximate $K_{\nu}(z)$ by the leading-order contribution on the RHS of Eq.~(\ref{ec283}), an approximation whose validity is conditioned on
\begin{equation}
2\mathfrak{m} \cosh(\theta/2) \sqrt{\ul\tau^2 + x^2} \gg a.
\label{ec284}
\end{equation}
Employing this approximation, and using the \emph{identity} (which underlies the expression 2.580.1 in Ref.~\cite{GR07}):
\begin{equation}
a + b \cos(\varphi) + c \sin(\varphi) = (a+p) - 2 p \sin^2(\psi), \label{ec285}
\end{equation}
where
\begin{equation}
\varphi = 2\psi + \vartheta,\;\;\; \tan(\vartheta) = \frac{c}{b},\;\;\; p = b \sec(\vartheta), \label{ec286}
\end{equation}
one can express $\t{\ul{\mathcal{I}}}_{\varsigma,\varsigma'}^{(\mathfrak{m})}(\theta;k,i\ul\omega_n)$ fully analytically in terms of the elliptic integrals of the first and second kind, $F(\psi,\kappa)$ and $E(\psi,\kappa)$ respectively (items 8.111.2 and 8.111.3 in Ref.~\cite{GR07}). We note in passing that although the real part of $\vartheta = \tan^{-1}(c/b)$ is determined up to the constant $n \pi$, where $n=0,\pm 1, \dots$, this ambiguity in the value of $\vartheta$ is of \emph{no} consequence to the validity of the identity in Eq.~(\ref{ec285}). We do not present this closed expression for $\t{\ul{\mathcal{I}}}_{\varsigma,\varsigma'}^{(\mathfrak{m})}(\theta;k,i\ul\omega_n)$ here essentially for the fact that evaluation of elliptic integrals for complex arguments and moduli is not computationally less time-consuming than evaluation of the integral with respect to $\varphi$ in Eq.~(\ref{ec275}) (additionally, the full closed expression for $\t{\ul{\mathcal{I}}}_{\varsigma,\varsigma'}^{(\mathfrak{m})}(\theta;k,i\ul\omega_n)$ is both very extensive and involves some detailed prescriptions for singling out the appropriate branches of the many-valued functions $F(\psi,\kappa)$ and $E(\psi,\kappa)$).

It should be noted that since $\mathfrak{m}$ diminishes exponentially for $V \downarrow V_{\rm c}(t,t')$, Eqs.~(\ref{e22}) and (\ref{e35}), the validity of the condition in Eq.~(\ref{ec284}) for all $\theta$ implies that as $V \downarrow V_{\rm c}(t,t')$, the approximation for $\t{G}_{\h{\ul{\Sc a}},\h{\ul{\Sc b}}}(k;z)$ based on the leading-order term in Eq.~(\ref{ec283}) is accurate in principle only for exponentially small neighbourhoods of $k=0$ and $z=0$.

\subsubsection{The leading-order term in the asymptotic series expansion of $G(k;z)$ for $\vert z\vert \to\infty$}
\label{sac.6d}

The \emph{exact} leading-order term in the asymptotic series expansion of the \emph{exact} $G(k;z)$ is $1/z$ (in the units where $\hbar = 1$), where the coefficient $1$ arises as a consequence of the following three properties (Sec.~B.3 in Ref.~\cite{BF07a}): (i) the completeness of the intermediate states used in the calculation of $G(k;z)$ (compare with the decomposition of the identity operator $\h{I}$ in Eq.~(\ref{ec246})), (ii) the anti-commutation relation $[\h{a}_k, \h{a}_k^{\dag}]_- = 1$ for canonical fermion operators, Eq.~(\ref{ec4}), and (iii) the normalisation $\langle\Psi_{N_{\rm e}}\vert \Psi_{N_{\rm e}}\rangle = 1$. Equivalently, with
\begin{equation}
A(k;\omega) \doteq \mp\frac{1}{\pi} \im[G(k;\omega \pm i 0^+)], \;\;\; \omega \in \mathds{R}, \label{ec287}
\end{equation}
the single-particle spectral function, the $1$ in the above-indicated $1/z$ is exactly the same constant as the $1$ on the RHS of the following exact sum rule:
\begin{equation}
\int_{-\infty}^{\infty} \rd \omega\; A(k;\omega) = 1, \;\; \forall k \in \mathrm{1BZ}. \label{ec288}
\end{equation}
It follows that within the framework of our formalism, where we have replaced the sum with respect of $n$ on the RHS of Eq.~(\ref{ec247}) by the underlying summand corresponding to $n=2$, the coefficient of the leading-order asymptotic term of $G(k;z)$ corresponding to $\vert z\vert \to\infty$ is less than unity. To appreciate this fact, since $A(k;\omega) \ge 0$ for all $\omega$ (a consequence of the assumed stability of the GS of the system under consideration), restriction of the last-mentioned \emph{infinite} sum to its summand corresponding to $n=2$, amounts to the imposition of a specific (soft) cut-off on $A(k;\omega)$, which in general leads to the integral in Eq.~(\ref{ec288}) taking a value less than unity.

Due to the complexity of the full analysis of the asymptotic behaviour of the Green function $G(k;z)$, for $\vert z\vert \to \infty$, as calculated here, Eq.~(\ref{e33}), below we only present an outline of the full analysis, which we do not claim to be rigorous. The aim of this investigation is to show that the Green function $G(k;z)$ as calculated in this paper indeed decays to leading order like $C/z$ for $\vert z\vert \to\infty$, where, following our above considerations, $C$ is a real constant satisfying $0 < C < 1$ (the crucial property $C>0$ should be noted). To this end, below we consider the leading-order term in the asymptotic series expansion of $\t{\ul{\mathcal{I}}}_{\varsigma,\varsigma'}^{(\mathfrak{m})}(\theta;k,i\ul\omega_n)$ corresponding to $\vert i\ul\omega_n\vert \to \infty$. In this connection, note that since $1/z$ is analytic everywhere in the complex $z$ plane outside $z=0$, the direction along which we let $z$ approach the point of infinity of the $z$ plane is of no significance to our present leading-order calculations.

For $\varphi \not= \mp\pi/2$ one readily verifies that $\partial f_{\nu}^{(\lambda)}/\partial\lambda$ decays to leading-order like $1/(i\ul\omega_n)^2$ as $\vert i\ul\omega_n\vert \to \infty$. It follows that the leading-order term of the asymptotic series expansion of $\t{\ul{\mathcal{I}}}_{\varsigma,\varsigma'}^{(\mathfrak{m})}$ for $\vert i\ul\omega_n\vert \to \infty$ originates from the contributions to the integral with respect to $\varphi$ in Eq.~(\ref{ec275}) as arising from the small neighbourhoods of $\mp\pi/2$ inside $[-\pi/2,\pi/2]$. This is directly plausible, since, following the first expression in Eq.~(\ref{ec274}), $\varphi \to \mp\pi/2$ corresponds to $\ul\tau\to 0$ for any finite value of $\varrho$. Expanding $\cos(\varphi)$ and $\sin(\varphi)$ around $\varphi = \mp\pi/2$, from the expression in Eq.~(\ref{ec278}) one deduces that
\begin{equation}
\frac{\alpha}{\chi} \sim \frac{\ul\omega_n}{\chi} (\pm\varphi +\frac{\pi}{2}) \pm \frac{k}{\chi}\;\;\; \mbox{\rm for}\;\;\; \varphi \to \mp\frac{\pi}{2}, \label{ec289}
\end{equation}
from which one infers the following leading-order asymptotic expression for $\vert i\ul\omega_n\vert \to \infty$:
\begin{widetext}
\begin{equation}
\t{\ul{\mathcal{I}}}_{\varsigma,\varsigma'}^{(\mathfrak{m})}(\theta;k,i\ul\omega_n) \sim \frac{-4}{\chi^2} \frac{\partial}{\partial\lambda}\, \Big[C_1 \int_{-\pi/2}^{-\pi/2 + \chi/\vert i\ul\omega_n\vert} + C_2 \int_{\pi/2 - \chi/\vert i\ul\omega_n\vert}^{\pi/2} \Big]\, \rd \varphi\; \e^{i\nu\varphi}\, f_{\nu}^{(\lambda)}(\theta;k,i\ul\omega_n;\varphi), \label{ec290}
\end{equation}
\end{widetext}
where $C_1$ and $C_2$ are constants of the order of unity. Since $\vert f_{\nu}^{(\lambda)} \vert$ is bounded and non-vanishing for $\varphi$ over the intervals of integration indicated, and since the lengths of these intervals is equal to $\chi/\vert i\ul\omega_n\vert$, one immediately observes that indeed the RHS of Eq.~(\ref{ec290}) decays like $\chi/(i\ul\omega_n)$ for $\vert i\ul\omega_n\vert \to \infty$. This in turn implies that $\t{\ul{\mathcal{I}}}_{\varsigma,\varsigma'}^{(\mathfrak{m})}$ to leading order decays like $1/(i\ul\omega_n\chi)$ for $\vert i\ul\omega_n\vert \to \infty$. We have verified the correctness of this result numerically.

With reference to Eq.~(\ref{ec281}), one observes that the processes of effecting the two limits $\vert\theta\vert \to \infty$ and $\vert i\ul\omega_n\vert \to \infty$ do \emph{not} commute: our considerations in the previous paragraph have revealed that for any finite value of $\theta$ (and therefore of $\chi$), the $1/\chi^2$ in Eq.~(\ref{ec281}) is replaced by $1/(i\ul\omega_n\chi)$ as $\vert i\ul\omega_n\vert$ approaches $\infty$. This ambiguity is bypassed through the observation that for any finite $\vert i\ul\omega_n\vert$, the integrand of the integral in Eq.~(\ref{ec272}) decays like $\vert\theta\vert^{1/2} \e^{-\vert\theta\vert/2}$ for $\vert\theta\vert\to \infty$, Eqs.~(\ref{ec281}) and (\ref{ed19}), so that to exponential accuracy the upper bound of the integral on the RHS of Eq.~(\ref{ec272}) can be replaced by a \emph{finite} (large) constant $\theta_0$ (in this connection, one should note that, following Eq.~(\ref{ec279}), $\chi$ increases exponentially with $\theta$ as $\vert\theta\vert\to \infty$). With reference the conclusion arrived at in the previous paragraph, on replacing the upper bound of the integral on the RHS of Eq.~(\ref{ec272}) by a \emph{finite} constant, the conclusion that for $\vert i\ul\omega_n\vert \to\infty$ the corresponding $\t{G}_{\h{\ul{\Sc a}},\h{\ul{\Sc b}}}^+(k;i\ul\omega_n)$ decays like $1/(i\ul\omega_n)$ is immediate. With reference to Eq.~(\ref{ec239}), we thus conclude that for $\vert z\vert\to \infty$ the corresponding $G(k;z)$, Eq.~(\ref{e33}), indeed to leading order decays like $C/z$, where $C >0$. Our numerically-calculated $G(k;\omega \pm i 0^+)$ (see Fig.~\ref{f2}) conform with this observation; explicitly, our numerical calculations show that for $\omega_1 \le \vert\omega\vert \le \omega_2$ this function is accurately described by a function decaying like $1/\omega^{\upalpha}$, where $\upalpha$ increases monotonically towards $1$ for $\omega_1 \ll \omega_2$ as $\omega_1,\omega_2 \to \infty$.

\section{The closed form of $\vert\mathcal{G}(\theta)\vert$ for $\upxi \to \infty$}
\label{sad}

The leading-order expression for $\mathcal{G}(\theta)$ in the asymptotic region corresponding to $\upxi \to\infty$ is presented in Eq.~(\ref{ec260}). Here we determine the closed form of the absolute value of this expression. To this end, we define
\begin{equation}
\mathcal{J}(\theta) \doteq \int_0^{\infty} \frac{\rd t}{t}\, \frac{\sin^2(\theta t/\pi)\, \e^{-t}}{\tanh(2t) \cosh(t)}.
\label{ed1}
\end{equation}

Making use of the expressions for $\tanh(2 t)$ and $\cosh(t)$ in terms of $y \doteq \e^{2t}$ and $y^{-1}$, followed by a partial-fraction expansion of the resulting expression, one deduces the following identities:
\begin{widetext}
\begin{eqnarray}
\frac{\e^{-t}}{\tanh(2t) \cosh(t)}
&\equiv& \frac{1}{2}\, \frac{1}{y-1} - \frac{1}{2}\,\frac{1}{y+1} +\frac{2 y}{(y+1)^2} +\frac{1}{2} \Big( \frac{y}{1-y^{-1}} - \frac{y}{1+ y^{-1}} -2\Big) \nonumber\\
&\equiv& \frac{\e^{-t}}{4 \sinh(t)} - \frac{\e^{-t}}{4 \cosh(t)} + \frac{1}{1+\cosh(2 t)} + \frac{\e^{-2 t}}{2 \sinh(2 t)}.\label{ed2}
\end{eqnarray}
\end{widetext}
One has
\begin{equation}
\int_0^{\infty} \frac{\rd t}{t}\, \frac{\sin^2(\theta t/\pi)\, \e^{-t}}{\sinh(t)} = \ln\!\Big(\sqrt{\frac{\sinh(\theta)}{\theta}}\Big), \label{ed3}
\end{equation}
where we have used the analytic continuation of expression 3.553.1 in Ref.~\cite{GR07} for $a <1$ to $a$ along the imaginary axis (explicitly, to $a=i\theta/\pi$, where $\theta \in \mathds{R}$). From Eq.~(\ref{ed3}) one immediately obtains that
\begin{equation}
\int_0^{\infty} \frac{\rd t}{t}\, \frac{\sin^2(\theta t/\pi)\, \e^{-2 t}}{\sinh(2 t)} = \ln\!\Big(\sqrt{\frac{\sinh(\theta/2)}{\theta/2}}\Big). \label{ed4}
\end{equation}
Further,
\begin{equation}
\int_0^{\infty} \frac{\rd t}{t}\, \frac{\sin^2(\theta t/\pi)\, \e^{-t}}{\cosh(t)} = \ln\!\Big(\sqrt{\frac{\theta}{2} \coth\big(\frac{\theta}{2} \big)}\Big), \label{ed5}
\end{equation}
where we have used the analytic continuation of expression 3.555.2 in Ref.~\cite{GR07} for $a <\frac{1}{2}$ to $a$ along the imaginary axis.

In order to obtain the closed expression for the integral corresponding to the third term on the RHS of Eq.~(\ref{ed2}), we introduce the following auxiliary integral
(this would not have been necessary, had the expression 3.557.1 in Ref.~\cite{GR07} been correct):
\begin{equation}
\Upsilon(p) \doteq \int_0^{\infty} \frac{\rd t}{t}\, \frac{\e^{-p t} -1}{1+ \cosh(t)}.  \label{ed6}
\end{equation}
One has
\begin{equation}
\int_0^{\infty} \frac{\rd t}{t}\, \frac{\sin^2(\theta t/\pi)}{1+\cosh(2 t)} = -\frac{1}{4} \Big(\Upsilon\big(\frac{i \theta}{\pi}\big) + \Upsilon\big(- \frac{i \theta}{\pi}\big)\Big). \label{ed7}
\end{equation}

For determining the closed expression for $\Upsilon(p)$, it proves convenient to consider $\partial \Upsilon(p)/\partial p$ and subsequently obtain $\Upsilon(p)$ through integrating the latter function. One obtains (see expression 3.541.8 in Ref.~\cite{GR07})
\begin{equation}
\frac{\partial}{\partial p} \Upsilon(p) = -\int_0^{\infty} \rd t\; \frac{\e^{- 2 p t}}{\cosh^2(t)} = 1 -2 p\, \upbeta(p), \label{ed8}
\end{equation}
where (Sec.~8.37 in Ref.~\cite{GR07})
\begin{equation}
\upbeta(z) \doteq \frac{1}{2} \Big(\psi\big(\frac{z+1}{2}\big) - \psi\big(\frac{z}{2}\big) \Big), \label{ed9}
\end{equation}
in which $\psi(z)$ is the \emph{digamma} function (Sec.~8.36 in Ref.~\cite{GR07} and Sec.~6.3 in Ref.~\cite{AS72})
\begin{equation}
\psi(z) \doteq \frac{\rd}{\rd z} \ln\big( \Gamma(z)\big). \label{ed10}
\end{equation}
In the following we shall encounter the \emph{polygamma} function $\psi^{(m)}(z)$ defined according to (Sec.~10.2 in Ref.~\cite{AW01})
\begin{equation}
\psi^{(m)}(z) \doteq \frac{{\rm d}^{m}}{\rd z^{m}} \psi(x). \label{ed11}
\end{equation}
Thus $\psi(z) \equiv \psi^{(0)}(z)$. One has
\begin{equation}
\psi^{(m-1)}(z) \doteq \int_0^z \rd z'\; \psi^{(m)}(z'). \label{ed12}
\end{equation}
For completeness, we point out that the digamma and polygamma functions in Ref.~\cite{AW01} are defined in terms of \[z! \doteq \Gamma(z+1) \equiv z \Gamma(z),\] whereby the polygamma function $F_m(z)$ in Ref.~\cite{AW01} is identical to $\psi^{(m)}(z+1)$. Further, although $\psi^{(m)}(z)$ is defined for integer values of $m$, this function can be analytically continued into the complex $\nu$ plane whereby one can deal with $\psi^{(\nu)}(z)$, where $\nu\in \mathds{C}$ \cite{SM04}.

Since, following Eq.~(\ref{ed6}), $\Upsilon(0)=0$, on integrating both sides of Eq.~(\ref{ed8}) with respect to $p$ over $[0,p]$, one obtains that
\begin{widetext}
\begin{equation}
\Upsilon(p) = p -2 p \Big(\! \ln\big(\Gamma\big(\frac{p+1}{2}\big)\big) - \ln\big(\Gamma\big(\frac{p}{2}\big)\big) \Big) + 4 \Big(\psi^{(-2)}\big(\frac{p+1}{2}\big) - \psi^{(-2)}\big(\frac{p}{2}\big)\Big) - \ln\big(2^{5/6} \pi \mathfrak{G}^6\big), \label{ed13}
\end{equation}
where $\mathfrak{G} \doteq \e^{1/12 -\zeta'(-1)} \equiv 1.282\,427\,129\,\dots$ is the Glaisher number. In arriving at the expression in Eq.~(\ref{ed13}), we have determined the integral $\int_0^p \rd p'\, p' \upbeta(p')$ through applying integration by parts, making use of the defining expressions in Eqs.~(\ref{ed9}) and (\ref{ed10}), and of the exact result
\begin{equation}
\int_0^p \rd p'\; \Big\{\ln\!\big(\Gamma\big(\frac{p'+1}{2}\big)\big) - \ln\!\big(\Gamma\big(\frac{p'}{2}\big)\big)\Big\} = 2 \Big(\psi^{(-2)}\big(\frac{p+1}{2}\big) - \psi^{(-2)}\big(\frac{p}{2}\big)\Big) -\frac{1}{2} \ln\big( 2^{5/6} \pi \mathfrak{G}^6\big). \label{ed14}
\end{equation}

Collecting the closed forms of the relevant integrals of the terms on the RHS of Eq.~(\ref{ed2}), one arrives at
\begin{equation}
\mathcal{J}(\theta) = \frac{1}{8} \ln\!\Big(\frac{8 \sinh(\theta) \sinh^3(\theta/2)}{\theta^4 \cosh(\theta/2)}\Big)-\frac{1}{4} \Big(\Upsilon\big(\frac{i \theta}{\pi}\big) + \Upsilon\big(- \frac{i \theta}{\pi}\big)\Big). \label{ed15}
\end{equation}
One thus has
\begin{equation}
\mathcal{G}(\theta) \sim i\,\t{\mathcal{C}}_1\, \mathcal{U}(\theta) \sinh(\theta/2)\, \e^{-\mathcal{J}(\theta)} = i\,\t{\mathcal{C}}_1\, \mathcal{U}(\theta) \Big(\frac{\theta^4 \sinh^5(\theta/2)\cosh(\theta/2)}{8 \sinh(\theta)}\Big)^{1/8}\! \exp\!\Big[\frac{1}{4} \Big(\Upsilon\big(\frac{i \theta}{\pi}\big) + \Upsilon\big(- \frac{i \theta}{\pi}\big)\Big) \Big]\;\;\; \mbox{\rm for}\;\;\; \upxi \to\infty. \label{ed16}
\end{equation}

Since
\begin{equation}
\frac{1}{4} \Big(\Upsilon\big(\frac{i \theta}{\pi}\big) + \Upsilon\big(- \frac{i \theta}{\pi}\big)\Big) \sim \left\{ \begin{array}{ll} \displaystyle -\frac{\ln(2)}{2\pi^2}\, \theta^2, & \theta \to 0,\\ \\
\displaystyle -\frac{1}{8} \ln(\theta^2) + \frac{1 + \ln\big(\frac{2^{4/3} \pi}{\mathfrak{G}^{12}}\big)}{4}, & \theta \to \pm\infty, \end{array}\right. \label{ed17}
\end{equation}
with
\begin{equation}
\Big(\frac{\theta^4 \sinh^5(\theta/2)\cosh(\theta/2)}{8 \sinh(\theta)}\Big)^{1/8} \sim \left\{ \begin{array}{ll}\displaystyle \frac{1}{2}\, \vert\theta\vert, & \theta \to 0, \\ \\
\displaystyle \frac{1}{2}\,\vert\theta\vert^{1/2}\, \e^{\vert\theta\vert/4}, & \theta \to \pm\infty, \end{array} \right. \label{ed18}
\end{equation}
\end{widetext}
it follows that
\begin{equation}
\vert\mathcal{G}(\theta)\vert \sim \left\{ \begin{array}{ll} \displaystyle \frac{1}{2} \vert\t{\mathcal{C}}_1\vert\, \vert\theta\vert, & \theta \to 0,\\ \\
\displaystyle \frac{(\e \pi)^{1/4}}{2^{2/3} \mathfrak{G}^3} \vert\t{\mathcal{C}}_1\vert\, \vert\theta\vert^{1/4}\, \e^{\vert\theta\vert/4}, & \theta \to \pm\infty, \end{array}\right. \label{ed19}
\end{equation}
where, following Eq.~(\ref{ec261}) and (\ref{ec263}), one has
\begin{equation}
\t{\mathcal{C}}_1 \sim 1.265\,187\,350\,\dots \;\;\; \mbox{\rm for}\;\;\; \upxi \to\infty. \label{ed20}
\end{equation}
Recall that $\vert\mathcal{U}(\theta)\vert = 1$ for $\theta \in \mathds{R}$. The cusp in $\vert\mathcal{G}(\theta)\vert$ at $\theta=0$ is entirely due to the function in Eq.~(\ref{ed18}), so that it is also present in $\mathcal{G}(\theta)$ at $\theta=0$. Explicit calculation shows that for $\vert\theta\vert \lesssim 2$ ($\vert\theta\vert \gtrsim 2$)the upper (lower) leading-order expression in Eq.~(\ref{ed19}) relatively accurately describes the behaviour of the exact $\vert\mathcal{G}(\theta)\vert$.

For completeness, we point out that in our numerical calculations we have employed the following approximation for $\psi^{(-2)}(z)$:
\begin{equation}
\psi^{(-2)}(z) \approx \vartheta_{\eta}(1-\vert z\vert)\, f_1(z) + \vartheta_{\eta}(\vert z\vert -1)\, f_2(z), \label{ed21}
\end{equation}
where
\begin{equation}
\vartheta_{\eta}(x) \doteq \frac{1}{\exp(-x/\eta) + 1}, \label{ed22}
\end{equation}
and $f_1(z)$ and $f_2(z)$ are the asymptotic series expansions of $\psi^{(-2)}(z)$ corresponding to $z\to 0$ and $z \to \infty$ respectively. The Fermi function $\vartheta_{\eta}(x)$, in which $0 < \eta \ll 1$, serves to prevent discontinuity in the function on the RHS of Eq.~(\ref{ed21}) on the circle $\vert z\vert =1$ (in practice, we have used $\eta = 10^{-3}$). For our calculations, we have equated $f_1(z)$ and $f_2(z)$ with the relevant \emph{fifteenth}-order asymptotic series expansions of $\psi^{(-2)}(z)$, which we do not reproduce here; these can be readily obtained with the aid of \emph{Mathematica}. We note that both series expansions involve $\ln(z)$ so that they are not Taylor expansions around $z=0$ and $z=\infty$. We further note that the approximation in Eq.~(\ref{ed21}) is suited for our present calculations (where $\theta \in \mathds{R}$) and should be used, if at all, with care in other applications.

\section{On the behaviour of $\sqrt{Z_n(s)}$}
\label{sae}

\begin{figure}[tb!]
\includegraphics[angle=0, width=0.43\textwidth]{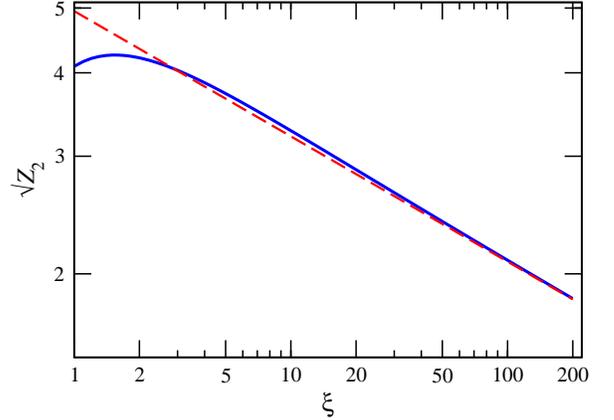}
\caption{(Colour online) The log-log plot of $\sqrt{Z_2(s)}$ for $s= \pm {\bm\upbeta}/4$ and $\mathfrak{m} =1$. \emph{Solid line} represents the numerically calculated $\sqrt{Z_2(s)}$ and the \emph{broken line} the expression $A/\upxi^{3/16}$, where $A=4.947\,289\,\dots$ (we have deduced both this value and the power $3/16$ from our numerically-calculated $\sqrt{Z_2(s)}$; numerically, we reproduce the exponent $3/16$ to an accuracy of $1$ part in $10^{7}$). For arbitrary $\mathfrak{m}$, the corresponding $\sqrt{Z_2(s)}$ is obtained by multiplying the data presented here by $\mathfrak{m}^{d(s,2)}$, where $d(s,2) = {\bm\upbeta}^2/8 + 1/(2 {\bm\upbeta}^2)$ for $s= \pm {\bm\upbeta}/4$, Eq.~(\protect\ref{ec257}), in which ${\bm\upbeta}^2 = 2 K = \upxi/(\upxi+1)$. For $\upxi\to\infty$ one has $d(\pm {\bm\upbeta}/4,2) \sim 5/8$, Eq.~(\protect\ref{ec258}). }\label{f3}
\end{figure}

As we indicated in appendix \ref{sac}, in Ref.~\cite{LZ01} LZ have \emph{conjectured} an expression for $\sqrt{Z_n(s)}$ (see Eq.~(3.1) in Ref.~\cite{LZ01}). We have reproduced this expression in Eq.~(\ref{ec256}) for the specific case of $n=2$ and $s = \pm {\bm\upbeta}/4$. In Fig.~\ref{f3} we display this function for the specific case of $\mathfrak{m}=1$ as calculated numerically; in the underlying calculation, we have taken full account of the dependence of $\mathcal{C}_1$, Eq.~(\ref{ec252}), on $\upxi$. The value for $\mathfrak{m}$ that is relevant for the case of $\Delta \downarrow 1$ is deduced from the expressions in Eqs.~(\ref{e22}) and (\ref{e35}) (see Eqs.~(\ref{ec190}) and (\ref{ec226})). In Eq.~(\ref{e35}), $u$ is the renormalized Fermi velocity at $k=k_{\Sc f}$ in the limit $\Delta\downarrow 1$; according to Eq.~(\ref{ec123}), in this limit one has $u = \frac{\pi}{2}\, v_{\Sc f}$ (Eqs.~(\ref{ec40}) and (\ref{ec126})). In this connection, we point out that in Ref.~\cite{LZ97} (Eq.~(12) herein)
\begin{equation}
\mathfrak{m} = \frac{2 \Gamma\big(\frac{\upxi}{2}\big)}{\sqrt{\pi} \Gamma\big(\frac{1}{2} +\frac{\upxi}{2}\big)}\, \left(\frac{\pi \vert {\bm\mu}\vert\, \Gamma\big(\frac{1}{\upxi+1}\big)}{\Gamma\big(\frac{\upxi}{\upxi+1}\big)}\right)^{ (\upxi+1)/2}, \label{ee1}
\end{equation}
where ${\bm\mu}$ is given in Eq.~(\ref{ec229}); we have replaced the ${\bm\mu}$ in the original expression by $\vert {\bm\mu}\vert$, since sign of ${\bm\mu}$ can be altered by means of the transformation in Eq.~(\ref{ec80}). From the expression in Eq.~(\ref{ee1})  one readily obtains that (items 6.1.33 and 6.1.40 in Ref.~\cite{AS72})
\begin{equation}
\mathfrak{m} \sim \frac{2\,\sqrt{2 \e}}{\e^{\gamma}}\, \big(\pi \vert {\bm\mu}\vert \upxi\big)^{\upxi/2}\;\;\; \mbox{\rm as}\;\;\; \upxi \to\infty, \label{ee2}
\end{equation}
where $\gamma = 0.577\,215\,\dots$ is the Euler constant. This stronger than factorial growth in $\mathfrak{m}$ for $\upxi\to\infty$ is what one expects in the strong-coupling (SC) regime $\mathrm{z}_{\perp} \ll  -\mathrm{z}_{\parallel}$ (Sec.~\ref{sac.5a}). To appreciate this fact, one has to consider the RG result in Eq.~(\ref{ec217}) (recall that $\mathfrak{m} = \mathcal{M}/(2 u)$) and take into account that from Eqs.~(\ref{ec198}), (\ref{ec95}), (\ref{ec228}) and (\ref{ec253}) one has
\begin{equation}
-\frac{1}{\mathrm{z}_{\parallel}^0} \equiv  \frac{1}{2-d} \equiv \frac{1}{2 \big(1 - \upbeta^2/(8\pi)\big)} \equiv \frac{1}{2 (1-{\bm\upbeta}^2)} \equiv \frac{\upxi+1}{2}. \label{ee3}
\end{equation}
With reference to Eq.~(\ref{ec217}), note that (Eq.~(\ref{ec194})) \[ \mathrm{z}_{\perp}^0 \doteq \mathrm{z}_{\perp}(l=0) = \sqrt{8 A}\, \mathrm{z}(l=0),\] where $\mathrm{z}(l=0) \equiv \mathrm{z}$ is the dimensionless coupling constant presented in Eq.~(\ref{ec191}); one observes that indeed $\mathrm{z}_{\perp}^0$ is indeed proportional to $\vert {\bm\mu}\vert$, Eq.~(\ref{ec229}).

From Fig.~\ref{f3} it is evident that for a constant $\mathfrak{m}$ (i.e. independent of $\upxi$) $\sqrt{Z_2(\pm {\bm\upbeta}/4)}$ decreases like $1/\upxi^{3/16}$ for $\upxi\to\infty$; more precisely, for $\upxi \gtrsim 3$, $\sqrt{Z_2(\pm {\bm\upbeta}/4)}|_{\mathfrak{m}=1}$ turns out to be relatively accurately described by the power low $A/\upxi^{3/16}$, with $A=4.947\,289\,\dots$~. Our numerical calculations reveal that for a constant $\mathfrak{m}$, $\sqrt{Z_1(\pm {\bm\upbeta}/2)}$ \emph{increases} like $\upxi^{3/16}$ for $\upxi\to\infty$. Interestingly, from the explicit expression in Eq.~(4.24) of Ref.~\cite{LZ01}, one immediately obtains that, for $\mathfrak{m}$ constant, $\sqrt{Z_4(0)}$ decays like $1/\upxi$ for $\upxi\to \infty$. It follows that unless $\mathfrak{m}$ is made to vary with an appropriate power of $\upxi$, the \emph{conjectured} expression for the normalization constant $\sqrt{Z_n(s)}$ is either unbounded or vanishing in the \emph{limit} $\upxi=\infty$; in both cases, $\upxi=\infty$ amounts to a singular point of the theory.
\hfill $\square$

\end{appendix}


\bibliographystyle{apsrev}

\end{document}